\documentclass[journal,onecolumn,letterpaper,12pt]{IEEEtran}
\usepackage{cite}
\usepackage{amsmath,amssymb,amsthm}
\usepackage{mathrsfs}
\usepackage{algorithm}
\usepackage[noEnd=false,indLines=true,spaceRequire=false,italicComments=false,commentColor=black,
 beginLComment={},endLComment={}]{algpseudocodex}
\usepackage{booktabs}
\usepackage{array,multirow}
\newcolumntype{C}[1]{>{\centering\arraybackslash}m{#1}}
\newcommand{\papertablestyle}{%
  \small
  \renewcommand{\arraystretch}{1.08}%
  \setlength{\tabcolsep}{4pt}%
  \setlength{\heavyrulewidth}{0.8pt}%
  \setlength{\lightrulewidth}{0.5pt}%
  \setlength{\cmidrulewidth}{0.3pt}%
  \setlength{\aboverulesep}{1.4pt}%
  \setlength{\belowrulesep}{1.4pt}%
}
\usepackage{graphicx}
\usepackage{siunitx}
\usepackage{xcolor}
\definecolor{designretained}{HTML}{176B70}
\definecolor{designcoupling}{HTML}{A84416}
\usepackage{tikz}
\usetikzlibrary{arrows.meta,positioning}
\tikzset{algpxIndentLine/.style={draw=black!55,line width=0.3pt}}
\usepackage{pgfplots}
\pgfplotsset{compat=1.18}
\usepackage{url}
\usepackage{hyperref}
\hypersetup{
 colorlinks=true,
 linkcolor=blue,
 citecolor=blue,
 urlcolor=blue,
 bookmarksnumbered=true,
 pdftitle={Deterministic Polar Subcode Design via Exact Low-Weight Spectrum Characterization},
 pdfauthor={Junhua You, Shaohua Wu, Aolin Liu, Yajing Deng, and Qinyu Zhang},
 pdfsubject={Deterministic dynamic frozen expression design for polar subcodes},
 pdfkeywords={polar subcodes, dynamic frozen bits, weight distribution, minimum distance, list decoding}
}
\newcommand{\F}{\mathbb F_2}
\newcommand{\cI}{\mathcal I}
\newcommand{\cF}{\mathcal F}
\newcommand{\cW}{\mathcal W}
\newcommand{\EE}{\mathbb E}
\newcommand{\PP}{\mathbb P}
\newcommand{\ind}[1]{\mathbf 1_{\{#1\}}}
\DeclareMathOperator{\wt}{wt}
\DeclareMathOperator{\rk}{rank}
\DeclareMathOperator{\spn}{span}
\newcommand{\pc}{w_2}
\DeclareRobustCommand{\proc}[1]{\ifmmode\text{\normalfont\scshape #1}\else{\normalfont\scshape #1}\fi}
\theoremstyle{definition}
\newtheorem{theorem}{Theorem}
\newtheorem{proposition}{Proposition}

\newtheorem{corollary}{Corollary}
\newtheorem{example}{Example}

\algrenewcommand\algorithmicindent{0.85em}
\algrenewcomment[1]{\hfill{\scriptsize\ttfamily //~#1}}
\algrenewcommand\alglinenumber[1]{\scriptsize #1:}
\algrenewcommand\textproc[1]{\proc{#1}}
\algrenewcommand\algorithmicrequire{\textbf{Input:}}
\algrenewcommand\algorithmicensure{\textbf{Output:}}
\begin{document}
\title{Deterministic Polar Subcode Design via
Exact Low-Weight Spectrum Characterization}
\author{Junhua~You, Shaohua~Wu, Aolin~Liu, Yajing~Deng, and Qinyu~Zhang%
\thanks{J. You, S. Wu, A. Liu, Y. Deng, and Q. Zhang are with Harbin Institute of Technology (Shenzhen), Shenzhen, China (e-mail: \nolinkurl{youjunhua@stu.hit.edu.cn}, \nolinkurl{hitwush@hit.edu.cn}, \nolinkurl{liuaolin@stu.hit.edu.cn}, \nolinkurl{hitdengyj@163.com}, \nolinkurl{zqy@hit.edu.cn}).}%
\thanks{S. Wu, A. Liu, and Q. Zhang are also with Pengcheng Laboratory, Shenzhen, China. Corresponding author: Shaohua Wu (\nolinkurl{hitwush@hit.edu.cn}).}}
\markboth{MANUSCRIPT FOR REVIEWING}{You \MakeLowercase{\textit{et al.}}: Deterministic Polar Subcode Design}
\maketitle
\begin{abstract}
We develop a deterministic framework for improving polar-subcode
minimum distance through information-set selection and dynamic
frozen expression design. Expression design faces two difficulties.
Changing an expression can introduce new low-weight codewords, and
excluding different codewords can impose conflicting requirements on
the same coefficients. To address both difficulties, we first fix the
information set and release the editable frozen constraints to form a
\textit{common supercode} containing every candidate code.
Its low-weight codewords give necessary and sufficient conditions for
one coefficient assignment to exclude every codeword of the specified
weights. To apply these conditions at longer lengths, we use the polar
decomposition. At each split, a one in the left codeword gives two
complementary output bits whose combined weight is always one. We can
therefore eliminate the corresponding variables from the parity checks
and count their compatible assignments to obtain the exact low-weight
spectrum. With both constituent codes fixed, the same reduced checks
also simplify the coefficient search. Assignments that give identical
check values across all relevant messages share one representative,
from which the entire group can be recovered. We give sufficient
conditions for choosing the coefficients sequentially without backtracking.
Finally, we use the outcome of expression design on one information set
to guide exchanges of information positions for the next, linking the
two design steps.
In experiments spanning block lengths from $64$ to $\num{1024}$
and several code rates, the proposed design meets the target minimum
distances in minutes and achieves
competitive frame error rates under successive cancellation list decoding.
\end{abstract}
\section{Introduction}
Polar subcodes \cite{trifonov2013dynamic,trifonov2015polar} introduce
causal linear constraints, called dynamic frozen expressions, at the input
of the polar transform \cite{arikan2009polarization}. For a given information set, these
expressions provide freedom to improve the weight spectrum while
retaining the structure used by successive cancellation (SC) decoding.
The rate profile specifies the information set and is called the
\textit{profile} hereafter. Our objective
is minimum-distance improvement through profile selection and
expression design. We first characterize feasible coefficients
for a fixed profile and then use this characterization to guide
reliability-aware profile construction.

Distance improvement requires eliminating successive low-weight
layers. For each trial distance, we seek \textit{coefficient configurations}
that exclude every nonzero codeword below that distance. Two features
make this problem nontrivial.
Changing an expression can introduce codewords absent from the
current code. Moreover, exclusions that are individually feasible
may be incompatible because the same coefficients occur in many
codeword conditions. A complete description must account for both
the codewords that can arise and their shared dependence on the
design variables.

\subsection{Related Work}
\paragraph{Dynamic constraints and code design}
Canonical dynamic frozen expressions and distance-constrained
generator-row design are developed in
\cite{miloslavskaya2020short}. A \textit{common-supercode} approach uses a code containing all candidate
subcodes to identify low-weight codewords for dynamic-constraint selection
\cite{trifonov2020optimized}. Recursive precoded-polar construction
\cite{miloslavskaya2021recursive} combines prescribed subcodes and
supercodes with complete low-weight lists and generator extension
in decreasing information-position order. The subcode restricts the design space,
whereas the supercode enables distance testing through shorter
codes. These methods establish the connection between
low-weight codewords and expression design on which we build.
Compact specifications also separate frozen-set and expression
design, combining a structured frozen-set family with deterministic
expression generation from a binary sequence
\cite{miloslavskaya2024compact}. Randomly generated constraints
have also been selected against codewords below twice the
supercode minimum distance \cite{oreshin2024improved}.

Related constructions combine spectral design with additional code structure.
Row-merged polar codes combine minimum-weight enumeration with
greedy repetition-precoder design and profile modification
\cite{zunker2025rowmerged}. Nested precoded codes use
target-distance generator extensions with backtracking
\cite{coskun2025nested}, while deep polar codes eliminate selected
leading-row cosets through structured inner-row freezing
\cite{han2025deep}. Restrictions on dynamic dependencies also
control frozen positions under affine permutations
\cite{chernikov2026permutation}. Expressions derived from extended
Bose--Chaudhuri--Hocquenghem (eBCH) codes can also be represented
by equivalent parity checks at final list selection
\cite{liu2026ebchchecks}.

For polarization-adjusted convolutional (PAC) codes \cite{arikan2019sequential}, approximate
spectra guide profile design \cite{seyedmasoumian2022approximate}.
Other methods jointly optimize profiles and convolutional responses
\cite{abdullah2023afer,abdullah2023scl} or constrain profile design
to reduce sequential-decoding complexity \cite{moradi2024search}. In particular,
\cite{liu2026methodology} connects minimum-weight analysis,
efficient evaluation, and deterministic PAC optimization.
We build on this connection between weight structure and code
design, allowing general causal expressions in place of a common
convolutional response.

\paragraph{Weight structure and exact enumeration}
Structural descriptions explain how low-weight codewords are
formed and support both counting and design. Rowshan et al.\
\cite{rowshan2023formation} describe how minimum-weight codewords
arise in polar and PAC codes. Bardet et al.\ \cite{bardet2016monomial}
introduced the decreasing monomial-code framework. Within this
framework, \cite{dragoi2024closed} counts codewords at 1.5 times the
minimum distance in closed form. For decreasing polar codes,
\cite{ye2024belowtwice} uses a polynomial representation to count
codewords below twice the minimum distance, and \cite{dragoi2025group}
characterizes Type-I codewords in this range using affine-group actions.
Minimum-weight descriptions
for arbitrary information sets also support shortened-polar
construction \cite{kankanamge2026shortened}. The applicability of
these descriptions depends on the information-set or precoding
structure and on the weight range being characterized. Ensemble spectrum analysis
\cite{chiu2023design} provides a complementary description,
averaged over coefficient choices.

Exact algorithms address the spectrum of an individual code.
Polar-coset recursion computes weight enumerating functions, including
dynamic constraints \cite{yao2021deterministic}. Complete
bounded-weight enumeration for punctured, shortened, and precoded
polar codes combines constituent lists by syndrome matching and
accounts for multiplicities through ranks
\cite{miloslavskaya2022partial}. Tree intersection counts codewords
at the minimum row weight of the underlying polar code
\cite{zunker2023tree}. Other reductions group equal-spectrum cosets
of decreasing monomial codes \cite{liu2025complexity} or use
bounded-weight recursion \cite{ellouze2024low}.
Parity-consistent decomposition combines
constituent spectra and reduces representations through
equal-spectrum transformations \cite{liu2026pcd}, with recursive
weight-threshold pruning also developed for ordinary polar codes
\cite{liu2026pcdjournal}. SC-Fano traversal provides another
method for low-weight enumeration without a partial-order restriction on the
profile \cite{zhou2025fano}. Our polar-quotient enumeration draws on
polar-coset recursion \cite{yao2021deterministic} and syndrome matching
across constituents \cite{miloslavskaya2022partial}. For each left
codeword, we test the right coordinates that can change the weight and
recover all compatible assignments of the remaining coordinates by solving
the parity checks.

\paragraph{Profiles and finite-list decoding}
The weight spectrum does not alone determine performance under
successive cancellation list (SCL) decoding \cite{tal2015list}. The analysis in
\cite{coskun2022information} relates conditional entropy to typical
list-size requirements during SCL decoding. Spectrum and decoding
considerations are also distinguished in \cite{chiu2023design}.
Correct-path pruning and erroneous final selection are treated
separately in sparse pre-transform design \cite{choi2025sparse}
and BLER bound analysis \cite{liu2026blerbounds}.
\textit{Profile} design consequently combines
several criteria: path-probability weights for PAC codes
\cite{liu2025pathprobability}, reliability and minimum-weight
multiplicity for short polar codes \cite{you2026urllc}, spectral
and decoding-complexity criteria
\cite{frozen2025design,piao2025construction}, and SC and
ensemble-averaged maximum-likelihood performance estimates
\cite{miloslavskaya2025pareto}. PAC designs also use target-distance
and estimated-list-size constraints \cite{mrani2026nearml}, or
capacity-guided bit reassignment \cite{guo2026capacity}.
For profiles defining decreasing monomial codes, causal unit
upper-triangular pre-transformations preserve the minimum distance
of the unprecoded code \cite{zunker2025rowmerged,dragoi2026algebraic}.
To improve on the unprecoded distance, we therefore need an information
set whose unprecoded code is not a decreasing monomial code, together
with expression coefficients that attain the higher distance.

\subsection{Scope and Contributions}
We study binary, linear, unpunctured polar subcodes based on the
binary Ar{\i}kan transform, with arbitrary information sets.

\begin{enumerate}
\item \emph{Exact distance feasibility with shared coefficients.}
For a fixed information set, we give necessary and sufficient
conditions for one assignment of the dynamic frozen coefficients
to meet a prescribed minimum-distance target. A common supercode
covers every codeword that could occur under the permitted changes,
including low-weight codewords absent from the starting code
(Section~\ref{sec:elimination}).

\item \emph{Lossless reduction of the coefficient search.}
With both constituent codes fixed, we search one representative
for each group of assignments that give the same reduced checks
and recover all feasible assignments from these representatives.
We prove that these checks use the fewest independent linear
observations needed to determine the
truncated spectrum over the full space of coefficients linking the
two codes. The polar decomposition also enables complete low-weight
enumeration for construction and verification
(Sections~\ref{sec:elimination} and~\ref{sec:guarantees}).

\item \emph{Reliability-guided construction with feedback.}
We use the distance conditions to construct polar subcodes on
reliability-guided information sets and give sufficient conditions
for assigning the coefficients sequentially without backtracking.
Proved obstructions guide the release of fixed coefficients, and
feedback from constituent construction prioritizes exchanges of
information positions. Once the target is met, \textit{decoding-prefix
protection} adjusts early frozen expressions to improve path retention
under SCL decoding while preserving every codeword through a chosen
weight (Sections~\ref{sec:guarantees} and~\ref{sec:reliability-design}).
\end{enumerate}

Across code lengths from $64$ to $\num{1024}$ and several rates, the
proposed constructions raise the minimum distance of the corresponding
unprecoded codes to the design targets and achieve competitive frame
error rates (FERs) under SCL decoding. A controlled comparison
shows that early frozen expressions can lower FER even with the
information set and low-weight codewords held fixed.

Section~\ref{sec:model} defines the coefficient model and distance
objective. Section~\ref{sec:elimination} develops the exact
characterization, spectrum invariance, and complete enumeration.
Section~\ref{sec:guarantees} establishes coefficient recovery and
reduction, affine-domain preservation, and completion guarantees.
Section~\ref{sec:reliability-design} adds initial-domain selection
and obstruction feedback to fixed-profile construction, profile
selection, and expression refinement, and
analyzes the complexity.
Section~\ref{sec:example} presents the spectral and FER comparisons.
Section~\ref{sec:conclusion} concludes the paper.
The supplementary material includes implementations of the algorithms and
worked examples, together with the code configurations and data used in
Section~\ref{sec:example}.\footnote{The companion repository is available at
\url{https://github.com/barryu9/polar-subcode-design}.}

\section{Polar Subcodes and the Design Problem}
\label{sec:model}
We describe polar subcodes by their information positions and
frozen-expression coefficients, then formulate the distance-design problem.
\subsection{Polar Transform and Notation}
Using the polar transform of \cite{arikan2009polarization},
let $N=2^{n}$ for a nonnegative integer $n$ and let
$\mathbf{G}_{N}=\mathbf{F}^{\otimes n}$, where $\otimes$ is the Kronecker product and
$\mathbf{F}=\left[\begin{smallmatrix}1&0\\1&1\end{smallmatrix}\right]$.
There is no bit reversal, and $\mathbf{G}_{N}^{-1}=\mathbf{G}_{N}$.
Encoding is $\mathbf{x}=\mathbf{u}\mathbf{G}_{N}$, with input $\mathbf{u}\in\F^{N}$ and codeword
$\mathbf{x}\in\F^{N}$ written as rows. All linear algebra is over the binary field $\F$.
Bold lowercase Latin and bold Greek letters denote vectors. Bold uppercase
Latin letters denote matrices, and calligraphic or script letters denote sets
and codes. Scalar entries use ordinary type. Polar-transform
coordinates start at zero. The symbol $0$ also denotes a zero vector or matrix
when its dimensions are clear. Weights and counts are integers. Superscript $\mathrm{T}$ denotes
transposition, and $\mathbf I_{r}$ is the $r\times r$ identity matrix.
Write $[r]=\{0,\ldots,r-1\}$, $[i,j)=\{i,\ldots,j-1\}$,
and $\wt(\mathbf{x})$ for Hamming weight. The set
$\operatorname{supp}(\mathbf{x})$ contains its nonzero coordinate
indices. $\rk$ and $\spn$ denote binary rank and span.
The indicator $\ind{E}$ is one when
statement $E$ holds and zero otherwise. Row $\mathbf{g}_{i}$ of $\mathbf{G}_{N}$ has
weight $2^{\pc(i)}$, where $\pc(i)$ counts the ones in the
binary expansion of index $i$.
For $N\ge2$, let $h=N/2$ and write $\mathbf u=(\mathbf u_{\mathrm L},\mathbf u_{\mathrm R})$
with two length-$h$ input halves. The polar transform satisfies
\[
 \begin{gathered}
 \mathbf{G}_{2h}=\begin{bmatrix}\mathbf{G}_{h}&0\\\mathbf{G}_{h}&\mathbf{G}_{h}\end{bmatrix},\\
 (\mathbf{u}_{\mathrm{L}},\mathbf{u}_{\mathrm{R}})\mathbf{G}_{2h}
 =((\mathbf{u}_{\mathrm{L}}+\mathbf{u}_{\mathrm{R}})\mathbf{G}_{h},\mathbf{u}_{\mathrm{R}}\mathbf{G}_{h}).
 \end{gathered}
\]
For a vector, an index-set subscript restricts its coordinates.
For a matrix, $\mathbf{A}_{\mathcal{J},:}$ and $\mathbf{A}_{:,\mathcal{J}}$ select rows and columns,
respectively. A single set subscript abbreviates column selection.
The image and kernel of a matrix are taken on column vectors.
We use $\bigsqcup$ for a disjoint union, and $\PP$ and $\EE$
for probability and expectation. The code $\operatorname{RM}(r,n)$
is the binary Reed--Muller (RM) code of order $r$ and length $2^n$
\cite{muller1954boolean,reed1954class}.
In the polar-row coordinates used here,
$\operatorname{RM}(r,n)=\spn\{\mathbf g_i:\pc(i)\ge n-r\}$
\cite[Sec.~II-B1]{rowshan2023formation}.

An $[N,K]$ code has length $N$ and dimension $K$.
We write $[N,K,d]$ when its minimum distance is exactly $d$,
and $[N,K,\ge d]$ when only a lower bound is asserted.

\subsection{Fixed Profile and Dynamic Frozen Expressions}
Fix the information set $\cI\subseteq[N]$, $|\cI|=K$, and the frozen set
$\cF=[N]\setminus\cI$.
For $j\in\cF$, let $\mathcal{I}_{j}=\cI\cap[0,j)$,
$k_{j}=|\mathcal{I}_{j}|$.
A canonical dynamic frozen expression
\cite{trifonov2015polar}, \cite[Sec.~III-B]{miloslavskaya2020short}
and its check row are
\begin{equation}
 u_{j}=\mathbf{u}_{\mathcal{I}_{j}}\mathbf{s}_{j},\qquad
 \mathbf{h}_{j}(\mathbf{s})=\mathbf{e}_{j}+\sum_{i\in \mathcal{I}_{j}}s_{ji}\mathbf{e}_{i} ,
 \label{eq:canonical}
\end{equation}
where $\mathbf{s}_{j}\in\F^{k_{j}}$ is a column with components $s_{ji}$
indexed by increasing $i\in\mathcal{I}_{j}$, and $\mathbf{e}_{j}$ is a standard basis row.
The \textit{expression support} $\{i\in \mathcal{I}_{j}:s_{ji}=1\}$ uniquely
specifies $\mathbf{s}_{j}$ over $\F$.
The case $\mathbf{s}_{j}=0$ is the static frozen constraint $u_{j}=0$.
When indexing the coefficients of a frozen expression, we call an
information position a \textit{source}.
Stacking the rows, labeled by their frozen positions,
gives $\mathbf{H}(\mathbf{s})\in\F^{(N-K)\times N}$ and the code
\begin{equation}
 \begin{aligned}
 \mathcal{C}_{\mathbf{s}}&=\{\mathbf{u}\mathbf{G}_{N}:\mathbf{H}(\mathbf{s})\mathbf{u}^{\mathrm{T}}=0\},\\
 \mathbf{s}&\in\Omega_{\cI}:=\F^{p_{\cI}},\qquad p_{\cI}=\sum_{j\in\cF}k_{j}.
 \end{aligned}
 \label{eq:code}
\end{equation}
Here $\mathbf{s}$, formed by concatenating the $\mathbf{s}_{j}$ in increasing $j$,
is the full \textit{coefficient configuration}. A code design
is specified by the pair $(\cI,\mathbf{s})$.
List $\cI=\{i_{0}<\cdots<i_{K-1}\}$ and
$\cF=\{j_{0}<\cdots<j_{N-K-1}\}$.
Physical information position $i_{r}$ corresponds to message
coordinate $r$. The coefficient matrix $\mathbf{S}\in\F^{(N-K)\times K}$
has entries
$S_{\ell r}=s_{j_{\ell} i_{r}}$ for $i_{r}<j_{\ell}$, and zero
otherwise. With columns temporarily ordered as $(\cI,\cF)$,
the input generator and check matrices are
\begin{equation}
 \mathbf{B}(\mathbf{s})=[\,\mathbf I_{K}\ \mathbf{S}^{\mathrm{T}}\,],\qquad \mathbf{H}(\mathbf{s})=[\,\mathbf{S}\ \mathbf I_{N-K}\,].
 \label{eq:systematic-input}
\end{equation}
Restore the natural column order before multiplying by $\mathbf{G}_{N}$.
For the information column $\mathbf{m}\in\F^{K}$, $\mathbf{u}=\mathbf{m}^{\mathrm{T}}\mathbf{B}(\mathbf{s})$.
Thus $\mathbf{B}(\mathbf{s})\mathbf{H}(\mathbf{s})^{\mathrm{T}}=0$, $\dim \mathcal{C}_{\mathbf{s}}=K$, and the codeword
generator is $\mathbf{B}(\mathbf{s})\mathbf{G}_{N}$. The systematic input representation
and invertibility of $\mathbf{G}_{N}$ ensure that distinct \textit{coefficient
configurations} give distinct codes.
Setting $\mathbf{s}=0$, equivalently $\mathbf{S}=0$, gives the unprecoded polar
code on $\cI$.

\begin{figure*}[!t]
\centering
\includegraphics[width=\textwidth]{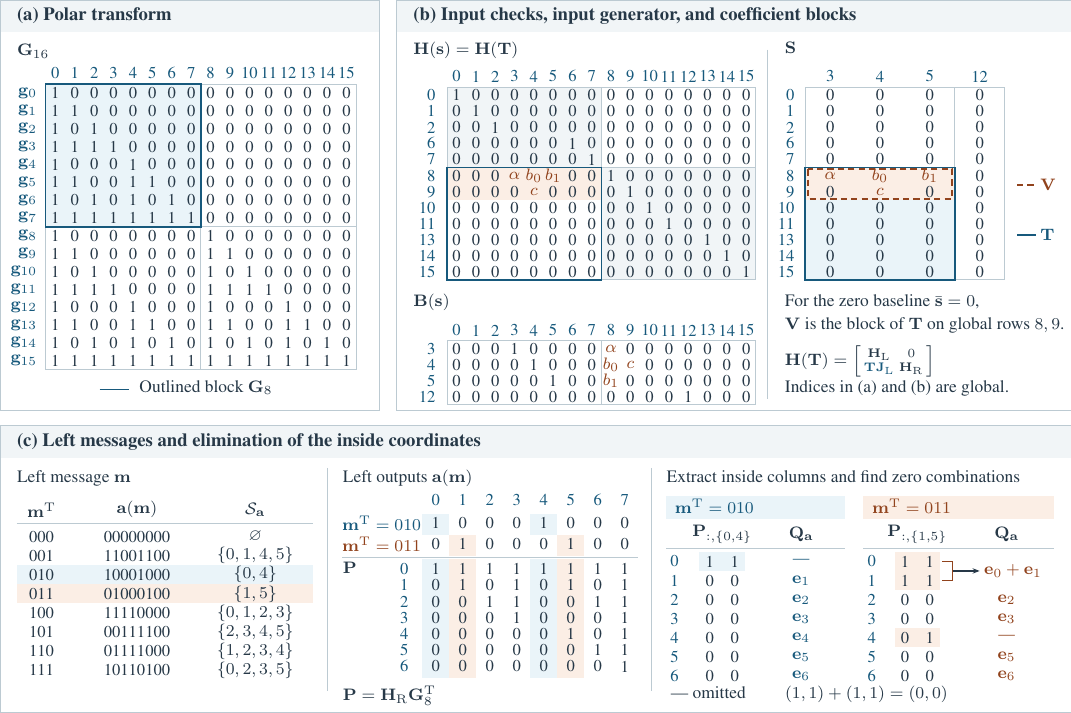}
\caption{Matrices for the length-16 running example.
(a) The outlined block is $\mathbf G_8$.
(b) Input matrices $\mathbf B(\mathbf s),\mathbf H(\mathbf s)$ used in Example~\ref{ex:canonical}, and blocks used in the later polar split: $\mathbf T$ links left information bits to right frozen bits, and its two edited rows form the zero-baseline increment $\mathbf V$.
(c) The right-codeword check matrix $\mathbf P$ and row combinations $\mathbf Q_{\mathbf a}$ canceling its inside columns. The ones of the left-constituent codeword $\mathbf a(\mathbf m)$ select those columns. Each $\mathbf e_i$ is the length-7
unit row at packed check index $i$.}
\label{fig:n16-data}
\end{figure*}

\begin{example}[A length-16 running example]
\label{ex:canonical}
Take $N=16$ and $\cI=\{3,4,5,12\}$. The two variable
frozen expressions are
\[
 u_8=\alpha u_3+b_0u_4+b_1u_5,\qquad u_9=c\,u_4,
 \qquad \alpha,b_0,c,b_1\in\F.
\]
All other frozen bits are zero. Fig.~\ref{fig:n16-data}(b) shows
$\mathbf H(\mathbf s)$ and $\mathbf B(\mathbf s)$ in natural
coordinate order. Multiplying $\mathbf B(\mathbf s)$ by
$\mathbf G_{16}$ in Fig.~\ref{fig:n16-data}(a) replaces each
unit row $\mathbf e_i$ by $\mathbf g_i$. For instance, the first
row $\mathbf e_3+\alpha\mathbf e_8$ gives
$\mathbf g_3+\alpha\mathbf g_8$.
Examples~\ref{ex:canonical}--\ref{ex:toy-completion} use this family.
Fig.~\ref{fig:n16-data}(c) gives the projections used in Example~\ref{ex:shared-observations}.
\end{example}

\subsection{Distance Objective and Cutoff-Preserving Reduction}
For a nonzero linear code $\mathcal{C}$, $d_{\min}(\mathcal{C})$ is the smallest
Hamming weight of a nonzero codeword.
For the zero code, we set $d_{\min}(\{0\})=+\infty$.
The leading position of a nonzero input is its first nonzero
coordinate index. For every $\mathcal H\subseteq[i+1,N)$,
the polar coset bound \cite[Cor.~5]{rowshan2023formation} gives
\[
 \wt\!\left(\mathbf{g}_{i}+\sum_{j\in\mathcal H}\mathbf{g}_{j}\right)
 \ge\wt(\mathbf{g}_{i})=2^{\pc(i)}.
\]
Equality can hold for different choices of $\mathcal H$. For example,
both $\mathbf g_3$ and $\mathbf g_3+\mathbf g_7$ in $\mathbf G_{16}$
have weight four.
For a nonempty profile, the leading position of any nonzero input $\mathbf u$ satisfying $\mathbf H(\mathbf s)\mathbf u^{\mathrm T}=0$ lies in $\cI$, so
\begin{equation}
 d_{\min}(\mathcal{C}_{\mathbf{s}})\ge d_{\mathrm P}
 :=\min_{i\in\cI}2^{\pc(i)} .
 \label{eq:profile-bound}
\end{equation}
The unprecoded code contains $\mathbf g_i$ for each $i\in\cI$, so its minimum
distance is $d_{\mathrm P}$.
If $0\notin\cI$, causality forces $u_0=0$. Every remaining polar row
has even weight, so their sums also have even weight.
For a binary code $\mathcal{C}\subseteq\F^N$, write
$A_{w}(\mathcal{C})=|\{\mathbf{x}\in \mathcal{C}:\wt(\mathbf{x})=w\}|$ and $A_{w}(\mathbf{s})=A_{w}(\mathcal{C}_{\mathbf{s}})$.
The sequence $(A_{w}(\mathcal{C}))_{w=0}^{N}$ is its \textit{weight spectrum}.
Its \textit{weight enumerating function} (WEF) is
$W_{\mathcal C}(X)=\sum_{w=0}^{N}A_w(\mathcal C)X^w$, with formal variable $X$.
For an integer cutoff $0\le D\le N$, the vector
$\boldsymbol\Sigma_D(\mathcal C)=(A_1(\mathcal C),\ldots,A_D(\mathcal C))$
is the truncated spectrum through $D$, and the truncated WEF is
$W_{\mathcal C,\le D}(X)=\sum_{w=0}^{D}A_w(\mathcal C)X^w$.
For a linear code $A_0(\mathcal C)=1$.
For a nonzero code $\mathcal C$ with minimum distance $d_{\min}$,
$A_{d_{\min}}(\mathcal C)$ is its minimum-weight multiplicity
(also called the \textit{error coefficient}).
A weight layer is the set of codewords of a
specified weight, and may be empty.

For a nonempty fixed profile, the design objective is
\begin{equation}
 d_{\mathrm{opt}}(\cI)=\max_{\mathbf{s}\in\Omega_{\cI}}d_{\min}(\mathcal{C}_{\mathbf{s}}).
 \label{eq:distance-optimization}
\end{equation}
For any profile and integer trial distance $d$, define the feasible set
\begin{equation}
 \mathscr F_{d}(\cI)=\{\mathbf{s}\in\Omega_{\cI}:A_{w}(\mathbf{s})=0,\ 1\le w<d\}.
 \label{eq:feasible-set}
\end{equation}
We abbreviate it as $\mathscr F_{d}$ when the profile is fixed.
The nesting $\mathscr F_{d+1}\subseteq\mathscr F_d$ expresses
distance improvement as a sequence of feasibility problems.

An admissible domain $\mathcal S\subseteq\Omega_{\cI}$
specifies the \textit{coefficient configurations} allowed during a design step.
It can fix some coefficients or impose relations among them.
For example, the family in Example~\ref{ex:canonical} fixes every
coefficient except $\alpha,b_0,c,b_1$ to zero.
A configuration is admissible if it belongs to $\mathcal S$.
Fix a trial $2\le d_{\star}\le N$ and put $D=d_{\star}-1$.
An admissible configuration is feasible for this trial if
\begin{equation}
 \mathbf{s}\in\mathcal S\cap\mathscr F_{d_\star}(\cI).
 \label{eq:target}
\end{equation}
Proven infeasibility at $d_{\star}$ in a strict subset
$\mathcal S\subsetneq\Omega_{\cI}$ rules out the target within $\mathcal S$.
To infer $d_{\mathrm{opt}}(\cI)<d_\star$, the restriction must preserve existence: $\mathscr F_{d_\star}(\cI)\ne\varnothing$ implies $\mathcal S\cap\mathscr F_{d_\star}(\cI)\ne\varnothing$.
To test the distance target, let $\cW\subseteq\{1,\ldots,D\}$
be the forbidden weights whose elimination remains to be enforced.
Any weight in $\{1,\ldots,D\}$ omitted from $\cW$ must already
be excluded throughout $\mathcal S$. Once $d_{\min}\ge d_{\star}$ is established, a verified codeword
of weight $d_{\star}$, called a \textit{witness}, establishes equality.

By the polar coset bound, an input producing a codeword of weight
at most $D$ must start at an information position with row weight
at most $D$. Define
\begin{equation}
 \begin{aligned}
 i_D&=\min\{i\in\cI:2^{\pc(i)}\le D\},\\
 \mathcal E_D&=\{(j,i):j\in\cF,\ i\in\cI,\ i_D\le i<j\},
 \end{aligned}
 \label{eq:cutoff-coordinates}
\end{equation}
with $i_D=N$ when the set defining the minimum is empty.
\begin{proposition}[Cutoff-preserving reduction]
\label{prop:reduction}
\textit{Coefficient configurations} agreeing on $\mathcal E_{D}$ have exactly the same
codewords of positive weight at most $D$.
In $\Omega_{\cI}$, the coefficients $s_{ji}$ with \textit{source} $i<i_D$
(those outside $\mathcal E_D$) may therefore
be set to zero without changing any attainable truncated spectrum.
\end{proposition}
\begin{IEEEproof}
Let $\mathbf{x}=\mathbf{u}\mathbf{G}_{N}\in \mathcal{C}_{\mathbf{s}}$ with $0<\wt(\mathbf{x})\le D$, and let
$i=\min\operatorname{supp}(\mathbf{u})$. Causality and the polar coset
bound give
\[
 i\in\cI,\qquad 2^{\pc(i)}\le\wt(\mathbf{x})\le D,
 \qquad i\ge i_{D},\quad \mathbf{u}_{[0,i_{D})}=0.
\]
For configurations $\mathbf{s},\mathbf{s}'$ agreeing on $\mathcal E_{D}$, every frozen
check therefore satisfies
\[
 (\mathbf{h}_{j}(\mathbf{s})+\mathbf{h}_{j}(\mathbf{s}'))\mathbf{u}^{\mathrm{T}}
   =\sum_{\substack{k\in \mathcal{I}_{j}\\k<i_{D}}}(s_{jk}+s'_{jk})u_{k}=0.
\]
Hence $\mathbf{H}(\mathbf{s}')\mathbf{u}^{\mathrm{T}}=\mathbf{H}(\mathbf{s})\mathbf{u}^{\mathrm{T}}=0$. This proves inclusion of the
cutoff codeword sets. Interchanging $\mathbf{s},\mathbf{s}'$ proves equality.
\end{IEEEproof}

\begin{example}[Cutoff and coefficient constraints]
\label{ex:cutoff-qualification}
In Example~\ref{ex:canonical}, take $D=3$.
Among the information rows in Fig.~\ref{fig:n16-data}(a), only row 4
has weight at most $D$, so $i_D=4$.
An input with positive output weight at most $D$ must therefore
have $u_3=0$.
Proposition~\ref{prop:reduction} therefore omits $\alpha$ while
retaining $b_0,b_1,c$.
A \textit{source} $i>i_D$ can still affect the cutoff spectrum, even
when its own generator row has weight greater than $D$.
For instance, take
$u_4=u_5=1$, $u_3=u_{12}=0$, and $b_0=0,c=1$.
The output is $\mathbf g_4+\mathbf g_5+\mathbf g_9+b_1\mathbf g_8$.
Taking the symmetric difference of the row supports in
Fig.~\ref{fig:n16-data}(a) gives $\{0,5,8,9\}$ for $b_1=0$.
For $b_1=1$, adding $\mathbf g_8$ toggles positions $0,8$,
leaving $\{5,9\}$.
In both cases the input starts at \textit{source} $4=i_D$, while
$u_5=1$. Thus the row-weight bound at the later \textit{source} 5
does not exclude this input, and its coefficient $b_1$ cannot
be discarded from the cutoff calculation.
Now impose the additional coefficient constraint $\alpha=b_1$
on the same family. The retained choice $b_1=1$ must then be
completed with $\alpha=1$. Setting $\alpha=0$ would leave the
admissible domain, even though it preserves the cutoff codeword set.
Thus coefficients irrelevant to the cutoff may be omitted from
the spectral calculation, but their recovered values must still
satisfy every imposed coefficient constraint.
\end{example}

\section{Exact Low-Weight Characterization and Enumeration}
\label{sec:elimination}
We first characterize the shared coefficients that exclude every
potential low-weight codeword, including those introduced by coefficient
changes. The polar decomposition then gives a reduced spectrum
representation, its invariant coefficient directions, and a complete
enumeration procedure.

\subsection{Complete Conditions on Shared Coefficients}
Let $\mathcal{J}=\{j_{1}<\cdots<j_{b}\}\subseteq\cF$ index the frozen
expressions allowed to change, and fix a baseline configuration $\bar{\mathbf{s}}$.
Write $\mathbf{H}_{\mathcal{J},:}(\bar{\mathbf{s}})$ for those check rows and
$\mathbf{H}_{\bar{\mathcal{J}},:}(\bar{\mathbf{s}})$ for the unchanged rows,
where $\bar{\mathcal{J}}=\cF\setminus \mathcal{J}$. Releasing the checks in $\mathcal J$
gives a \textit{common supercode} containing every candidate code.
To form the \textit{complete low-weight directory} $\mathcal U_{\mathcal J}(\cW)$,
enumerate the inputs of this supercode with output weight in $\cW$.
A first nonzero bit at a frozen position cannot equal a sum of earlier,
zero information bits. We therefore retain only inputs whose leading
position belongs to $\cI$ and use the same directory notation after
this filter. With $\widetilde{\mathcal U}_{\mathcal J}$ denoting the
full input space of the common supercode, we have
\begin{equation}
 \begin{aligned}
 \widetilde{\mathcal U}_{\mathcal{J}}&=\{\mathbf{u}\in\F^{N}:\mathbf{H}_{\bar{\mathcal{J}},:}(\bar{\mathbf{s}})\mathbf{u}^{\mathrm{T}}=0\},\\
 \mathcal U_{\mathcal{J}}(\cW)&=\{\mathbf{u}\in\widetilde{\mathcal U}_{\mathcal{J}}:\wt(\mathbf{u}\mathbf{G}_{N})\in\cW,\;
 \min\operatorname{supp}(\mathbf u)\in\cI\}.
 \end{aligned}
 \label{eq:relaxation}
\end{equation}
The remaining $N-K-b$ checks retain their identity columns and hence full row rank,
giving $\dim\widetilde{\mathcal U}_{\mathcal{J}}=K+b$.
In particular, the directory covers inputs of low-weight codewords
introduced by coefficient changes, not just those present at the baseline.

Let $\mathcal E\subseteq\{(j,i):j\in \mathcal{J},\ i\in\cI,\ i<j\}$
be the editable pairs and
$\mathcal I_{\mathrm{e}}=\{i_{1}<\cdots<i_{e}\}$ their information \textit{sources}.
The coefficient increments relative to the fixed baseline
$\bar{\mathbf s}$ form the shared matrix
\begin{equation}
 \mathbf{V}=[\mathbf{v}_{i_{1}}\ \cdots\ \mathbf{v}_{i_{e}}]\in\F^{b\times e},
 \qquad (\mathbf{v}_{i})_{t}=s_{j_{t}i}+\bar{s}_{j_{t}i},
 \label{eq:shared-array}
\end{equation}
with zero increments outside $\mathcal E$.
Let $\mathscr D$ be the set of admissible increment matrices, including
any imposed coefficient relations. With a zero baseline, the restriction
$\alpha=b_1$ in Example~\ref{ex:cutoff-qualification} requires the two
corresponding entries of $\mathbf V$ to be equal.
For each \textit{directory} input put
\begin{equation}
 \mathbf{q}(\mathbf{u})=(u_{i_{1}},\ldots,u_{i_{e}}),\qquad
 \boldsymbol{\beta}(\mathbf{u})=\mathbf{H}_{\mathcal{J},:}(\bar{\mathbf{s}})\mathbf{u}^{\mathrm{T}}\in\F^{b}.
 \label{eq:common-model}
\end{equation}
The \textit{source pattern} $\mathbf q(\mathbf u)$ is the input-bit row
on the editable \textit{sources}.
The column $\boldsymbol\beta(\mathbf u)$ is the
\textit{baseline check residual} on the released rows.
Let $\mathcal Q=\{\mathbf q(\mathbf u):\mathbf u\in\mathcal U_{\mathcal J}(\cW)\}$
contain all \textit{source patterns} occurring in the \textit{directory}. For
$\mathbf{q}\in\mathcal Q$, $\boldsymbol{\beta}\in\F^{b}$, and $w\in\cW$, define
\begin{equation}
 \begin{aligned}
 \mathcal B(\mathbf{q})&=\{\boldsymbol{\beta}(\mathbf{u}):\mathbf{u}\in\mathcal U_{\mathcal{J}}(\cW),\ \mathbf{q}(\mathbf{u})=\mathbf{q}\},\\
 M_{w}(\mathbf{q},\boldsymbol{\beta})&=|\{\mathbf{u}\in\mathcal U_{\mathcal{J}}(\cW):\mathbf{q}(\mathbf{u})=\mathbf{q},\;
 \boldsymbol{\beta}(\mathbf{u})=\boldsymbol{\beta},\ \wt(\mathbf{u}\mathbf{G}_{N})=w\}|.
 \end{aligned}
 \label{eq:pattern-directory}
\end{equation}
Inputs with the same \textit{source pattern} receive the same check increment
$\mathbf V\mathbf q^{\mathrm T}$. Thus $\mathcal B(\mathbf q)$ records
the residual values to avoid, while $M_w$ retains their multiplicities
for exact counting. Unrecorded multiplicities are zero.

\begin{theorem}[Complete elimination on shared coefficients]
\label{thm:shared-elimination}
Let $\mathbf{V}\in\mathscr D$ and let $\mathbf{s}$ be the \textit{coefficient configuration}
obtained from $\bar{\mathbf{s}}$ by the increments in $\mathbf{V}$.
Then $\mathcal{C}_{\mathbf{s}}$ contains no codeword of weight in $\cW$ if and only if
\begin{equation}
 \mathbf{V}\mathbf{q}^{\mathrm{T}}\notin\mathcal B(\mathbf{q})\qquad\text{for every }\mathbf{q}\in\mathcal Q.
 \label{eq:source-rejection}
\end{equation}
For each $w\in\cW$, the number of weight-$w$ codewords is
\begin{equation}
 \begin{aligned}
 A_{w}(\mathbf{s})
 &=\sum_{\substack{\mathbf{u}\in\mathcal U_{\mathcal{J}}(\cW)\\\wt(\mathbf{u}\mathbf{G}_{N})=w}}
                  \ind{\boldsymbol{\beta}(\mathbf{u})=\mathbf{V}\mathbf{q}(\mathbf{u})^{\mathrm{T}}}\\
 &=\sum_{\mathbf{q}\in\mathcal Q}M_{w}(\mathbf{q},\mathbf{V}\mathbf{q}^{\mathrm{T}}).
 \end{aligned}
 \label{eq:pattern-count}
\end{equation}
In particular, taking $\cW=\{1,\ldots,d_{\star}-1\}$ makes
\eqref{eq:source-rejection} necessary and sufficient for
$d_{\min}(\mathcal{C}_{\mathbf{s}})\ge d_{\star}$.
\end{theorem}
\begin{IEEEproof}
We first identify which \textit{directory} inputs generate codewords in $\mathcal{C}_{\mathbf{s}}$.
For $\mathbf{u}\in\mathcal U_{\mathcal{J}}(\cW)$, the unchanged checks
$\mathbf H_{\bar{\mathcal J},:}(\bar{\mathbf s})\mathbf u^{\mathrm T}$
vanish. Hence
\[
 \begin{aligned}
 \mathbf{H}_{\mathcal{J},:}(\mathbf{s})\mathbf{u}^{\mathrm{T}}
 &=\mathbf{H}_{\mathcal{J},:}(\bar{\mathbf{s}})\mathbf{u}^{\mathrm{T}}+\sum_{i\in\mathcal I_{\mathrm{e}}}\mathbf{v}_{i} u_{i}\\
 &=\boldsymbol{\beta}(\mathbf{u})+\mathbf{V}\mathbf{q}(\mathbf{u})^{\mathrm{T}},\\
 \mathbf{u}\mathbf{G}_{N}\in \mathcal{C}_{\mathbf{s}}
 &\Longleftrightarrow \mathbf{H}(\mathbf{s})\mathbf{u}^{\mathrm{T}}=0
 \Longleftrightarrow \boldsymbol{\beta}(\mathbf{u})=\mathbf{V}\mathbf{q}(\mathbf{u})^{\mathrm{T}}.
 \end{aligned}
\]
Conversely, every $\mathbf{x}\in \mathcal{C}_{\mathbf{s}}$ with $\wt(\mathbf{x})\in\cW$ has a unique
input $\mathbf{u}=\mathbf{x}\mathbf{G}_{N}^{-1}\in\widetilde{\mathcal U}_{\mathcal{J}}$.
Its leading position lies in $\cI$ by causality, so it belongs to
$\mathcal U_{\mathcal{J}}(\cW)$ by \eqref{eq:relaxation}.

By this coverage and the membership criterion, all weights in $\cW$
are absent exactly when $\mathbf V\mathbf q^{\mathrm T}$ avoids
$\mathcal B(\mathbf q)$ for every $\mathbf q\in\mathcal Q$,
proving \eqref{eq:source-rejection}.

For a fixed $w\in\cW$, invertibility of $\mathbf{G}_{N}$ shows that
counting surviving inputs counts codewords without repetition.
The membership equivalence gives the first sum in
\eqref{eq:pattern-count}. Grouping its terms by $\mathbf{q}(\mathbf{u})$ and
$\boldsymbol{\beta}(\mathbf{u})$ gives
\[
 \begin{aligned}
 A_{w}(\mathbf{s})
 &=\sum_{\mathbf{q}\in\mathcal Q}\sum_{\boldsymbol{\beta}\in\F^{b}}
       M_{w}(\mathbf{q},\boldsymbol{\beta})\ind{\boldsymbol{\beta}=\mathbf{V}\mathbf{q}^{\mathrm{T}}}\\
 &=\sum_{\mathbf{q}\in\mathcal Q}M_{w}(\mathbf{q},\mathbf{V}\mathbf{q}^{\mathrm{T}}),
 \end{aligned}
\]
which proves the second equality. Finally, a linear code has
minimum distance at least $d_{\star}$ exactly when it contains
no codeword of weight $1,\ldots,d_{\star}-1$.
\end{IEEEproof}

In particular, $0\in\mathcal B(0)$ is an obstruction that no editable coefficient can remove, because $\mathbf V0=0$ for every $\mathbf V$.
\label{subsec:expression-views}
With one editable row, exclusion requires
$\mathbf V\mathbf q(\mathbf u)^{\mathrm T}
=1+\boldsymbol\beta(\mathbf u)$ for every \textit{directory} input,
a binary linear system in that row's coefficients. With several
editable rows, each forbidden input must violate at least one
updated frozen check $\mathbf h_j(\mathbf s)\mathbf u^{\mathrm T}=0$, $j\in\mathcal J$.

\begin{example}[Direct enumeration and grouping of the complete directory]
\label{ex:direct-directory}
Use the four-coefficient family of Example~\ref{ex:canonical} with
$d_\star=4$ and $D=3$. From the zero-coefficient baseline, release
the checks at $\mathcal J=\{8,9\}$. The free input positions
$\{3,4,5,8,9,12\}$ of the \textit{common supercode} index even-weight
rows of $\mathbf G_{16}$, so $\mathcal W=\{2\}$ suffices. Summing the
selected rows for the $2^6$ free assignments gives the following
nine inputs with weight-2 outputs before the leading-position filter. Each pair of support lists
is in matching order. The last two columns give
$\mathbf q(\mathbf u)=(u_3,u_4,u_5)$ and
$\boldsymbol\beta(\mathbf u)^{\mathrm T}=(u_8,u_9)$ as bit strings.
\[
 \begin{array}{c|c|cc}
 \operatorname{supp}(\mathbf u)&\operatorname{supp}(\mathbf x)
     &\mathbf q&\boldsymbol\beta^{\mathrm T}\\\hline
 \{8\},\ \{8,12\}&\{0,8\},\ \{4,12\}&000&10\\
 \{8,9\}&\{1,9\}&000&11\\\hline
 \{4\},\ \{4,12\}&\{0,4\},\ \{8,12\}&010&00\\
 \{4,8\},\ \{4,8,12\}&\{4,8\},\ \{0,12\}&010&10\\
 \{4,5\}&\{1,5\}&011&00\\
 \{4,5,8,9\}&\{5,9\}&011&11
 \end{array}
\]
The first three inputs start at a released frozen position and are removed by the leading-position filter, leaving six inputs in $\mathcal U_{\mathcal J}(\{2\})$. Equivalently, their $\mathbf q=0$ gives $\mathbf V\mathbf q^{\mathrm T}=0$, which cannot cancel their nonzero residual $\boldsymbol\beta$.
For $\mathbf q=010$, the residuals $00,10$ each occur twice.
Multiplying $\mathbf V$ from Fig.~\ref{fig:n16-data}(b) by
$\mathbf q^{\mathrm T}$ gives $(b_0,c)^{\mathrm T}$,
which matches one of them exactly when $c=0$.
For $\mathbf q=011$, the residuals $00,11$ each occur once.
The increment $(b_0+b_1,c)^{\mathrm T}$ matches one exactly when
$b_0+b_1=c$. Theorem~\ref{thm:shared-elimination} gives
\[
 \begin{aligned}
 A_2&=2\times\ind{c=0}+\ind{b_0+b_1=c},\\
 d_{\min}\ge4&\ \Longleftrightarrow\ c=1,\ b_1=b_0.
 \end{aligned}
\]
The four feasible configurations are $(\alpha,b_0,c,b_1)
=(\alpha,b_0,1,b_0)$ with $(\alpha,b_0)\in\F^2$.
The message with only $u_{12}=1$
produces the weight-4 row $\mathbf g_{12}$ in Fig.~\ref{fig:n16-data}(a)
for each, proving $d_{\min}=4$.
\end{example}

For larger polar subcodes the \textit{directory} can be too large to
handle input by input. We now fix the coefficients within each half
and vary only the coefficients linking the halves. For a given left
message this lets us group right-half check values that give the same
extension counts.

\subsection{Polar Quotients and the Exact Spectrum}
\label{subsec:polar-quotients}
For a polar split $N=2h$, we first remove from the right-half checks
the contributions of coordinates whose values do not affect the full
codeword weight. We then represent the shared coefficients linking the
two halves by the checks that remain across messages. These two maps
form the \textit{polar quotient representation}.
\subsubsection{From constituent checks to right-code cosets}
Using the polar split of Section~\ref{sec:model}, write
$\mathbf u\mathbf G_{2h}=(\mathbf a+\mathbf x_{\mathrm R},\mathbf x_{\mathrm R})$,
where $\mathbf a=\mathbf u_{\mathrm L}\mathbf G_h$ and
$\mathbf x_{\mathrm R}=\mathbf u_{\mathrm R}\mathbf G_h$ are length-$h$ rows.

Fix the coefficients $s_{ji}$ for which $i$ and $j$ lie in the same half.
Let $k_{\mathrm L},k_{\mathrm R}$
be the numbers of left and right information positions, and
$f=h-k_{\mathrm R}$. We call the dependence of right frozen bits on
left information bits coupling. The \textit{outer coupling matrix}
$\mathbf T=(s_{ji})\in\F^{f\times k_{\mathrm L}}$ is the block of
$\mathbf S$ with rows $j\in\cF\cap[h,2h)$ and columns
$i\in\cI\cap[0,h)$, in increasing order. These entries are causal
because left positions precede right positions.
Setting $\mathbf T=0$ gives the left and right constituent codes
$\mathcal C_{\mathrm L},\mathcal C_{\mathrm R}$, with canonical input check matrices
$\mathbf H_{\mathrm L},\mathbf H_{\mathrm R}$, respectively.
The left information column $\mathbf m\in\F^{k_{\mathrm L}}$ uniquely
determines $\mathbf u_{\mathrm L}$ and $\mathbf a(\mathbf m)$.
Let $\mathbf J_{\mathrm L}$ select the left information coordinates.
In natural input order, the full check matrix $\mathbf H(\mathbf s)$ is
\begin{equation}
 \mathbf{H}(\mathbf{T})=\begin{bmatrix}
       \mathbf{H}_{\mathrm{L}}&0\\ \mathbf{T}\mathbf{J}_{\mathrm{L}}&\mathbf{H}_{\mathrm{R}}
      \end{bmatrix},\qquad \mathbf{J}_{\mathrm{L}}\mathbf{u}_{\mathrm{L}}^{\mathrm{T}}=\mathbf{m}.
 \label{eq:assembled-check}
\end{equation}
The block row $[\,\mathbf T\mathbf J_{\mathrm L}\ \mathbf H_{\mathrm R}\,]$
in \eqref{eq:assembled-check} contains the checks for the right frozen
positions. These checks impose
$\mathbf H_{\mathrm R}\mathbf u_{\mathrm R}^{\mathrm T}=\mathbf T\mathbf m$.
Since $\mathbf u_{\mathrm R}=\mathbf x_{\mathrm R}\mathbf G_h$, define
$\mathbf P=\mathbf H_{\mathrm R}\mathbf G_h^{\mathrm T}\in\F^{f\times h}$.
This is the full-row-rank parity-check matrix of $\mathcal C_{\mathrm R}$
in codeword coordinates. The right half $\mathbf x_{\mathrm R}$ of a
coupled codeword satisfies $\mathbf P\mathbf x_{\mathrm R}^{\mathrm T}=\mathbf T\mathbf m$,
giving the code $\mathcal C_{\mathbf s}$, now written as $\mathcal C(\mathbf T)$
with $\mathbf H_{\mathrm L}$ and $\mathbf H_{\mathrm R}$ fixed:
\begin{equation}
 \mathcal{C}(\mathbf{T})=\bigsqcup_{\mathbf{m}\in\F^{k_{\mathrm{L}}}}
       \{(\mathbf{a}(\mathbf{m})+\mathbf{x}_{\mathrm R},\mathbf{x}_{\mathrm R}):\mathbf{P}\mathbf{x}_{\mathrm R}^{\mathrm{T}}=\mathbf{T}\mathbf{m}\}.
 \label{eq:outer-union}
\end{equation}
Thus $\mathbf{x}_{\mathrm R}$ lies in a right-code coset, and in
$\mathcal C_{\mathrm R}$ itself exactly when $\mathbf T\mathbf m=0$.
Addition of the two halves recovers $\mathbf a(\mathbf m)$ and $\mathbf m$,
so the union is disjoint.

\subsubsection{Separating weight from internal consistency}
For $\mathbf{m}\ne0$, write $\mathbf{a}=\mathbf{a}(\mathbf{m})$, $w_{\mathbf{a}}=\wt(\mathbf{a})$,
$\mathcal{S}_{\mathbf{a}}=\operatorname{supp}(\mathbf{a})$, and $\mathcal{O}_{\mathbf{a}}=[h]\setminus \mathcal{S}_{\mathbf{a}}$.
In a full codeword with left message $\mathbf m$, the bits at
$j$ and $h+j$ are $a_j+x_{\mathrm R,j}$ and $x_{\mathrm R,j}$.
If $j\in\mathcal S_{\mathbf a}$, then $a_j=1$, so this bit pair is
either $(1,0)$ or $(0,1)$. Its contribution to the Hamming weight
of the full codeword is therefore exactly one, independently of
$x_{\mathrm R,j}$. If $j\in\mathcal O_{\mathbf a}$, then $a_j=0$,
so the pair is $(0,0)$ or $(1,1)$ and contributes Hamming weight
$2x_{\mathrm R,j}$, namely zero or 2.
Write the \textit{inside coordinates} as the column
$\boldsymbol\xi=(\mathbf x_{\mathrm R})_{\mathcal S_{\mathbf a}}^{\mathrm T}\in\F^{w_{\mathbf a}}$, and let
$\mathbf z=(\mathbf x_{\mathrm R})_{\mathcal O_{\mathbf a}}\in\F^{h-w_{\mathbf a}}$
denote the \textit{outside coordinates}. Fixing $\mathbf z$ determines
the total weight $w_{\mathbf a}+2\wt(\mathbf z)$ for every choice of
$\boldsymbol\xi$. Such an assignment gives a codeword exactly when
it satisfies the right-half check $\mathbf P\mathbf x_{\mathrm R}^{\mathrm T}=\mathbf T\mathbf m$.
Splitting this check into its inside and outside contributions gives
the equation that $\boldsymbol\xi$ must satisfy:
\begin{equation}
 \mathbf{P}_{\mathcal{S}_{\mathbf{a}}}\boldsymbol\xi
 =\mathbf{T}\mathbf{m}+\mathbf{P}_{\mathcal{O}_{\mathbf{a}}}\mathbf{z}^{\mathrm{T}}.
 \label{eq:extension}
\end{equation}
Flipping an inside bit adds its column of $\mathbf P$ to the syndrome.
If two residuals $\mathbf T\mathbf m+\mathbf P_{\mathcal O_{\mathbf a}}\mathbf z^{\mathrm T}$
in \eqref{eq:extension} differ by a sum of inside columns, flipping
the corresponding bits converts a solution for either one into a
solution for the other. These sums form
$\operatorname{im}\mathbf P_{\mathcal S_{\mathbf a}}$, so it is enough
to retain the syndrome classes in
$\F^f/\operatorname{im}\mathbf P_{\mathcal S_{\mathbf a}}$.

\subsubsection{A matrix representation of the syndrome quotient}
Given an assignment $\mathbf z$ to the \textit{outside coordinates}, we need
to decide whether any inside assignment $\boldsymbol\xi$ satisfies
\eqref{eq:extension}. To make this decision without solving for
$\boldsymbol\xi$, combine the rows of the right-half parity-check
equation $\mathbf P\mathbf x_{\mathrm R}^{\mathrm T}=\mathbf T\mathbf m$
so that the inside bits cancel. A binary row $\boldsymbol\omega$ of
length $f$ selects the checks to add. The inside bits cancel exactly
when $\boldsymbol\omega\mathbf P_{\mathcal S_{\mathbf a}}=0$.
The solutions form the left nullspace of
$\mathbf P_{\mathcal S_{\mathbf a}}$. By binary
elimination, find a basis and stack it as the rows of
$\mathbf Q_{\mathbf a}\in\F^{r_{\mathbf a}\times f}$.
Rank--nullity for the transposed system gives
$r_{\mathbf a}=f-\rk\mathbf P_{\mathcal S_{\mathbf a}}$.
Each row retains a check that the inside bits cannot change.

Since $\mathbf Q_{\mathbf a}\mathbf P_{\mathcal S_{\mathbf a}}=0$,
every syndrome change the inside bits can supply maps to zero under
$\mathbf Q_{\mathbf a}$. Thus
$\operatorname{im}\mathbf P_{\mathcal S_{\mathbf a}}\subseteq\ker\mathbf Q_{\mathbf a}$.
The $r_{\mathbf a}$ independent rows of $\mathbf Q_{\mathbf a}$
give
\[
 \dim\ker\mathbf Q_{\mathbf a}=f-r_{\mathbf a}
 =\rk\mathbf P_{\mathcal S_{\mathbf a}}
 =\dim\operatorname{im}\mathbf P_{\mathcal S_{\mathbf a}}.
\]
The image fills the kernel because they have the same dimension.
Thus $\mathbf Q_{\mathbf a}$ gives two syndromes the same value
exactly when the inside bits can supply their difference.
In particular, a residual can be supplied exactly when
$\mathbf Q_{\mathbf a}$ sends it to zero. This gives the
\textit{projected equation}
\begin{equation}
 \mathbf Q_{\mathbf a}\mathbf T\mathbf m
 =\mathbf Q_{\mathbf a}\mathbf P_{\mathcal O_{\mathbf a}}\mathbf z^{\mathrm T}.
 \label{eq:word-extensions}
\end{equation}
The left side $\mathbf Q_{\mathbf a}\mathbf T\mathbf m$ is the \textit{syndrome observation} induced by the current $\mathbf T$ for message $\mathbf m$.
The right side is the contribution of the outside assignment $\mathbf z$.
The \textit{syndrome quotient} $\F^f/\operatorname{im}\mathbf P_{\mathcal S_{\mathbf a}}$
removes the syndrome differences supplied by \textit{inside coordinates},
leaving $r_{\mathbf a}$ checks.
When $\mathbf z$ satisfies \eqref{eq:word-extensions}, the linear system
\eqref{eq:extension} for $\boldsymbol\xi$
has at least one solution. All its solutions are obtained by adding
vectors from $\ker\mathbf P_{\mathcal S_{\mathbf a}}$ to any one solution.
The kernel has dimension $k_{\mathbf a}=w_{\mathbf a}-\rk\mathbf P_{\mathcal S_{\mathbf a}}$,
so each such $\mathbf z$ admits $2^{k_{\mathbf a}}$ inside assignments.
They give distinct codeword right halves $\mathbf x_{\mathrm R}$
and hence distinct full codewords of the same weight $w_{\mathbf a}+2\wt(\mathbf z)$.
This is the multiplicity retained when eliminating the \textit{inside coordinates}.

\subsubsection{Counting extensions and excluding low weights}
For a fixed left-constituent codeword $\mathbf a$, we count assignments $\mathbf z$
to the \textit{outside coordinates} of $\mathbf x_{\mathrm R}$ by their Hamming
weight and their projected syndrome in \eqref{eq:word-extensions}.
For $\boldsymbol\sigma\in\F^{r_{\mathbf a}}$, define
\begin{equation}
 n_{\mathbf a,t}(\boldsymbol\sigma)=|\{\mathbf z\in\F^{h-w_{\mathbf a}}:
       \wt(\mathbf z)=t,\ \mathbf Q_{\mathbf a}\mathbf P_{\mathcal O_{\mathbf a}}\mathbf z^{\mathrm T}
       =\boldsymbol\sigma\}|.
 \label{eq:outside-syndrome-count}
\end{equation}
Set $n_{\mathbf a,t}=0$ for noninteger $t$ or $t\notin[0,h-w_{\mathbf a}]$.
For $D=d_\star-1$ and $w_{\mathbf a}\le D$, the weight bound gives
\begin{equation}
 \begin{aligned}
 t_{\mathbf a}&=\left\lfloor\frac{D-w_{\mathbf a}}2\right\rfloor,\\
 \mathcal B_{\mathbf a}(D)&=\{\mathbf Q_{\mathbf a}\mathbf P_{\mathcal O_{\mathbf a}}\mathbf z^{\mathrm T}:\wt(\mathbf z)\le t_{\mathbf a}\}.
 \end{aligned}
 \label{eq:outer-data}
\end{equation}
Here $t_{\mathbf a}$ is the largest Hamming weight of the outside coordinates $\mathbf z$ allowed by the total cutoff $D$. The \textit{forbidden set} $\mathcal B_{\mathbf a}(D)$ collects distinct syndromes admitting
low-weight extensions, whereas $n_{\mathbf a,t}$ retains multiplicities.
Both are fixed before $\mathbf T$ selects $\mathbf Q_{\mathbf a}\mathbf T\mathbf m$.
The zero-message branch has codewords $(\mathbf x_{\mathrm R},\mathbf x_{\mathrm R})$
with $\mathbf x_{\mathrm R}\in\mathcal C_{\mathrm R}$, contributing
$A_{w/2}(\mathcal C_{\mathrm R})$ at weight $w$, or zero for odd $w$.

\begin{example}[Hand calculation of the syndrome quotient]
\label{ex:shared-observations}
Revisit Example~\ref{ex:direct-directory} at $d_\star=4$ and $D=3$.
The local profiles are $\mathcal I_{\mathrm L}=\{3,4,5\}$
and $\mathcal I_{\mathrm R}=\{4\}$. Fig.~\ref{fig:n16-data}(b)
gives $\mathbf H(\mathbf T)$ and $\mathbf T$.
For $\mathbf m=(u_3,u_4,u_5)^{\mathrm T}$, the only nonzero messages
with $w_{\mathbf a}\le D$ in Fig.~\ref{fig:n16-data}(c) are $010$ and $011$.
Both give $t_{\mathbf a}=\lfloor(3-2)/2\rfloor=0$, so every outside bit
is zero and \eqref{eq:extension} reduces to
$\mathbf P_{\mathcal S_{\mathbf a}}\boldsymbol\xi=\mathbf T\mathbf m$.
Since the right constituent is unprecoded, deleting row 4 of
$\mathbf G_8^{\mathrm T}$ gives $\mathbf P$. Its packed check rows
$0,\ldots,6$ correspond to global frozen positions $8,9,10,11,13,14,15$.
Example~\ref{ex:direct-directory} releases only positions $8,9$.
In this family, its two-row increment matrix and residual therefore embed as
\[
 \mathbf T=\begin{bmatrix}\mathbf V\\0_{5\times3}\end{bmatrix},
 \qquad
 (u_8,u_9)^{\mathrm T}\longmapsto(u_8,u_9,0,0,0,0,0)^{\mathrm T}.
\]
The last five zero rows retain the other right frozen constraints.

For $010$, the inside set is $\mathcal S_{\mathbf a}=\{0,4\}$, so
$(\xi_0,\xi_1)=(x_{\mathrm R,0},x_{\mathrm R,4})$.
The selected columns of $\mathbf P$ give
\[
\begin{aligned}
\mathbf P_{\mathcal S_{\mathbf a}}\boldsymbol\xi
  &=(\xi_0+\xi_1)\mathbf e_0^{\mathrm T},\\
\mathbf T\mathbf m&=b_0\mathbf e_0^{\mathrm T}+c\mathbf e_1^{\mathrm T}.
\end{aligned}
\]
In Fig.~\ref{fig:n16-data}(c), the inside pair is $(1,1)$ on
check row 0 and $(0,0)$ on rows 1 through 6. Taking these six
zero rows gives the selectors $\mathbf e_1,\ldots,\mathbf e_6$ of
$\mathbf Q_{\mathbf a}$ and hence
$\mathbf Q_{\mathbf a}\mathbf T\mathbf m=(c,0,\ldots,0)^{\mathrm T}$,
with $r_{\mathbf a}=6$ and $k_{\mathbf a}=1$.
For $c=0$, the original equation $\xi_0+\xi_1=b_0$ gives two inside
solutions. For $c=1$, there is no inside solution.
Before imposing the right checks, the inside assignments produce the
residuals $0$ and $\mathbf e_0^{\mathrm T}$, each twice. These are precisely
the residuals $00,10$ of Example~\ref{ex:direct-directory} with five zeros
appended. Both project to zero, giving the single forbidden observation $0$.

For $011$, $\mathcal S_{\mathbf a}=\{1,5\}$ gives
$(\xi_0,\xi_1)=(x_{\mathrm R,1},x_{\mathrm R,5})$. Now
\[
\begin{aligned}
\mathbf P_{\mathcal S_{\mathbf a}}\boldsymbol\xi
  &=(\xi_0+\xi_1)(\mathbf e_0+\mathbf e_1)^{\mathrm T}
      +\xi_1\mathbf e_4^{\mathrm T},\\
\mathbf T\mathbf m&=(b_0+b_1)\mathbf e_0^{\mathrm T}
      +c\mathbf e_1^{\mathrm T}.
\end{aligned}
\]
The two inside pairs $(1,1)$ on rows 0 and 1 cancel when added.
Rows 2, 3, 5, and 6 already have zero pairs. Applying the same
row combinations from Fig.~\ref{fig:n16-data}(c) to $\mathbf T\mathbf m$
gives $\mathbf Q_{\mathbf a}\mathbf T\mathbf m
=(b_0+b_1+c,0,\ldots,0)^{\mathrm T}$, with $r_{\mathbf a}=5$ and
$k_{\mathbf a}=0$. When this projection vanishes, original check row 4
forces $\xi_1=0$ and row 1 gives $\xi_0=c$, the unique inside solution.
Otherwise there is no inside solution.
If all seven right checks were released, the inside assignments would
instead give the four residuals
\[
 \left\{0,\ (\mathbf e_0+\mathbf e_1)^{\mathrm T},\
 \mathbf e_4^{\mathrm T},\
 (\mathbf e_0+\mathbf e_1+\mathbf e_4)^{\mathrm T}\right\}.
\]
Example~\ref{ex:direct-directory} keeps the check $u_{13}=0$ fixed,
so the last two residuals, whose packed check coordinate 4 is one,
are excluded. The remaining two are exactly its residuals $00,11$
with five zeros appended. They also project to zero.
Thus the smaller directory is the part of the full-release residual set
whose last five coordinates vanish. Each coefficient configuration
selects one exact residual. When compatible, it has $2^{k_{\mathbf a}}$
inside solutions, namely $2$ for $010$ and $1$ for $011$, reproducing the multiplicities
in Example~\ref{ex:direct-directory}.
For $\mathbf m=0$, repeating row 4 of $\mathbf G_8$ gives weight 4.
Thus $c=1$ and $b_0+b_1+c=1$ eliminate all nonzero weights below 4.

\label{ex:enumeration}
For $\alpha=b_0=b_1=0,c=1$, raise the enumeration cutoff to $D=4$.
The message $\mathbf m^{\mathrm T}=010$ now has outside-coordinate weight limit
$t_{\mathbf a}=1$. Only the single-bit outside assignment at
right-half position 1 satisfies the projected equation. Its column
in $\mathbf P$ is $(\mathbf e_0+\mathbf e_1)^{\mathrm T}$, so
\eqref{eq:extension} reduces to $\xi_0+\xi_1=1$.
The two inside assignments give the full codewords
\[
 \begin{array}{c|c}
 \boldsymbol\xi^{\mathrm T}&
   \mathbf x=(\mathbf a+\mathbf x_{\mathrm R},\mathbf x_{\mathrm R})\\\hline
 (1,0)&(01001000,11000000)\\
 (0,1)&(11000000,01001000)
 \end{array}.
\]
Thus each accepted outside assignment must be extended by every
compatible inside assignment when enumerating codewords.
\end{example}

\subsubsection{Shared observations and the coefficient quotient}
For one message, only $\mathbf Q_{\mathbf a}\mathbf T\mathbf m$
enters the extension counts. These values cannot be chosen independently
for different messages, because all messages share $\mathbf T$.
Stack its columns
into $\boldsymbol{\theta}=\operatorname{vec}(\mathbf{T})\in\F^{fk_{\mathrm{L}}}$.
For $\mathbf m=(m_0,\ldots,m_{k_{\mathrm L}-1})^{\mathrm T}$,
we have $\mathbf Q_{\mathbf a}\mathbf T\mathbf m
=\sum_i m_i\mathbf Q_{\mathbf a}\mathbf t_i$, where $\mathbf t_i$ is column $i$.
The observation matrix thus has blocks $m_i\mathbf Q_{\mathbf a}$.
Only nonzero messages whose left-constituent codewords have weight at most $D$
can contribute to the truncated spectrum. Define their set by
\begin{equation}
 \mathcal M_D=\{\mathbf m\in\F^{k_{\mathrm L}}\setminus\{0\}:
                   \wt(\mathbf a(\mathbf m))\le D\}.
 \label{eq:critical-messages}
\end{equation}
We call the elements of $\mathcal M_D$ \textit{critical messages}.
Let $n_{\mathrm c}=|\mathcal M_D|$ and list its messages as $\mathbf m_1,\ldots,\mathbf m_{n_{\mathrm c}}$ in a fixed order. Define
\begin{equation}
 \begin{aligned}
 \mathbf{L}_{\mathbf{m}}&=\mathbf{m}^{\mathrm{T}}\otimes \mathbf{Q}_{\mathbf{a}},\\
 \mathbf{L}_{D}&=[\,\mathbf L_{\mathbf m_1}^{\mathrm T}\ \cdots\
                 \mathbf L_{\mathbf m_{n_{\mathrm c}}}^{\mathrm T}\,]^{\mathrm T},
 \quad\rho=\rk\mathbf L_D.
 \end{aligned}
 \label{eq:joint-check-matrix}
\end{equation}
An empty stack has rank zero. The vector $\mathbf L_D\boldsymbol\theta$ stacks the \textit{syndrome observations} $\mathbf L_{\mathbf m}\boldsymbol\theta=\mathbf Q_{\mathbf a}\mathbf T\mathbf m$ for all $\mathbf m\in\mathcal M_D$. Their \textit{joint kernel} $\ker\mathbf L_D$ consists of coefficient changes invisible to every such message.
For fixed constituents, abbreviate the truncated spectrum
$\boldsymbol\Sigma_D(\mathcal C(\mathbf T))$ as
$\boldsymbol\Sigma_D(\boldsymbol\theta)$, viewing it as a function
of the outer coefficients $\boldsymbol\theta$.
The weight-zero count is fixed at one. The \textit{coefficient quotient}
identifies parameter vectors with the same joint observation.
The theorem below allows us to compute the observations first,
then recover the truncated spectrum:
$\boldsymbol\theta\mapsto\mathbf L_D\boldsymbol\theta
\mapsto\boldsymbol\Sigma_D(\boldsymbol\theta)$.
Here $\rho$ counts independent binary observation coordinates.
Different observations can give the same spectrum, so $2^{\rho}$
need not count distinct spectra.

\begin{theorem}[Exact spectrum and its shared linear quotient]
\label{thm:polar-spectrum}
Fix the constituent pair and let $D=d_{\star}-1$.
The following statements hold over the full outer coefficient
space $\F^{f\times k_{\mathrm{L}}}$.

(i) For every $\mathbf{T}$ and integer $0\le w\le N$,
\begin{equation}
 A_{w}(\mathcal{C}(\mathbf{T}))=A_{w/2}(\mathcal{C}_{\mathrm{R}})
 +\sum_{\substack{\mathbf{m}\ne0\\w_{\mathbf{a}}\le w}}
             2^{k_{\mathbf{a}}}n_{\mathbf{a},(w-w_{\mathbf{a}})/2}(\mathbf{Q}_{\mathbf{a}}\mathbf{T}\mathbf{m}).
 \label{eq:full-outer-spectrum}
\end{equation}
In particular, $d_{\min}(\mathcal{C}(\mathbf{T}))\ge d_{\star}$ if and only if
\begin{equation}
 2d_{\min}(\mathcal{C}_{\mathrm{R}})\ge d_{\star},\qquad
 \mathbf{Q}_{\mathbf{a}}\mathbf{T}\mathbf{m}\notin\mathcal B_{\mathbf{a}}(D)\quad(\mathbf{m}\ne0,\ w_{\mathbf{a}}\le D).
 \label{eq:outer-condition}
\end{equation}

(ii) The spectrum-preserving directions satisfy
\begin{equation}
 \{\boldsymbol{\eta}:\boldsymbol{\Sigma}_{D}(\boldsymbol{\theta}+\boldsymbol{\eta})=\boldsymbol{\Sigma}_{D}(\boldsymbol{\theta})
                           \text{ for every }\boldsymbol{\theta}\}
 =\{\boldsymbol{\eta}:\boldsymbol{\Sigma}_{D}(\boldsymbol{\eta})=\boldsymbol{\Sigma}_{D}(0)\}
              =\ker \mathbf{L}_{D} .
 \label{eq:spectrum-kernel}
\end{equation}

(iii) If a binary matrix $\mathbf{U}$ with $fk_{\mathrm{L}}$ columns has
the property that $\mathbf{U}\boldsymbol{\theta}$ determines $\boldsymbol{\Sigma}_{D}(\boldsymbol{\theta})$
for every $\boldsymbol{\theta}$, then $\rk \mathbf{U}\ge\rho$.
Equality is attained by taking a row basis of $\mathbf{L}_{D}$.
\end{theorem}
\begin{IEEEproof}
\textbf{(i) Spectrum and distance.}
For $\mathbf m\ne0$, the weight decomposition above gives
\begin{equation}
 \wt(\mathbf a+\mathbf x_{\mathrm R},\mathbf x_{\mathrm R})
 =w_{\mathbf a}+2\wt(\mathbf z).
 \label{eq:relative-weight}
\end{equation}
Weight $w$ therefore requires $\wt(\mathbf z)=t=(w-w_{\mathbf a})/2$.
By the \textit{projected equation},
$n_{\mathbf a,t}(\mathbf Q_{\mathbf a}\mathbf T\mathbf m)$ assignments
$\mathbf z$ of this weight admit inside solutions. Each has $2^{k_{\mathbf a}}$
inside solutions, and different assignments $\mathbf z$ give different
$\mathbf x_{\mathrm R}$. Their product is the contribution of $\mathbf m$.

For $\mathbf m=0$, the codewords $(\mathbf x_{\mathrm R},\mathbf x_{\mathrm R})$
with $\mathbf x_{\mathrm R}\in\mathcal C_{\mathrm R}$ contribute
$A_{w/2}(\mathcal C_{\mathrm R})$. Summing the disjoint branches in
\eqref{eq:outer-union} gives \eqref{eq:full-outer-spectrum}.

For the distance condition, the zero-message branch contributes a
nonzero codeword below $d_\star$ exactly when
$2d_{\min}(\mathcal C_{\mathrm R})<d_\star$.
Every extension of a nonzero $\mathbf m$ is nonzero because
$\mathbf a(\mathbf m)\ne0$. By \eqref{eq:relative-weight},
only messages with $w_{\mathbf a}\le D$ can contribute below $d_\star$.
For each such message, \eqref{eq:outer-data} collects precisely the
projected syndromes of the assignments $\mathbf z$ with $\wt(\mathbf z)\le t_{\mathbf a}$.
A codeword of weight at most $D$ therefore exists in that branch
exactly when $\mathbf Q_{\mathbf a}\mathbf T\mathbf m\in\mathcal B_{\mathbf a}(D)$.
Excluding these codewords from both types of branches gives
\eqref{eq:outer-condition}.

\textbf{(ii) Spectrum-preserving directions.}
If $\boldsymbol{\eta}\in\ker \mathbf{L}_{D}$, then
\[
 \mathbf{L}_{\mathbf{m}}(\boldsymbol{\theta}+\boldsymbol{\eta})=\mathbf{L}_{\mathbf{m}}\boldsymbol{\theta}
 \qquad\text{for every critical }\mathbf{m}.
\]
All observations appearing in \eqref{eq:full-outer-spectrum}
for $w\le D$ are unchanged. Thus
$\boldsymbol{\Sigma}_{D}(\boldsymbol{\theta}+\boldsymbol{\eta})=\boldsymbol{\Sigma}_{D}(\boldsymbol{\theta})$ for every $\boldsymbol{\theta}$,
and in particular $\boldsymbol{\Sigma}_{D}(\boldsymbol{\eta})=\boldsymbol{\Sigma}_{D}(0)$.
For the converse, assume $\boldsymbol{\Sigma}_{D}(\boldsymbol{\eta})=\boldsymbol{\Sigma}_{D}(0)$.
Suppose for contradiction that $\boldsymbol{\eta}=\operatorname{vec}(\Delta \mathbf{T})\notin\ker \mathbf{L}_{D}$.
Consider the smallest affected left weight: at that weight the
change can only remove lifts of $\mathbf z=0$, and no other message
can offset the loss. To see this, let
\[
 w_{\Delta}=\min\{w_{\mathbf{a}}:\mathbf{m}\ne0,\ w_{\mathbf{a}}\le D,\ \mathbf{Q}_{\mathbf{a}}\Delta \mathbf{T}\mathbf{m}\ne0\}.
\]
This set is nonempty because $\mathbf{L}_{D}\boldsymbol{\eta}\ne0$.
In the difference between $A_{w_{\Delta}}(\mathcal{C}(0))$ and
$A_{w_{\Delta}}(\mathcal{C}(\Delta \mathbf{T}))$, the zero-left-message contributions $A_{w_\Delta/2}(\mathcal C_{\mathrm R})$ cancel.
Terms with $w_{\mathbf{a}}<w_{\Delta}$ cancel because $\mathbf{Q}_{\mathbf{a}}\Delta \mathbf{T}\mathbf{m}=0$.
Terms with $w_{\mathbf{a}}>w_{\Delta}$ cannot contribute by
\eqref{eq:relative-weight}. For $w_{\mathbf{a}}=w_{\Delta}$,
the only outside assignment is $\mathbf{z}=0$, so
$n_{\mathbf{a},0}(\boldsymbol{\sigma})=\ind{\boldsymbol{\sigma}=0}$.
Consequently,
\begin{equation}
 A_{w_{\Delta}}(\mathcal{C}(0))-A_{w_{\Delta}}(\mathcal{C}(\Delta \mathbf{T}))
   =\sum_{\substack{\mathbf{m}:w_{\mathbf{a}}=w_{\Delta}\\\mathbf{Q}_{\mathbf{a}}\Delta \mathbf{T}\mathbf{m}\ne0}}2^{k_{\mathbf{a}}}>0 .
 \label{eq:first-spectrum-change}
\end{equation}
Since $w_{\Delta}\le D$, the positive difference contradicts
$\boldsymbol{\Sigma}_{D}(\boldsymbol{\eta})=\boldsymbol{\Sigma}_{D}(0)$.
Thus preservation at zero forces kernel membership. Kernel membership
implies preservation at every $\boldsymbol\theta$, as shown above, and
preservation at every $\boldsymbol\theta$ includes preservation at zero.
This proves \eqref{eq:spectrum-kernel}.

\textbf{(iii) Minimal linear representation.}
Suppose $\mathbf{U}\boldsymbol{\theta}$ determines $\boldsymbol{\Sigma}_{D}(\boldsymbol{\theta})$ for every
$\boldsymbol{\theta}$. For any $\boldsymbol{\eta}\in\ker \mathbf{U}$ and every $\boldsymbol{\theta}$,
\[
 \mathbf{U}(\boldsymbol{\theta}+\boldsymbol{\eta})=\mathbf{U}\boldsymbol{\theta}
 \quad\Longrightarrow\quad
 \boldsymbol{\Sigma}_{D}(\boldsymbol{\theta}+\boldsymbol{\eta})=\boldsymbol{\Sigma}_{D}(\boldsymbol{\theta}).
\]
Part (ii) therefore gives $\ker \mathbf{U}\subseteq\ker \mathbf{L}_{D}$.
By rank--nullity,
\[
 \rk \mathbf{U}=fk_{\mathrm{L}}-\dim\ker \mathbf{U}
 \ge fk_{\mathrm{L}}-\dim\ker \mathbf{L}_{D}=\rho.
\]
A matrix $\mathbf{U}$ formed from a row basis of $\mathbf{L}_{D}$ attains the
bound. Every row of $\mathbf{L}_{D}$ is a combination of rows of $\mathbf{U}$,
so $\mathbf{U}\boldsymbol{\theta}$ recovers $\mathbf{L}_{D}\boldsymbol{\theta}$ and then the spectrum.
\end{IEEEproof}

The stacked map $\mathbf L_D$ retains the checks that inside bits
cannot absorb. For the fixed constituent pair, a feasible configuration
$\boldsymbol\theta$, and a protection cutoff $w_\star\ge d_\star$,
Theorem~\ref{thm:polar-spectrum}(ii) gives two domains:
$\boldsymbol\theta+\ker\mathbf L_{d_\star-1}$ preserves the distance target,
whereas $\boldsymbol\theta+\ker\mathbf L_{w_\star}$ also preserves every count
through $w_\star$, including $A_{d_\star}$.
Individual codewords may change. Proposition~\ref{prop:reduction}
instead preserves the codeword set through $w_\star$ when configurations
agree on $\mathcal E_{w_\star}$.

For the family of Example~\ref{ex:shared-observations} at $D=3$,
the observation pair $(c,b_0+b_1+c)$ takes the values
$00,01,10,11$, giving $A_2=3,2,1,0$, respectively.
Each pair leaves $\alpha$ and $b_0$ free.
Thus two observation bits are necessary and sufficient.

Continue Example~\ref{ex:shared-observations} at $D=4$.
Compare $b_0=b_1=0$ with $b_0=b_1=1$ at $\alpha=0,c=1$.
Flipping $b_0,b_1$ together adds $(u_4+u_5)\mathbf P_{:,4}$
to the right-code syndrome $\mathbf T\mathbf m$. The left-constituent codeword has
$a_4=u_4+u_5$, since the length-8 rows for \textit{sources} 4 and 5
contain position 4 and row 3 does not. Whenever the offset
changes, position 4 is therefore inside, and
the inside column $\mathbf P_{:,4}$ absorbs the change. All projected
observations are unchanged, so the two designs have the same counts
through $D$. For $\mathbf m=(0,1,0)^{\mathrm T}$ and $u_{12}=0$,
the output support changes from $\{1,4,8,9\}$ to $\{0,1,4,9\}$.
The first codeword's unique input has $u_8=0$, whereas the second
design requires $u_8=u_4+u_5=1$ for that input.

To relate the common-supercode conditions of Theorem~\ref{thm:shared-elimination}
to the fixed-constituent distance condition of Theorem~\ref{thm:polar-spectrum}(i),
retain the coefficients within each half, set the baseline outer block
to zero, and release the right frozen
checks $\mathcal J=\cF\cap[h,2h)$. If $f=0$, there are no outer
coefficients and \eqref{eq:outer-condition} applies directly.
For $f>0$, editing all left-information coefficients on these rows
gives $\mathbf V=\mathbf T$ and
$\mathbf q=\mathbf m^{\mathrm T}$ in \eqref{eq:shared-array}.
The baseline residual is
$\boldsymbol\beta=\mathbf H_{\mathrm R}\mathbf u_{\mathrm R}^{\mathrm T}
=\mathbf P\mathbf x_{\mathrm R}^{\mathrm T}$.

For a \textit{critical message} $\mathbf m$, let $\mathcal B_{\mathbf m}^{\mathrm{raw}}(D)$
be the set $\mathcal B(\mathbf m^{\mathrm T})$ of \eqref{eq:pattern-directory} for this choice of released checks and forbidden weights $1,\ldots,D$. It contains the baseline residuals from that message branch.
Before restoring the right checks, the \textit{inside coordinates} are free
and the weight bound constrains only the \textit{outside coordinates}.
Thus
\begin{equation}
 \begin{aligned}
 \mathcal B_{\mathbf{m}}^{\mathrm{raw}}(D)
 &=\operatorname{im}\mathbf{P}_{\mathcal{S}_{\mathbf{a}}}
       +\{\mathbf{P}_{\mathcal{O}_{\mathbf{a}}}\mathbf{z}^{\mathrm{T}}:\wt(\mathbf{z})\le t_{\mathbf{a}}\}\\
 &=\{\boldsymbol\beta\in\F^f:
          \mathbf Q_{\mathbf a}\boldsymbol\beta\in\mathcal B_{\mathbf a}(D)\}.
 \end{aligned}
 \label{eq:outer-raw-quotient}
\end{equation}
The equality follows from
$\ker\mathbf Q_{\mathbf a}=\operatorname{im}\mathbf P_{\mathcal S_{\mathbf a}}$:
projection erases exactly the contributions available from inside bits.
For each critical message, avoiding the unprojected set
in \eqref{eq:outer-raw-quotient} is equivalent to
$\mathbf Q_{\mathbf a}\mathbf T\mathbf m\notin\mathcal B_{\mathbf a}(D)$.
All \textit{critical messages} must use the same $\mathbf T$.
For $\mathbf m=0$, the increment vanishes. A low-weight input in
this branch survives exactly when its residual is zero, in which
case it produces a repeated nonzero right codeword.
Excluding these inputs is equivalent to
$2d_{\min}(\mathcal C_{\mathrm R})\ge d_\star$.
Example~\ref{ex:shared-observations} also shows how the smaller release
of Example~\ref{ex:direct-directory} embeds in this full-check description.

The fixed right constituent also gives
\begin{equation}
 d_{\min}(\mathcal{C}(\mathbf{T}))\le2d_{\min}(\mathcal{C}_{\mathrm{R}}),\qquad
 A_{2t}(\mathcal{C}(\mathbf{T}))\ge A_{t}(\mathcal{C}_{\mathrm{R}})
 \label{eq:right-only-limits}
\end{equation}
when $\mathcal{C}_{\mathrm{R}}\ne\{0\}$. These bounds concern the fixed pair,
not the full profile. For $D'\ge D$, the observation maps and, for existing
messages with $w_{\mathbf a}\le D$, the \textit{forbidden sets} satisfy
\[
 \begin{gathered}
 \ker \mathbf{L}_{D'}\subseteq\ker \mathbf{L}_{D},\qquad
 \mathcal B_{\mathbf{a}}(D)\subseteq\mathcal B_{\mathbf{a}}(D'),\\
 A_{w}(\mathcal{C}_{\mathrm{L}})=0\ (D<w\le D')
 \ \Longrightarrow\ \ker \mathbf{L}_{D'}=\ker \mathbf{L}_{D}.
 \end{gathered}
\]
Increasing $D$ can increase the outside-coordinate weight limits
$t_{\mathbf a}$, while the existing matrices $\mathbf Q_{\mathbf a}$ remain fixed.
Only newly included left messages add blocks to $\mathbf L_D$.
If there are none, $\mathbf L_D$ and its kernel remain unchanged,
but the \textit{forbidden sets} may still grow and reduce the feasible set.
Thus a higher distance target reuses the existing projections but requires
updated forbidden sets, any new message blocks, and a renewed check of
the right-constituent distance condition in \eqref{eq:outer-condition}.
Eliminating only the original minimum-weight layer does not establish
the higher target.

\subsection{Exact Low-Weight Codeword Enumeration}
\label{sec:enumeration}
We construct the \textit{complete low-weight directory} by reversing
the quotient reduction: test the \textit{outside coordinates}, then
recover every compatible assignment of the \textit{inside coordinates}.
The lifting multiplicities also support spectrum queries without
producing each codeword individually.

\subsubsection{Recursive codeword generation}
Let $\mathbf H$ be the fixed canonical input-check matrix to be
enumerated, and let $N$ denote its power-of-two column count in the current
recursive call. Define
\[
 \proc{List}(\mathbf H,D)
 =\{\mathbf u\in\F^N:\mathbf H\mathbf u^{\mathrm T}=0,\
                 \wt(\mathbf u\mathbf G_N)\le D\},
\]
emitted in a fixed order, including zero when $D\ge0$.
A causal check system is first put in canonical form by substitution.
At each call, $\mathcal I_{\mathrm L}$ denotes the free positions of the current left check matrix $\mathbf H_{\mathrm L}$, and $\mathbf J_{\mathrm L}$ selects them. These may include frozen positions released in the original problem.

\begin{algorithm}[!t]
\caption{Complete input enumeration with an output-weight cutoff}
\label{alg:directory}
\small
\begin{algorithmic}[1]
\Require Canonical causal checks $\mathbf H$ of power-of-two length, cutoff $D$
\Ensure Each $\mathbf u$ with $\mathbf H\mathbf u^{\mathrm T}=0$ and $\wt(\mathbf u\mathbf G_N)\le D$, once
\Function{List}{$\mathbf H,D$}
 \State $N\gets$ number of columns of $\mathbf H$
 \If{$D<0$} \State \Return \EndIf
 \If{$N=1$}
  \State \textbf{yield} each $\mathbf u\in\F^1$ with $\mathbf H\mathbf u^{\mathrm T}=0$ and $\wt(\mathbf u)\le D$
  \State \Return
 \EndIf
 \State $h\gets N/2$; obtain $\mathbf H_{\mathrm L},\mathbf H_{\mathrm R},\mathbf T,\mathbf J_{\mathrm L}$ from \eqref{eq:assembled-check}
 \ForAll{$\mathbf u_{\mathrm R}$ emitted by $\Call{List}{\mathbf H_{\mathrm R},\lfloor D/2\rfloor}$}
  \State \textbf{yield} $(0,\mathbf u_{\mathrm R})$ \Comment{zero-left branch, (\ref{eq:outer-union})}
 \EndFor
 \If{$\mathcal I_{\mathrm L}=\varnothing$ or the bound \eqref{eq:profile-bound} exceeds $D$} \State \Return \EndIf
 \State $\mathbf P\gets\mathbf H_{\mathrm R}\mathbf G_h^{\mathrm T}$
 \ForAll{nonzero $\mathbf u_{\mathrm L}$ emitted by $\Call{List}{\mathbf H_{\mathrm L},D}$}
  \State $\mathbf m\gets\mathbf J_{\mathrm L}\mathbf u_{\mathrm L}^{\mathrm T}$; $\mathbf a\gets\mathbf u_{\mathrm L}\mathbf G_h$
  \State $\mathcal S_{\mathbf a}\gets\operatorname{supp}(\mathbf a)$; $\mathcal O_{\mathbf a}\gets[h]\setminus\mathcal S_{\mathbf a}$
  \State $t_{\mathbf a}\gets\lfloor(D-w_{\mathbf a})/2\rfloor$ \Comment{outside radius, (\ref{eq:outer-data})}
  \State Eliminate $\mathbf P_{\mathcal S_{\mathbf a}}$; retain a left-nullspace basis $\mathbf Q_{\mathbf a}$, kernel basis $\mathbf K_{\mathcal S_{\mathbf a}}$, and row operations
  \ForAll{$\mathbf z$ satisfying \eqref{eq:word-extensions} and $\wt(\mathbf z)\le t_{\mathbf a}$, in fixed order}
    \State Recover an inside solution $\boldsymbol\xi_0$ of \eqref{eq:extension} using the retained row operations
    \ForAll{$\boldsymbol\lambda\in\F^{k_{\mathbf a}}$ in fixed binary order}
     \State $\boldsymbol\xi\gets\boldsymbol\xi_0+\mathbf K_{\mathcal S_{\mathbf a}}\boldsymbol\lambda$
     \State Set $(\mathbf x_{\mathrm R})_{\mathcal S_{\mathbf a}}\gets\boldsymbol\xi^{\mathrm T}$, $(\mathbf x_{\mathrm R})_{\mathcal O_{\mathbf a}}\gets\mathbf z$
     \State \textbf{yield} $(\mathbf u_{\mathrm L},\mathbf x_{\mathrm R}\mathbf G_h)$
    \EndFor
  \EndFor
 \EndFor
\EndFunction
\end{algorithmic}
\end{algorithm}

To build $\mathcal U_{\mathcal J}(\cW)$ in \eqref{eq:relaxation}, run
\proc{List} with the retained checks
$\mathbf H_{\bar{\mathcal J},:}(\bar{\mathbf s})$ and a cutoff covering $\cW$.
The frozen positions $\mathcal J$ are free in this call.
From its output, retain inputs whose codeword weights lie in $\cW$
and whose leading positions lie in $\mathcal I$.

At $N=2h$, all quantities refer to the current block form
\eqref{eq:assembled-check}, with $\mathbf T$ fixed and
$\mathbf m=\mathbf J_{\mathrm L}\mathbf u_{\mathrm L}^{\mathrm T}$.
For each $\mathbf z$ with $\wt(\mathbf z)\le t_{\mathbf a}$ passing \eqref{eq:word-extensions},
\textit{inside-coordinate lifting} recovers every solution
$\boldsymbol\xi$ of \eqref{eq:extension}. This restores the variables
eliminated by the quotient. Place $\boldsymbol\xi^{\mathrm T},\mathbf z$
in their original coordinates to obtain $\mathbf x_{\mathrm R}$,
then recover $\mathbf u_{\mathrm R}=\mathbf x_{\mathrm R}\mathbf G_h$.
The coordinate definitions also apply to $\mathbf a=0$, with
$\mathcal S_{\mathbf a}=\varnothing$ and an empty inside column.
Thus the recursion is
\begin{equation}
 \thickmuskip=2mu \medmuskip=2mu \thinmuskip=2mu\relax
 \proc{List}(\mathbf H,D)=
 \bigsqcup_{\substack{\mathbf u_{\mathrm L}\in\\ \proc{List}(\mathbf H_{\mathrm L},D)}}
 \bigl\{(\mathbf u_{\mathrm L},\mathbf x_{\mathrm R}\mathbf G_h):
       \wt(\mathbf z)\le t_{\mathbf a},\,
       (\mathbf x_{\mathrm R})_{\mathcal S_{\mathbf a}}=\boldsymbol\xi^{\mathrm T},\,
       (\mathbf x_{\mathrm R})_{\mathcal O_{\mathbf a}}=\mathbf z,\,
       \mathbf P_{\mathcal S_{\mathbf a}}\boldsymbol\xi
       =\mathbf T\mathbf m+\mathbf P_{\mathcal O_{\mathbf a}}\mathbf z^{\mathrm T}\bigr\}.
 \label{eq:word-recursion}
\end{equation}
The base case $N=1$ tests each binary input directly against
$\mathbf H\mathbf u^{\mathrm T}=0$ and $\wt(\mathbf u)\le D$.
A negative cutoff returns the empty set. Empty coordinate sets
and rank-zero systems are included in these rules.

\begin{proposition}[Enumeration completeness]
\label{prop:directory}
For every power-of-two length $N$, canonical causal check matrix
$\mathbf H$ with $N$ columns, and integer cutoff $D$, recursion
\eqref{eq:word-recursion}, with the stated base cases, lists
each input satisfying $\mathbf H\mathbf u^{\mathrm T}=0$ and
$\wt(\mathbf u\mathbf G_N)\le D$ exactly once.
\end{proposition}
\begin{IEEEproof}
For $D<0$ both sets are empty. For $N=1$, the direct test lists
all valid inputs. Assume the claim holds at length $h$.

\textbf{Valid extensions.}
For a nonzero left input, Theorem~\ref{thm:polar-spectrum}(i)
makes the projected test \eqref{eq:word-extensions} equivalent
to solvability of \eqref{eq:extension}. For each accepted
$\mathbf z$, choose an inside solution $\boldsymbol\xi_0$ and
a matrix $\mathbf K_{\mathcal S_{\mathbf a}}$ whose columns form
a basis of $\ker\mathbf P_{\mathcal S_{\mathbf a}}$. All solutions are
\[
 \boldsymbol\xi=\boldsymbol\xi_0+\mathbf K_{\mathcal S_{\mathbf a}}\boldsymbol\lambda,
 \qquad\boldsymbol\lambda\in\F^{k_{\mathbf a}}.
\]
The independent basis columns make these lifts distinct.
Each reconstructed input satisfies both block rows of \eqref{eq:assembled-check}, and
\[
 \wt(\mathbf u\mathbf G_{2h})
 =w_{\mathbf a}+2\wt(\mathbf z)
 \le w_{\mathbf a}+2t_{\mathbf a}\le D.
\]
For the zero left input, $\mathbf m=0$ and the right input
belongs to $\proc{List}(\mathbf H_{\mathrm R},\lfloor D/2\rfloor)$.
Its repeated codeword satisfies the checks and cutoff by induction.

\textbf{Completeness and uniqueness.}
Conversely, let $\mathbf u=(\mathbf u_{\mathrm L},\mathbf u_{\mathrm R})$
satisfy the checks and cutoff. Then
\[
 \mathbf H_{\mathrm L}\mathbf u_{\mathrm L}^{\mathrm T}=0,\qquad
 \wt(\mathbf u_{\mathrm L}\mathbf G_h)\le\wt(\mathbf u\mathbf G_{2h})\le D.
\]
Induction ensures that $\mathbf u_{\mathrm L}$ is emitted.
If it is zero, the right input is emitted by the repeated-code
branch. Otherwise, $\mathbf x_{\mathrm R}=\mathbf u_{\mathrm R}\mathbf G_h$
uniquely determines $\mathbf z$ and $\boldsymbol\xi$ on their
disjoint coordinate sets. Its weight implies
$\wt(\mathbf z)\le t_{\mathbf a}$, and
$\mathbf P\mathbf x_{\mathrm R}^{\mathrm T}=\mathbf T\mathbf m$ implies
\eqref{eq:extension}. Hence its \textit{outside coordinates} pass the
projected test and its inside vector is one of the recovered lifts.
These coordinates and their independent kernel coefficients are
unique, so no input is omitted or repeated.
\end{IEEEproof}

\subsubsection{Enumeration and spectrum queries}
After processing the zero-left branch, Algorithm~\ref{alg:directory}
stops if the left free positions are empty or their profile bound
\eqref{eq:profile-bound} exceeds $D$. Otherwise, it enumerates
the nonzero left inputs with $\wt(\mathbf a)\le D$.

For each such input, assign the \textit{outside coordinates} in a fixed
order. Discard a branch if its assigned weight exceeds $t_{\mathbf a}$,
or if the target $\mathbf Q_{\mathbf a}\mathbf T\mathbf m$ minus the
assigned column sum lies outside the span of the unassigned columns
of $\mathbf Q_{\mathbf a}\mathbf P_{\mathcal O_{\mathbf a}}$.
A branch rejected by either test has no valid continuation. Trying both values on
the remaining branches and lifting every accepted assignment therefore
preserves Proposition~\ref{prop:directory}.

When only the spectrum is needed, count each accepted $\mathbf z$ as
$2^{k_{\mathbf a}}$ codewords of weight $w_{\mathbf a}+2\wt(\mathbf z)$
without listing the lifts. For construction, list the lifts, retain
the inputs specified by \eqref{eq:relaxation}, and group them by
\eqref{eq:pattern-directory}.

\section{Coefficient Reduction and Construction Guarantees}
\label{sec:guarantees}
Section~\ref{sec:elimination} supplies complete elimination conditions,
the \textit{complete low-weight directory} $\mathcal U_{\mathcal J}(\cW)$,
and the joint observation $\mathbf L_D\boldsymbol\theta$ for fixed constituents.
We now use these results to retain the coefficient choices needed
for source-wise construction, recover all configurations, and
characterize domains consisting entirely of feasible configurations.
We then derive completion guarantees and count lower bounds.
When these bounds are inconclusive, the same conditions support exact
traversal.

By \eqref{eq:outer-condition}, the fixed-constituent distance claims
below require the right-constituent distance condition
$2d_{\min}(\mathcal C_{\mathrm R})\ge d_\star$.
Once it holds, we choose $\mathbf T$ to exclude low-weight extensions
of the nonzero left messages.

\subsection{Coefficient Recovery and Source-Compatible Reduction}
\label{subsec:coefficient-recovery}
\subsubsection{From feasible observations to coefficient configurations}
Example~\ref{ex:shared-observations} shows that the four coefficients
affect weight-2 exclusion only through $c$ and $b_0+b_1+c$.
We first choose coefficient representatives realizing every
attainable observation, then recover all the feasible
configurations found in Example~\ref{ex:direct-directory}.

Under this right-constituent distance condition, let
$\mathcal X$ be the set of coefficient vectors meeting
$d_{\min}(\mathcal C(\mathbf T))\ge d_\star$.
The \textit{joint observation} $\mathbf L_D\boldsymbol\theta$, with
$\boldsymbol\theta=\operatorname{vec}(\mathbf T)$, must belong to
\begin{equation}
 \mathcal Y_D=\operatorname{im}\mathbf L_D\cap
 \prod_{\mathbf m\in\mathcal M_D}
       \bigl(\F^{r_{\mathbf a}}\setminus\mathcal B_{\mathbf a}(D)\bigr).
 \label{eq:feasible-observation-image}
\end{equation}
Membership in $\operatorname{im}\mathbf L_D$ means that one shared
$\mathbf T$ produces the entire joint observation.
Membership in the Cartesian product means that every message block
$\mathbf Q_{\mathbf a}\mathbf T\mathbf m$ avoids its own
\textit{forbidden set} $\mathcal B_{\mathbf a}(D)$, in the order of
\eqref{eq:joint-check-matrix}. Both requirements must hold.
An empty product contains the empty tuple.

To solve $\mathbf L_D\boldsymbol\theta=\mathbf y$ for
$\mathbf y\in\mathcal Y_D$, retain $\rho$ independent columns of
$\mathbf L_D$ and set the other coefficients to zero.
Every attainable observation is a unique sum of the retained columns.
Their coefficients therefore give one representative per observation.
Let $\mathcal V$ denote this representative space.
Every original configuration is recovered by adding a vector in
$\ker\mathbf L_D$. This is \textit{coefficient lifting}.
Such a subspace, giving unique representatives,
is called a linear complement of $\ker\mathbf L_D$.

\begin{example}[Coefficient reduction and recovery in the running example]
\label{ex:kernel-spectrum}
For $d_\star=4$ and $D=3$, reuse the two projections from
Example~\ref{ex:shared-observations}. Write
$\mathbf T=[\,\mathbf t_0\ \mathbf t_1\ \mathbf t_2\,]$ in
\textit{source} order $3,4,5$. The two \textit{critical messages}
$\mathbf m^{\mathrm T}=010,011$ give the joint observation
\[
 \mathbf L_3\operatorname{vec}(\mathbf T)=
 \begin{bmatrix}
 \mathbf Q_{\mathbf a(010)}\mathbf t_1\\
 \mathbf Q_{\mathbf a(011)}(\mathbf t_1+\mathbf t_2)
 \end{bmatrix}.
\]
With both constituents fixed, all 21 entries of $\mathbf T$ can vary.
Choose $\mathbf t_1$ first and then $\mathbf t_2$ to give any
desired sum $\mathbf t_1+\mathbf t_2$.
The two projection inputs therefore vary independently. The ranks
6 and 5 from Example~\ref{ex:shared-observations} give
$\operatorname{rank}\mathbf L_3=6+5=11$.

The four-parameter family of Example~\ref{ex:canonical} is the domain
\[
 \begin{aligned}
 \boldsymbol\gamma&=(\alpha,b_0,c,b_1)^{\mathrm T},
       \quad\boldsymbol\theta=\mathbf R\boldsymbol\gamma,\\
 \mathbf R&=(\mathbf I_{21})_{:,\{0,7,8,14\}}
       \in\F^{21\times4}.
 \end{aligned}
\]
The selected coordinates belong to the column-stacked coefficient
vector $\boldsymbol\theta$, with seven entries per column of $\mathbf T$.
Thus $\mathbf R$ fixes the other 17 coefficients to zero.
The restricted observation matrix $\mathbf L_3\mathbf R$ has nine
identically zero rows. Omitting these observation rows gives
\[
 \begin{bmatrix}c\\b_0+c+b_1\end{bmatrix}
 =\begin{bmatrix}0&0&1&0\\0&1&1&1\end{bmatrix}
  \boldsymbol\gamma.
\]
In this $2\times4$ matrix, the columns multiplying $c$ and $b_1$
are $(1,1)^{\mathrm T}$ and $(0,1)^{\mathrm T}$, respectively.
Thus changing $c$ flips both observations, whereas changing either $b_0$ or $b_1$
flips only the second. These columns are independent, so the map has rank 2.
Setting $\alpha=b_0=0$ gives a unique representative for each
pair in $\F^2$.

Example~\ref{ex:shared-observations} requires the pair
$(1,1)^{\mathrm T}$. At $\alpha=b_0=0$, the equations $c=1$ and
$c+b_1=1$ give the representative $(0,0,1,0)^{\mathrm T}$.
Recovery within this family uses $\ker(\mathbf L_3\mathbf R)$.
The omitted zero rows impose no equations, so setting both remaining
observation increments to zero gives
\[
 \Delta c=0,\qquad \Delta b_0+\Delta c+\Delta b_1=0.
\]
Thus $\Delta\alpha,\Delta b_0$ are free and
$\Delta b_1=\Delta b_0$. Taking
$(\Delta\alpha,\Delta b_0)=(1,0)$ and $(0,1)$ gives the kernel
basis $(1,0,0,0)^{\mathrm T}$ and $(0,1,0,1)^{\mathrm T}$, respectively.
Adding their span to the representative recovers exactly
\[
 \{(\alpha,b_0,1,b_0)^{\mathrm T}:\alpha,b_0\in\F\}.
\]
These are the four feasible configurations obtained by
direct enumeration in Example~\ref{ex:direct-directory}.
\end{example}

\subsubsection{Retaining the choices needed at each source}
\begin{figure}[!t]
\centering
\includegraphics[width=4.05in]{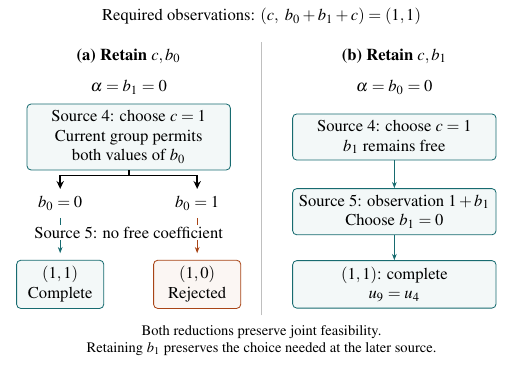}
\caption{Source-wise assignment in Example~\ref{ex:kernel-spectrum}.}
\label{fig:source-choices}
\end{figure}

In Example~\ref{ex:kernel-spectrum}, the columns multiplying $b_0$
and $b_1$ are equal. Hence retaining either pair $c,b_0$ or $c,b_1$
gives a basis of the same observation image.
Fig.~\ref{fig:source-choices} shows why these bases differ when
coefficients are assigned in source order: only the second leaves
a free coefficient for the last source.

For columnwise construction, we assign retained coefficients in increasing \textit{source} order,
$\mathbf t_0,\mathbf t_1,\ldots$, holding earlier columns fixed.
A message whose last nonzero coordinate is $i$ depends only on
$\mathbf t_0,\ldots,\mathbf t_i$. Its exclusion must hold when
$\mathbf t_i$ is assigned, since later columns cannot change it.

Group \textit{critical messages} by their last nonzero coordinate:
\begin{equation}
 \mathcal L_i=\{\mathbf m\in\mathcal M_D:m_i=1,\ m_j=0\ (j>i)\}.
 \label{eq:last-source-groups}
\end{equation}
Here $i$ indexes message coordinates in information-set order.
In Example~\ref{ex:kernel-spectrum},
$\mathbf m=(u_3,u_4,u_5)^{\mathrm T}$. Their last nonzero information
bits give the two groups below.
\[
 \begin{array}{c|c|c}
 \mathbf m^{\mathrm T}&\text{last nonzero bit}&\text{condition}\\\hline
 (0,1,0)&u_4&c=1\\
 (0,1,1)&u_5&b_0+b_1+c=1
 \end{array}
\]
Fig.~\ref{fig:source-choices} shows the forward assignments for
$d_\star=4$, with $\alpha=0$ in both representative spaces.
In Fig.~\ref{fig:source-choices}(a), retaining $c,b_0$ fixes $b_1=0$.
The current group at \textit{source} 4 requires $c=1$ for
$\mathbf m^{\mathrm T}=010$ but accepts either value of $b_0$.
The choice $b_0=1$ gives $b_0+b_1+c=0\ne1$ for
$\mathbf m^{\mathrm T}=011$; no coefficient remains adjustable at
\textit{source} 5, so this choice must be revised to $b_0=0$.
In Fig.~\ref{fig:source-choices}(b), retaining $c,b_1$ fixes $b_0=0$:
after choosing $c=1$ at \textit{source} 4, choose $b_1=0$ at
\textit{source} 5. Both spaces contain a feasible configuration,
but the latter retains the second message's adjustable coefficient
at the later \textit{source}.

For any $\mathbf m\in\mathcal L_i$,
\[
 \mathbf Q_{\mathbf a}\mathbf T\mathbf m
 =\mathbf Q_{\mathbf a}\Bigl(\sum_{j<i}m_j\mathbf t_j+\mathbf t_i\Bigr).
\]
With earlier columns fixed, the sum is known and $\mathbf t_i$
provides the remaining choices. A reduction is
\textit{source-compatible} if the retained entries of $\mathbf t_i$
still provide all these joint choices for $\mathcal L_i$.

Each column of $\mathbf T$ collects coefficients from one source; each column of $\mathbf L_D$ instead records the effect of one coefficient $T_{ti}$ on all observations. Before assigning values, scan these coefficient columns of $\mathbf L_D$ in
decreasing \textit{source} order, with a fixed order within each
\textit{source}. Retain a column exactly when it is independent of those
already retained. This is \textit{reverse-source basis selection}.
In the restricted map of Example~\ref{ex:kernel-spectrum}, retain
$b_1$ at \textit{source} 5, then discard the identical $b_0$ column
and retain the independent $c$ column at \textit{source} 4.
Discarding the zero $\alpha$ column at \textit{source} 3 leaves
the coordinates $c,b_1$ used in Fig.~\ref{fig:source-choices}(b).

\begin{proposition}[Exact recovery and source-compatible reduction]
\label{prop:global-completion}
Fix the constituent pair, let $D=d_\star-1$, and assume
$2d_{\min}(\mathcal C_{\mathrm R})\ge d_\star$.

\textbf{(i) Recovery from representatives.} For every linear complement $\mathcal V$ of $\ker\mathbf L_D$
in $\F^{fk_{\mathrm L}}$,
\begin{equation}
 \begin{aligned}
 \mathcal X&=\{\boldsymbol\theta:\mathbf L_D\boldsymbol\theta\in\mathcal Y_D\}
             =(\mathcal X\cap\mathcal V)+\ker\mathbf L_D,\\
 |\mathcal X|&=2^{fk_{\mathrm L}-\rho}|\mathcal Y_D|.
 \end{aligned}
 \label{eq:joint-support-count}
\end{equation}
Each feasible observation has one representative in $\mathcal V$.

\textbf{(ii) Reduction for sequential assignment.} Let
$\mathcal P_i\subseteq[f]$ record the row positions of the coefficients
retained in $\mathbf t_i$ by \textit{reverse-source basis selection},
and put $p_i=|\mathcal P_i|$.
Vectors supported on these entries form a complement as in (i), with
\begin{equation}
 \sum_i p_i=\rho,\qquad
 \rk\bigl((\mathbf Q_{\mathbf a})_{\mathcal P_i}\bigr)=r_{\mathbf a}
       \quad(\mathbf m\in\mathcal L_i).
 \label{eq:source-compatible-quotient}
\end{equation}
For any current column $\mathbf t_i$, one assignment supported on
$\mathcal P_i$ reproduces $\mathbf Q_{\mathbf a}\mathbf t_i$
simultaneously for every $\mathbf m\in\mathcal L_i$.

\textbf{(iii) Reduction on a restricted domain.} In
$\boldsymbol\theta=\boldsymbol\theta_0+\mathbf R\boldsymbol\gamma$,
$\boldsymbol\theta_0$ is a fixed configuration and the independent
columns of $\mathbf R$ span the admissible coefficient variations.
Let $\boldsymbol\gamma\in\F^p$ be the free parameters. Put
$\rho_{\mathbf R}=\rk(\mathbf L_D\mathbf R)$ and
$\mathcal Y_{\mathbf R}=(\mathbf L_D\boldsymbol\theta_0+
\operatorname{im}(\mathbf L_D\mathbf R))\cap\mathcal Y_D$.
There are $2^{p-\rho_{\mathbf R}}|\mathcal Y_{\mathbf R}|$ feasible
configurations. Restricting $\boldsymbol\gamma$ to a complement of
$\ker(\mathbf L_D\mathbf R)$ leaves $|\mathcal Y_{\mathbf R}|$
feasible representatives among $2^{\rho_{\mathbf R}}$ points,
preserving the feasible fraction
$|\mathcal Y_{\mathbf R}|/2^{\rho_{\mathbf R}}$.
For columnwise construction, let
$\mathcal D_i=\mathbf t_i^0+\operatorname{im}\mathbf B_i$,
where $\mathbf t_i^0\in\F^f$ is a fixed base column and
$\mathbf B_i$ has $f$ rows and linearly independent columns.
These column constraints are independent: any choices
$\mathbf t_i\in\mathcal D_i$ can be combined into an admissible $\mathbf T$.
Then take $\mathbf R=\operatorname{diag}(\mathbf B_0,\ldots,
\mathbf B_{k_{\mathrm L}-1})$ and let $\boldsymbol\theta_0$
stack the fixed base columns $\mathbf t_i^0$ in source order.
The parameter vector $\boldsymbol\gamma$ stacks the corresponding
blocks $\boldsymbol\gamma_i$, one entry per column of $\mathbf B_i$.
\textit{Reverse-source basis selection} on $\mathbf L_D\mathbf R$ preserves
each group's joint current-column observations within that domain.
With earlier columns fixed, the observation increments supplied by the current column range over $\operatorname{im}(\mathbf Q_{\mathbf a}\mathbf B_i)$, which need not be all of $\F^{r_{\mathbf a}}$. The full observation also includes the fixed base column and earlier-column contributions.
\end{proposition}
\begin{IEEEproof}
\textbf{(i) Recovery and counting.}
The first equality in \eqref{eq:joint-support-count} restates
\eqref{eq:outer-condition} using \eqref{eq:feasible-observation-image}.
Write the unique decomposition
$\boldsymbol\theta=\boldsymbol\zeta+\boldsymbol\kappa$,
with $\boldsymbol\zeta\in\mathcal V$ and
$\boldsymbol\kappa\in\ker\mathbf L_D$. Since
$\mathbf L_D\boldsymbol\theta=\mathbf L_D\boldsymbol\zeta$,
the two configurations are feasible together. Thus $\mathcal X$
is the disjoint union of $\boldsymbol\zeta+\ker\mathbf L_D$ over
$\boldsymbol\zeta\in\mathcal X\cap\mathcal V$.
The map $\mathbf L_D$ sends $\mathcal V$ bijectively onto its image,
so there are $|\mathcal Y_D|$ feasible representatives.
Rank--nullity gives $2^{fk_{\mathrm L}-\rho}$ configurations per
representative, proving \eqref{eq:joint-support-count}.

\textbf{(ii) Source-compatible coordinates.}
The retained columns form a basis of $\operatorname{im}\mathbf L_D$,
so $\sum_i p_i=\rho$ and setting all other coefficients to zero
defines a complement as in (i).

Index a column by its coefficient $T_{ti}$. A discarded column
is a sum of retained columns from \textit{sources} $j\ge i$.
Restrict this single column dependence to all observation blocks
in $\mathcal L_i$. The block for $\mathbf m\in\mathcal L_i$
of the column indexed by $T_{tj}$ is
\[
 m_j(\mathbf Q_{\mathbf a})_{:,t},\qquad
 m_i=1,\quad m_j=0\ (j>i).
\]
The later-source terms vanish on every block in this group.
Thus some coefficients $c_\ell\in\F$, independent of
$\mathbf m$, satisfy
\[
 (\mathbf Q_{\mathbf a})_{:,t}
 =\sum_{\ell\in\mathcal P_i}c_\ell
       (\mathbf Q_{\mathbf a})_{:,\ell}
 \qquad(\mathbf m\in\mathcal L_i).
\]
Thus each discarded coefficient can be replaced using retained
entries at \textit{source} $i$, with one replacement reproducing its
effects on the entire group. Applying these replacements to all
discarded entries proves the joint preservation claim.
In particular, the retained columns span the image of each
$\mathbf Q_{\mathbf a}$, whose rank is $r_{\mathbf a}$.
This gives the rank identities in \eqref{eq:source-compatible-quotient}.

\textbf{(iii) Restricted affine domains.}
The fixed observation $\mathbf L_D\boldsymbol\theta_0$ translates
the image of $\boldsymbol\gamma\mapsto\mathbf L_D\mathbf R\boldsymbol\gamma$.
Applying (i) to this linear map gives $2^{p-\rho_{\mathbf R}}$
parameter assignments per attainable observation. The independent
columns of $\mathbf R$ make these distinct \textit{coefficient configurations}.
Restricting to a complement of $\ker(\mathbf L_D\mathbf R)$ keeps
one representative per observation, dividing both the total and
feasible counts by $2^{p-\rho_{\mathbf R}}$ and preserving their ratio.
For the independently constrained columns, write
$\mathbf t_j=\mathbf t_j^0+\mathbf B_j\boldsymbol\gamma_j$.
For $\mathbf m\in\mathcal L_i$, $m_i=1$ and $m_j=0$ for $j>i$, so
\[
 \mathbf Q_{\mathbf a}\mathbf T\mathbf m
 =\mathbf Q_{\mathbf a}\Bigl(\sum_{j<i}m_j\mathbf t_j+\mathbf t_i^0\Bigr)
    +\mathbf Q_{\mathbf a}\mathbf B_i\boldsymbol\gamma_i.
\]
Once earlier columns are fixed, only the last term can change.
As in (ii), a discarded parameter column of $\mathbf L_D\mathbf R$
at \textit{source} $i$ is a sum of retained columns at sources $j\ge i$.
On every block in $\mathcal L_i$, the later terms vanish because $m_j=0$.
The same retained parameters at \textit{source} $i$ therefore
reproduce the variable observations for the entire group.
The base columns remain fixed, so recovered columns stay in $\mathcal D_i$.
\end{IEEEproof}

Proposition~\ref{prop:global-completion}(iii) gives
$\mathcal Y_{\mathbf R}\ne\varnothing$ as the condition for an affine
domain to contain a feasible configuration. Requiring every configuration
in that domain to be feasible gives the following stronger condition.
\begin{corollary}[Distance preservation in an affine domain]
\label{cor:affine-distance-domain}
Fix the constituents and assume
$2d_{\min}(\mathcal{C}_{\mathrm{R}})\ge d_{\star}$, with $D=d_{\star}-1$.
For the affine domain $\boldsymbol\theta=\boldsymbol\theta_0+\mathbf R\boldsymbol\gamma$
of Proposition~\ref{prop:global-completion}(iii), every configuration is feasible
if and only if
\begin{equation}
 (\mathbf{L}_{\mathbf{m}}\boldsymbol{\theta}_{0}+\operatorname{im}(\mathbf{L}_{\mathbf{m}}\mathbf{R}))
       \cap\mathcal B_{\mathbf{a}}(D)=\varnothing
 \quad(0<w_{\mathbf{a}}\le D).
 \label{eq:affine-domain-distance}
\end{equation}
\end{corollary}
\begin{IEEEproof}
Each message's affine attainable set is
$\mathbf L_{\mathbf m}\boldsymbol\theta_0+
\operatorname{im}(\mathbf L_{\mathbf m}\mathbf R)$.
Every parameter assignment satisfies every message condition exactly
when each such set avoids $\mathcal B_{\mathbf a}(D)$.
Together with the assumed right-constituent distance condition,
Theorem~\ref{thm:polar-spectrum}(i) gives \eqref{eq:affine-domain-distance}.
\end{IEEEproof}

In Example~\ref{ex:kernel-spectrum}, the full four-bit domain
contains feasible configurations because it reaches the observation
$(1,1)^{\mathrm T}$ with $A_2=0$. Fixing that joint observation makes
each message's attainable set a single value outside its forbidden set.
Every configuration in the resulting domain is therefore feasible.

\subsection{Conditional Completion Guarantees}
\label{subsec:support-feasibility}
\subsubsection{From joint feasibility to remaining choices}
Enumerating all elements of $\mathcal Y_D$ would require solving all exclusions.
Instead, we choose coefficients bit by bit and examine what the
unfixed bits can still supply. For the polar-quotient conditions, assume
$2d_{\min}(\mathcal C_{\mathrm R})\ge d_\star$.

Use the affine parameterization
$\boldsymbol\theta=\boldsymbol\theta_0+\mathbf R\boldsymbol\gamma$
of Proposition~\ref{prop:global-completion}(iii), with
$\boldsymbol\gamma\in\F^p$ and independent columns of $\mathbf R$.
A \textit{partial coefficient assignment} $\boldsymbol\pi$ fixes the first
$|\boldsymbol\pi|$ free parameter bits in a chosen order.
The notation $\boldsymbol\gamma\succeq\boldsymbol\pi$ means that
$\boldsymbol\gamma$ extends those fixed bits. Reordering the
parameters accommodates any specified partial assignment.
The partial assignment has a feasible completion exactly when
\[
 \{\mathbf L_D(\boldsymbol\theta_0+\mathbf R\boldsymbol\gamma):
      \boldsymbol\gamma\succeq\boldsymbol\pi\}\cap\mathcal Y_D
      \ne\varnothing.
\]
For any $\boldsymbol\gamma\succeq\boldsymbol\pi$, the joint observation
$\mathbf L_D\boldsymbol\theta$ stacks the individual values
$\mathbf L_{\mathbf m}\boldsymbol\theta=\mathbf Q_{\mathbf a}\mathbf T\mathbf m$.
For a fixed \textit{critical message} $\mathbf m\in\mathcal M_D$,
$\mathcal Y_{\mathbf m}(\boldsymbol\pi)$ collects these values as the
unfixed bits vary, including any attainable values in $\mathcal B_{\mathbf a}(D)$.
An assignment producing a value in $\mathcal B_{\mathbf a}(D)$
violates the exclusion condition for this message.
Feasibility therefore requires one common parameter assignment to satisfy
$\mathbf L_{\mathbf m}\boldsymbol\theta\notin\mathcal B_{\mathbf a}(D)$
for every \textit{critical message}, as imposed by $\mathcal Y_D$.
The individual observations share the unfixed parameters and cannot
be chosen independently.
The \textit{attainable syndrome set} and conditional mean are
\begin{equation}
 \begin{aligned}
 \mathcal Y_{\mathbf m}(\boldsymbol\pi)
 &=\{\mathbf L_{\mathbf m}(\boldsymbol\theta_0+\mathbf R\boldsymbol\gamma):
                  \boldsymbol\gamma\succeq\boldsymbol\pi\},\\
 \Phi(\boldsymbol\pi)
 &=\sum_{\substack{\mathbf m\ne0\\w_{\mathbf a}\le D}}
 \frac{|\mathcal B_{\mathbf a}(D)\cap\mathcal Y_{\mathbf m}(\boldsymbol\pi)|}
      {|\mathcal Y_{\mathbf m}(\boldsymbol\pi)|}.
 \end{aligned}
 \label{eq:general-conditional-mean}
\end{equation}
To compute $\mathcal Y_{\mathbf m}(\boldsymbol\pi)$, first add the
contributions of the fixed bits. Each remaining bit can add its
column of $\mathbf L_{\mathbf m}\mathbf R$, so all sums of these
columns give the remaining choices. In Example~\ref{ex:kernel-spectrum},
the restricted map shows that fixing $c=1$ settles the first
observation and leaves both values of the second attainable through $b_1$.

Each summand in $\Phi(\boldsymbol\pi)$ is the fraction of attainable values that
violate one message condition. Uniform unfixed bits give equal
probabilities to those values. Thus $\Phi$ is the expected number
of violated conditions. The \textit{strict mean bound} $\Phi(\boldsymbol\pi)<1$ therefore guarantees a completion violating none.
The conditional-expectation method finds such a completion by
choosing each next bit so that the mean does not increase
\cite{raghavan1988conditional}.

We can also apply the conditional-mean calculation to one \textit{source} group $\mathcal L_i$ at a time.
Later columns cannot change a group once its last \textit{source} is fixed.
For the groups $\mathcal L_i$ in \eqref{eq:last-source-groups}, use the
independently constrained columns $\mathbf t_i\in\mathcal D_i$ of
Proposition~\ref{prop:global-completion}(iii).
After fixing earlier columns, put
$\mathbf b_{\mathbf m}=\sum_{j<i}m_j\mathbf t_j$.
The current column $\mathbf t_i\in\mathcal D_i$ must satisfy
\begin{equation}
 \mathbf Q_{\mathbf a}(\mathbf t_i+\mathbf b_{\mathbf m})
 \notin\mathcal B_{\mathbf a}(D)
 \qquad(\mathbf m\in\mathcal L_i).
 \label{eq:source-allowed}
\end{equation}
For the full column space, define
\begin{equation}
 \lambda_i=\sum_{\mathbf m\in\mathcal L_i}
                2^{f-r_{\mathbf a}}|\mathcal B_{\mathbf a}(D)|.
 \label{eq:source-bound}
\end{equation}
For any fixed earlier columns, each summand counts choices
$\mathbf t_i\in\F^f$ violating one message condition.
Thus $\lambda_i$ bounds the rejected choices for $\mathcal L_i$,
counting overlaps repeatedly. The ratio $\lambda_i/2^f$ is the
initial mean number of violations under a uniform current column.
The next theorem uses $\lambda_i<2^f$ to guarantee extension
after every accepted assignment of the earlier columns.
When columns are linked, compute the attainable set of each message
in \eqref{eq:general-conditional-mean} using the same free parameters
$\boldsymbol\gamma$, rather than choosing columns independently.

\begin{theorem}[Conditional completion and source-wise extendability]
\label{thm:conditional-completion}
(i) On an injectively parameterized affine domain with the complete
fixed-constituent exclusion conditions \eqref{eq:outer-condition}, the number of satisfying configurations
extending any partial assignment $\boldsymbol\pi$ is at least
\begin{equation}
 \left\lceil2^{p-|\boldsymbol\pi|}[1-\Phi(\boldsymbol\pi)]_+\right\rceil,
 \qquad [t]_+=\max\{t,0\}.
 \label{eq:completion-count}
\end{equation}
If $\Phi(\boldsymbol\pi)<1$, choosing the smaller-mean child at
each step, with zero on ties, gives a satisfying assignment in
at most $p-|\boldsymbol\pi|$ further decisions without backtracking.
The same assertion holds on an injectively parameterized affine
coefficient domain for the complete \textit{common-supercode} conditions of
Theorem~\ref{thm:shared-elimination}, using the mean specified below.
This form also permits constituent coefficients to vary together
with the outer coefficients.

(ii) For the full outer coefficient space $\F^{f\times k_{\mathrm L}}$, suppose
$2d_{\min}(\mathcal C_{\mathrm R})\ge d_\star$ and $\lambda_i<2^f$
for every \textit{source} $i$. There are at least
\begin{equation}
 \prod_{i=0}^{k_{\mathrm L}-1}(2^f-\lambda_i)
 \label{eq:source-completion-count}
\end{equation}
distinct outer completions of distance at least $d_\star$.
Source-wise conditional means find one without intersource
backtracking. The \textit{source-compatible reduction} in
Proposition~\ref{prop:global-completion} bounds the number of binary
decisions by $\rho$.
\end{theorem}
\begin{IEEEproof}
\textbf{(i) Affine-image completion.}
Choose the $p-|\boldsymbol\pi|$ unfixed bits uniformly and let
\[
 Z(\boldsymbol\gamma)=\sum_{\substack{\mathbf m\ne0\\w_{\mathbf a}\le D}}
 \ind{\mathbf L_{\mathbf m}(\boldsymbol\theta_0+\mathbf R\boldsymbol\gamma)
                          \in\mathcal B_{\mathbf a}(D)}.
\]
Each indicator contributes zero or one. Thus $Z(\boldsymbol\gamma)$
counts the violated message conditions and takes nonnegative integer values.
It is zero exactly when all conditions hold.
Apply the equal-preimage argument of
Proposition~\ref{prop:global-completion}(iii) to each affine message
map after fixing $\boldsymbol\pi$. Uniform unfixed bits give a
uniform distribution on $\mathcal Y_{\mathbf m}(\boldsymbol\pi)$, so
\[
 \PP\{\mathbf L_{\mathbf m}\boldsymbol\theta\in\mathcal B_{\mathbf a}(D)
                                      \mid\boldsymbol\pi\}
 =\frac{|\mathcal B_{\mathbf a}(D)\cap\mathcal Y_{\mathbf m}(\boldsymbol\pi)|}
       {|\mathcal Y_{\mathbf m}(\boldsymbol\pi)|},\qquad
 \EE[Z\mid\boldsymbol\pi]=\Phi(\boldsymbol\pi).
\]
Linearity of expectation requires no independence between
violations. Since $\ind{Z\ge1}\le Z$,
\[
 \begin{aligned}
 \PP\{Z\ge1\mid\boldsymbol\pi\}&\le\Phi(\boldsymbol\pi),\\
 |\{\boldsymbol\gamma\succeq\boldsymbol\pi:Z(\boldsymbol\gamma)=0\}|
 &=2^{p-|\boldsymbol\pi|}\PP\{Z=0\mid\boldsymbol\pi\}\\
 &\ge2^{p-|\boldsymbol\pi|}(1-\Phi(\boldsymbol\pi)).
 \end{aligned}
\]
Taking the nonnegative integer lower bound and using injectivity
gives \eqref{eq:completion-count} for distinct configurations.
The two children partition the extensions into equal parts, so
\[
 \min_{b\in\{0,1\}}\Phi(\boldsymbol\pi b)
 \le\tfrac12\sum_{b\in\{0,1\}}\Phi(\boldsymbol\pi b)=\Phi(\boldsymbol\pi).
\]
Repeated minimum-mean choices yield a full assignment
$\boldsymbol\gamma^\star$ with
\[
 0\le Z(\boldsymbol\gamma^\star)=\Phi(\boldsymbol\gamma^\star)
       \le\Phi(\boldsymbol\pi)<1.
\]
The integer $Z(\boldsymbol\gamma^\star)$ is therefore zero.
There are exactly $p-|\boldsymbol\pi|$ remaining bit choices.

For the assertion allowing constituent coefficients to vary,
use the complete \textit{common-supercode} conditions of
Theorem~\ref{thm:shared-elimination}. Let $\mathbf V(\boldsymbol\gamma)$
parameterize the admissible increment matrices by an injective affine map.
For each $\mathbf q\in\mathcal Q$, the attainable set is
\[
 \mathcal Y_{\mathbf q}(\boldsymbol\pi)
 =\{\mathbf V(\boldsymbol\gamma)\mathbf q^{\mathrm T}:
     \boldsymbol\gamma\succeq\boldsymbol\pi\}.
\]
Compute $\Phi$ as in \eqref{eq:general-conditional-mean}, summing over
$\mathbf q\in\mathcal Q$ with $\mathcal Y_{\mathbf q}(\boldsymbol\pi)$
and $\mathcal B(\mathbf q)$. Since
$\mathbf V(\boldsymbol\gamma)\mathbf q^{\mathrm T}$ is affine in the
remaining bits, the same probability, counting, and bit-selection
arguments apply to these conditions.
Theorem~\ref{thm:shared-elimination} equates zero violations with
complete elimination of the forbidden weight layers.

\textbf{(ii) Source-wise extension.}
Fix any earlier columns satisfying groups $\mathcal L_{0},
\ldots,\mathcal L_{i-1}$.
For each $\mathbf m\in\mathcal L_i$ and
$\mathbf y\in\F^{r_{\mathbf a}}$, full row rank of
$\mathbf Q_{\mathbf a}$ in \eqref{eq:source-allowed} gives
\[
 |\{\mathbf{t}_{i}:\mathbf{Q}_{\mathbf{a}}(\mathbf{t}_{i}+\mathbf b_{\mathbf{m}})=\mathbf{y}\}|
 =2^{f-r_{\mathbf{a}}}.
\]
This count is independent of $\mathbf b_{\mathbf m}$.
Distinct forbidden syndromes have disjoint preimages.
Taking the union over $\mathbf{m}\in\mathcal L_{i}$ leaves at least
\[
 2^{f}-\left|\bigcup_{\mathbf{m}\in\mathcal L_{i}}
       \{\mathbf{t}_{i}:\mathbf{Q}_{\mathbf{a}}(\mathbf{t}_{i}+\mathbf b_{\mathbf{m}})\in\mathcal B_{\mathbf{a}}(D)\}\right|
 \ge2^{f}-\sum_{\mathbf{m}\in\mathcal L_{i}}
                    2^{f-r_{\mathbf{a}}}|\mathcal B_{\mathbf{a}}(D)|
       =2^{f}-\lambda_{i}>0.
\]
Every accepted assignment of the earlier columns therefore has
at least $2^f-\lambda_i$ choices for the next column.
Induction over the columns multiplies these lower bounds, giving
\eqref{eq:source-completion-count}.
For each accepted earlier assignment the initial group mean is
$\sum_{\mathbf{m}\in\mathcal L_{i}}|\mathcal B_{\mathbf{a}}(D)|/2^{r_{\mathbf{a}}}
=\lambda_i/2^f<1$.
Part (i) therefore constructs the next column. Later columns
cannot change an accepted group, since its messages have
$m_{j}=0$ for $j>i$. The right-constituent distance condition $2d_{\min}(\mathcal C_{\mathrm R})\ge d_\star$ and \eqref{eq:outer-condition} then prove the distance.

After \textit{source-compatible reduction},
\eqref{eq:source-compatible-quotient} gives
$\rk\bigl((\mathbf{Q}_{\mathbf{a}})_{\mathcal P_{i}}\bigr)=r_{\mathbf{a}}$. Each syndrome then has $2^{p_{i}-r_{\mathbf{a}}}$
preimages among the $2^{p_{i}}$ retained choices. Its violation
probability remains $|\mathcal B_{\mathbf{a}}(D)|/2^{r_{\mathbf{a}}}$,
so the group still satisfies the \textit{strict mean bound}.
Each \textit{source} needs at most $p_{i}$ decisions, with
$\sum_{i} p_{i}=\rho$.
\end{IEEEproof}

\subsubsection{Exact completion of the exclusion conditions}
\label{subsec:exact-solvers}
\label{subsec:outer-algorithm}
When the \textit{strict mean bounds} are inconclusive, we use an exact
parameter-assignment procedure, denoted $\proc{SolveDomain}$.
It assigns the free bits successively and uses attainable sets to
reject infeasible continuations.

For fixed constituents $\mathcal C_{\mathrm L},\mathcal C_{\mathrm R}$,
first use Algorithm~\ref{alg:directory} to check
$2d_{\min}(\mathcal C_{\mathrm R})\ge d_\star$.
If it fails, return $\bot$ for this pair. Otherwise, use the algorithm
to list the left codewords of the \textit{critical messages} $\mathbf m\in\mathcal M_D$.
These codewords give $\mathbf Q_{\mathbf a}$ and $\mathcal B_{\mathbf a}(D)$.
Reject a partial assignment $\boldsymbol\pi$ when
$\mathcal Y_{\mathbf m}(\boldsymbol\pi)\subseteq\mathcal B_{\mathbf a}(D)$
for any such message.

If coefficients within the constituents may also vary, use the
complete \textit{common-supercode} directory conditions of
Theorem~\ref{thm:shared-elimination}. Reject a partial assignment
when $\mathcal Y_{\mathbf q}(\boldsymbol\pi)\subseteq\mathcal B(\mathbf q)$
for any \textit{source pattern} $\mathbf q\in\mathcal Q$.
In either formulation, a feasible leaf supplies a value outside
the corresponding \textit{forbidden set} for every condition.
Thus the rejection test preserves its entire path.
At a full assignment each attainable set is a singleton.
Passing all tests is then equivalent to feasibility by
Theorems~\ref{thm:polar-spectrum} and \ref{thm:shared-elimination},
respectively.

Visit both children of each unrejected assignment in increasing
$(\Phi(\boldsymbol\pi b),b)$ order, where $b$ is the next bit.
The children partition the extensions. $\proc{SolveDomain}$ returns
the first accepted leaf or $\bot$ after a complete failed traversal.
Continuing past accepted leaves lists all feasible assignments in the
original domain, using joint parameters for linked columns.
For independent column domains $\mathcal D_i$, Algorithm~\ref{alg:outer}
reduces the domain and searches group by group. Its feasible leaves are
representatives. Insert zeros for discarded parameters, then add the
vectors in $\ker(\mathbf L_D\mathbf R)$ to recover all original feasible
configurations by Proposition~\ref{prop:global-completion}(iii).

To complete the outer matrix source by source, write each
$\mathbf t_i\in\mathcal D_i$ using the independent columns of $\mathbf B_i$
as in Proposition~\ref{prop:global-completion}.
Each column of $\mathbf B_i$ has one parameter bit in $\boldsymbol\gamma_i$.
After reduction, $p_i$ counts the retained columns of $\mathbf B_i$.
The unrestricted case has $\mathbf{t}_{i}^{0}=0$, $\mathbf{B}_{i}=\mathbf I_{f}$.
For a partial assignment $\boldsymbol\pi$ of the parameters
$\boldsymbol\gamma_i$ of column $\mathbf t_i$,
the attainable syndrome set $\mathcal Y_{\mathbf m}(\boldsymbol\pi)$ in \eqref{eq:general-conditional-mean} is
\begin{equation}
 \mathcal Y_{\mathbf m}(\boldsymbol\pi)=\{
 \mathbf Q_{\mathbf a}(\mathbf b_{\mathbf m}+\mathbf t_i^0
                 +\mathbf B_i\boldsymbol\gamma_i):
 \boldsymbol\gamma_i\succeq\boldsymbol\pi\},\qquad\mathbf m\in\mathcal L_i.
 \label{eq:source-mean-specialization}
\end{equation}
Write $\Phi_i$ for the sum in \eqref{eq:general-conditional-mean}
restricted to $\mathcal L_i$, using these images.
In Algorithm~\ref{alg:outer}, $\boldsymbol{\pi} b$ appends bit $b$ to
this column, $()$ denotes the empty assignment, and
$\mathcal B_{\mathbf{a}}$ abbreviates $\mathcal B_{\mathbf{a}}(D)$. A full
assignment $\boldsymbol{\pi}$ is interpreted as a column vector in $\mathbf{t}_{i}=\mathbf{t}_{i}^{0}+\mathbf{B}_{i}\boldsymbol{\pi}$.

\begin{algorithm}[!t]
\caption{Exact outer completion with conditional-mean ordering}
\label{alg:outer}
\small
\begin{algorithmic}[1]
\Require Fixed $\mathcal{C}_{\mathrm{L}},\mathcal{C}_{\mathrm{R}}$, cutoff $D$, independently constrained columns $(\mathbf{t}_{i}^{0},\mathbf{B}_{i})$
\Ensure An admissible $\mathbf{T}$ excluding weights $1,\ldots,D$, or $\bot$
\If{$\mathcal{C}_{\mathrm{R}}$ has a positive weight $\le\lfloor D/2\rfloor$}
 \State \Return $\bot$
\EndIf
\State Enumerate $\mathbf{a}(\mathbf{m})\ne0$ with $\wt(\mathbf{a}(\mathbf{m}))\le D$
\State $\mathbf P\gets\mathbf H_{\mathrm R}\mathbf G_h^{\mathrm T}$; initialize all groups $\mathcal L_i\gets\varnothing$
\ForAll{these messages $\mathbf{m}$}
 \State Form a left-nullspace basis $\mathbf Q_{\mathbf a}$ of $\mathbf P_{\mathcal S_{\mathbf a}}$; compute $t_{\mathbf a}$ by \eqref{eq:outer-data}; $\mathcal B_{\mathbf{a}}\gets\{0\}$
 \ForAll{$\mathcal{U}\subseteq \mathcal{O}_{\mathbf{a}}$ with $1\le|\mathcal{U}|\le t_{\mathbf{a}}$}
  \State $\mathcal B_{\mathbf{a}}\gets\mathcal B_{\mathbf{a}}\cup\{\sum_{j\in \mathcal{U}}\mathbf{Q}_{\mathbf{a}}\mathbf{P}_{:,j}\}$
 \EndFor
 \State Place $\mathbf{m}$ in $\mathcal L_{\max\{i:m_{i}=1\}}$ \Comment{last source, (\ref{eq:last-source-groups})}
\EndFor
\State Stack $\mathbf L_D$ by \eqref{eq:joint-check-matrix}; set $\mathbf{M}\gets \mathbf{L}_{D}\operatorname{diag}(\mathbf{B}_{0},\ldots,\mathbf{B}_{k_{\mathrm{L}}-1})$
\State $\mathcal P\gets\proc{ColumnBasis}(\mathbf{M})$ by \textit{reverse-source basis selection}
\State Retain the selected columns of each $\mathbf B_i$; set $p_i$ to its remaining column count
\State \Return $\Call{Complete}{0,(),()}$
\Function{Complete}{$i,(\mathbf{t}_{0},\ldots,\mathbf{t}_{i-1}),\boldsymbol{\pi}$}
 \If{$i=k_{\mathrm{L}}$} \State \Return $(\mathbf{t}_{0},\ldots,\mathbf{t}_{k_{\mathrm{L}}-1})$ \EndIf
 \State Compute $\mathcal Y_{\mathbf{m}}(\boldsymbol{\pi})$ for $\mathbf m\in\mathcal L_i$ using \eqref{eq:source-mean-specialization}
 \If{some $\mathcal Y_{\mathbf{m}}(\boldsymbol{\pi})\subseteq\mathcal B_{\mathbf{a}(\mathbf{m})}$}
  \State \Return $\bot$
 \EndIf
 \If{$|\boldsymbol{\pi}|=p_{i}$}
  \State $\mathbf{t}_{i}\gets \mathbf{t}_{i}^{0}+\mathbf{B}_{i}\boldsymbol{\pi}$
  \State \Return $\Call{Complete}{i+1,(\mathbf{t}_{0},\ldots,\mathbf{t}_{i}),()}$
 \EndIf
 \State Compute $\Phi_{i}(\boldsymbol{\pi}0),\Phi_{i}(\boldsymbol{\pi}1)$ exactly \Comment{group mean, (\ref{eq:general-conditional-mean})}
 \ForAll{$b\in\{0,1\}$ ordered by $(\Phi_{i}(\boldsymbol{\pi} b),b)$}
  \State $\mathbf{T}\gets\Call{Complete}{i,(\mathbf{t}_{0},\ldots,\mathbf{t}_{i-1}),\boldsymbol{\pi} b}$
  \If{$\mathbf{T}\ne\bot$} \State \Return $\mathbf{T}$ \EndIf
 \EndFor
 \State \Return $\bot$
\EndFunction
\end{algorithmic}
\end{algorithm}

The indices returned by $\proc{ColumnBasis}$ select local columns
of each $\mathbf B_i$, leaving $\mathbf t_i^0$ unchanged.
Fixed columns have $p_i=0$.

\begin{example}[From zero coefficients to a complete design]
\label{ex:toy-completion}
Consider the \textit{source-compatible} basis in
Fig.~\ref{fig:source-choices}(b), with $\alpha=b_0=0$ and $c,b_1$ free.
The right constituent satisfies $2d_{\min}(\mathcal C_{\mathrm R})=4=d_\star$.
Before either bit is assigned, each message
condition is violated for half of the four assignments, giving
$\Phi(())=1/2+1/2=1$. The \textit{strict mean bound} is therefore
inconclusive for the two messages together. In Algorithm~\ref{alg:outer},
however, each source group starts with mean $1/2$ and admits a choice
without backtracking. The table gives the violation counts for each bit.
\[
 \begin{array}{c|c|cc|c}
 \text{source position}&\text{initial mean}&
 \text{bit }0&\text{bit }1&\text{choice}\\\hline
 4&1/2&1&0&c=1\\
 5\text{ with }c=1&1/2&0&1&b_1=0
 \end{array}
\]
The unique feasible representative is $(\alpha,b_0,c,b_1)=(0,0,1,0)$,
with observation pair $(c,b_0+b_1+c)=(1,1)$.
It gives $u_9=u_4$ and sets all other frozen bits to zero.
The distance target is satisfied, and $\mathbf g_{12}$ is a weight-4 witness, giving $d_{\min}=4$. Adding the two kernel basis vectors of
Example~\ref{ex:kernel-spectrum} gives the four feasible assignments
$(\alpha,b_0,1,b_0)$ with $(\alpha,b_0)\in\F^2$.
\end{example}

\begin{example}[Counting bounds and source-wise completion at length 32]
\label{ex:n32-completion}
The running example shows how to recover coefficients and assign
them sequentially. We now illustrate why the two counting bounds
differ and how reduction preserves the choices used in completion.
Take two unprecoded length-16 constituents with
\[
 \mathcal I_{\mathrm L}=\{3,5\},\qquad
 \mathcal I_{\mathrm R}=\{3,5,9,10,11,12,13,14,15\}.
\]
The length-32 information set is
$\mathcal I=\mathcal I_{\mathrm L}\cup\{16+j:j\in\mathcal I_{\mathrm R}\}$.
All 14 entries of $\mathbf T\in\mathbb F_2^{7\times2}$ are free.
\begin{figure*}[!t]
\centering
\begin{tikzpicture}
\begin{axis}[width=.44\textwidth,height=5.0cm,
 xlabel={Trial distance $d_\star$},ylabel={Fraction of outer matrices},
 xmin=3.8,xmax=9.2,ymin=-.04,ymax=1.06,xtick={4,5,6,7,8,9},
 tick label style={font=\small},label style={font=\small},
 title={(a) Exact feasibility and a sufficient bound},title style={font=\small},
 legend style={font=\scriptsize,at={(.98,.98)},anchor=north east},
 grid=major,grid style={gray!20}]
\addplot[blue!65!black,thick,mark=*] coordinates {(4,1) (5,0.822265625) (6,0.822265625) (7,0.09375) (8,0.09375) (9,0)};
\addlegendentry{Exact feasible fraction}
\addplot[red!70!black,dashed,thick,mark=square*] coordinates {(4,1) (5,0.8125) (6,0.8125) (7,0) (8,0) (9,0)};
\addlegendentry{$[1-\Phi(())]_+$}
\node[anchor=west,font=\scriptsize,fill=white,inner sep=2pt]
 at (rel axis cs:.04,.42) {\shortstack[l]{$d_\star=6$, reduced space:\\[2pt]
 $\underbrace{416}_{\mathrm{(i)}}\le\underbrace{420}_{\mathrm{(ii)}}
 \le\underbrace{421}_{\mathrm{exact}}$}};
\end{axis}
\end{tikzpicture}\hfill
\begin{tikzpicture}
\begin{axis}[width=.44\textwidth,height=5.0cm,
 xlabel={Number of fixed parameters},ylabel={Current-source mean $\Phi_i(\boldsymbol{\pi})$},
 xmin=-.3,xmax=14.3,ymin=-.01,ymax=.20,xtick={0,2,4,6,8,10,12,14},
 ytick={0,.0625,.125},yticklabels={$0$,$1/16$,$1/8$},
 tick label style={font=\small},label style={font=\small},
 title={(b) Source-wise execution at $d_\star=6$},
 title style={font=\small},grid=major,grid style={gray!20},
 legend style={font=\scriptsize,at={(.98,.98)},anchor=north east}]
\addplot[blue!65!black,thick,mark=*] coordinates {(0,0.0625) (1,0.0625) (2,0.0625) (3,0.0625) (4,0) (5,0) (6,0) (7,0)};
\addlegendentry{Without reduction: 14}
\addplot[blue!65!black,thick,mark=*,forget plot] coordinates {(7,0.125) (8,0.125) (9,0.125) (10,0.0625) (11,0) (12,0) (13,0) (14,0)};
\draw[blue!65!black,densely dotted,->] (axis cs:7,0) -- (axis cs:7,0.125);
\node[blue!65!black,font=\scriptsize,anchor=east] at (axis cs:6.85,0.1) {$i=1$};
\addplot[red!70!black,dashed,thick,mark=square*,mark size=2pt] coordinates {(0,0.0625) (1,0) (2,0) (3,0) (4,0)};
\addlegendentry{Algorithm~\ref{alg:outer}: 9}
\addplot[red!70!black,dashed,thick,mark=square*,mark size=2pt,forget plot] coordinates {(4,0.125) (5,0.0625) (6,0) (7,0) (8,0) (9,0)};
\draw[red!70!black,densely dotted,->] (axis cs:4,0) -- (axis cs:4,0.125);
\node[red!70!black,font=\scriptsize,anchor=east] at (axis cs:3.85,0.1) {$i=1$};
\end{axis}
\end{tikzpicture}
\caption{Exact verification for the $N=32$ constituent pair with
$\mathcal I_{\mathrm L}=\{3,5\}$ and
$\mathcal I_{\mathrm R}=\{3,5,9,10,11,12,13,14,15\}$. (a) Feasible fractions over $2^{14}$
matrices and a sufficient bound, including the right-constituent distance obstruction
for $d_\star>8$. The inset gives the reduced-space bounds from
Theorem~\ref{thm:conditional-completion}(i), (ii), and the exact count.
(b) Current-source means along Algorithm~\ref{alg:outer} and the same
traversal without quotient reduction. Arrows mark the switch from
$i=0$ to $i=1$ and the new group mean. Both executions give exact
distance 6 without backtracking.}
\label{fig:toy-completion}
\end{figure*}
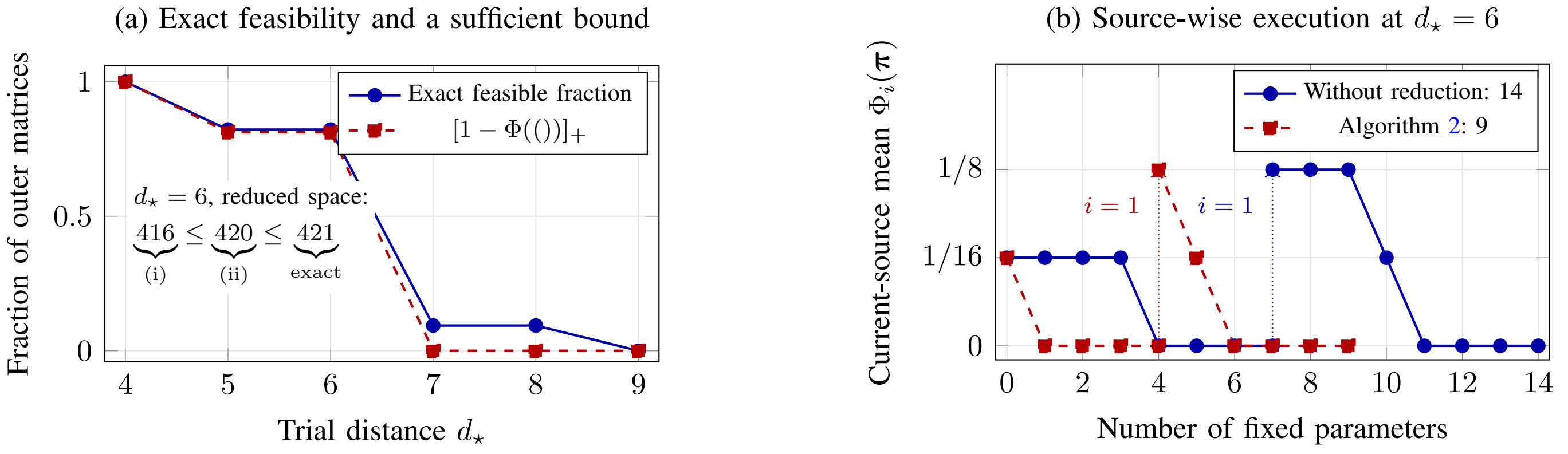

The three nonzero left messages produce weight-4 left-constituent codewords.
Each quotient has rank 4 and inside nullity one,
and $\mathcal B_{\mathbf a}(5)=\{0\}$. The first two message blocks
have joint rank 8, and the sum-message block adds one independent
observation, giving $\rho=9$. Increasing the target from 6 to 8 adds no critical left messages,
so both retain $\rho=9$ and $\dim\ker \mathbf{L}_{D}=5$.
Proposition~\ref{prop:global-completion} therefore replaces
$2^{14}$ matrices by $2^{9}$ representatives, each with 32 lifts.
Complete counts give
\[
 \begin{array}{c|rr|c}
 d_{\star}&\text{full feasible}&\text{reduced feasible}&\text{fraction}\\\hline
 6&\num{13472}&421&421/512\\
 8&\num{1536}&48&48/512
 \end{array}
\]
Direct \textit{common-supercode} enumeration checks $(A_{1},\ldots,A_{7})$
for all \num{16384} matrices: every member of each observation
class has the same truncated spectrum as its representative.
Distinct classes can nevertheless have the same spectrum.
Each feasible or infeasible class retains one representative, so
reduction preserves the feasible fraction: at target 6, for example,
the 421 feasible representatives recover all $421\cdot32=\num{13472}$
feasible matrices.
The differing feasible fractions arise from larger \textit{forbidden sets},
not a change in rank.

Fig.~\ref{fig:toy-completion}(a) compares the exact fraction
with the first bound in Theorem~\ref{thm:conditional-completion}.
At target 6, $\Phi(())=3/16$ sums the initial violation probabilities
over all \textit{critical messages}, whereas the \textit{source}
means are $1/16$ and $1/8$. In the reduced space, the \textit{source}
dimensions are 4 and 5, and these means are unchanged.
Thus the source-wise counting argument gives at least
$2^4(1-1/16)=15$ first-column choices and
$2^5(1-1/8)=28$ second-column choices after each accepted first column.
The two parts of the theorem therefore give
\[
 \underbrace{512(1-3/16)}_{416\;\text{(part (i))}}
 \ \le\ \underbrace{(16-1)(32-4)}_{420\;\text{(part (ii))}}
 \ \le\ \underbrace{421}_{\text{exact count}}.
\]
Part (i) counts $512\Phi(())=96$ violations
with multiplicity, whereas only $512-421=91$ representatives
are infeasible.
Exactly 15 retained first-column choices satisfy the first group.
Of these, 14 admit 28 second-column completions each, whereas
one admits 29. The uniform guarantee of 28 gives the product
bound 420, while the exact count is $14\cdot28+29=421$.
This bound holds for every accepted first column and does not
assume independence between groups.
Multiplication by 32 restores the full-space lower bounds
\num{13312} and \num{13440}, compared with \num{13472}
feasible matrices.
At $d_{\star}=8$, the cutoff $D=7$ gives $t_{\mathbf a}=1$.
For each message, the zero and single-bit outside assignments
give eight distinct forbidden syndromes out of 16, so
$\Phi(())=3(8/16)=3/2$. Part (i) gives a zero lower bound,
although \num{1536} matrices remain feasible.
At $d_{\star}=9$, no outer matrix meets the distance requirement:
the fixed right code has
distance 4 and supplies an unavoidable weight-8 codeword.
This is the fixed-right limit in \eqref{eq:right-only-limits},
not a profile-wide distance bound. Because these codes have even weights,
each odd target and the following even target have the same \textit{feasible set}.

Fig.~\ref{fig:toy-completion}(b) executes Algorithm~\ref{alg:outer}
at target 6, compared with the same traversal without reduction.
Within each source, basis selection scans destination positions in
decreasing order and assignment scans the retained positions in
increasing order.
Groups $i=0,1$ correspond to physical \textit{sources} 3 and 5.
Reduction changes their dimensions from $(7,7)$ to $(4,5)$
and preserves the initial means $1/16$ and $1/8$.
Each group mean is nonincreasing. At a group boundary, the
next mean starts anew while earlier groups remain satisfied.
Both initial group means are below one, so
Theorem~\ref{thm:conditional-completion} guarantees completion without backtracking.
Both executions assign all remaining free parameters, choosing zero on ties,
in 9 decisions with reduction and 14 without it.
Both yield $u_{18}=u_5$ and $u_{20}=u_3+u_5$,
with all other frozen bits zero. The completed code has
$d_{\min}=6$ and $A_6=12$, as verified by complete enumeration.
\end{example}

\section{Profile and Dynamic Frozen Expression Design}
\label{sec:reliability-design}
\begin{figure*}[!t]
\centering
\resizebox{\linewidth}{!}{\begin{tikzpicture}[
 x=1pt,y=1pt,>=Stealth,
 every node/.style={font=\footnotesize,align=center,inner sep=1pt},
 frame/.style={draw=black!50,rounded corners=3pt,line width=.6pt},
 flow/.style={->,line width=.8pt},
 feedback/.style={flow,draw=designcoupling},
 retained/.style={->,draw=black!50,densely dotted,line width=.65pt},
 note/.style={font=\scriptsize,fill=white,inner sep=1.2pt}
]
\path[use as bounding box] (0,-5) rectangle (506,144);
\draw[frame,dashed] (104,13) rectangle (505,129);
\node[font=\footnotesize\bfseries] at (48,137) {Profile selection};
\node[font=\footnotesize\bfseries] at (306,137)
  {Expression design at a fixed profile $\mathcal I$};
\draw[frame,fill=white] (3,61) rectangle (92,110);
\node[font=\footnotesize\bfseries] at (47.5,97) {Propose next\\unseen profile};
\node[font=\scriptsize] at (47.5,80) {Reliability and diversity};
\node[font=\scriptsize] at (47.5,68) {Necessary checks};
\draw[frame,fill=white] (112,61) rectangle (242,110);
\node[font=\footnotesize\bfseries] at (177,99) {Fixed-profile construction};
\node[font=\scriptsize] at (177,84) {Complete low-weight conditions};
\node[font=\scriptsize] at (177,71) {Reduce and complete as needed};
\draw[flow] (92,86) -- (112,86);
\node[note] at (102,94) {$\mathcal I$};
\draw[feedback] (177,110) -- (177,120) -- (47.5,120) -- (47.5,110);
\node[note,text=designcoupling] at (112,120) {Profile or class excluded};
\draw[feedback] (144,61) -- (144,55);
\draw[feedback] (221,55) -- (221,61);
\draw[frame,draw=designcoupling,fill=white] (108,17) rectangle (274,55);
\node[anchor=west,font=\scriptsize] at (111,46) {$2d_{\min}(\mathcal C_{\mathrm R})<d_\star$};
\node[anchor=west,font=\scriptsize] at (111,33) {Domain ruled out};
\node[anchor=west,font=\scriptsize] at (111,23) {$\mathcal E_{\mathrm{fix}}\ne\varnothing$};
\foreach \y in {46,28} \draw[flow,designcoupling,line width=.55pt] (174,\y) -- (181,\y);
\node[anchor=west,font=\scriptsize] at (185,46) {Redesign right code};
\node[anchor=west,font=\scriptsize] at (185,28) {Release $(j,i)=\min\mathcal E_{\mathrm{fix}}$};
\draw[flow,dashed,black!60] (112,71) -- (97,71) -- (97,48)
  -- (15,48) -- (15,61);
\node[note,text=black!75] at (56,48) {Unresolved suffix};
\node[font=\scriptsize,text=black!75] at (56,36) {Prioritize exchanges};
\draw[frame,draw=designretained,fill=designretained!3]
  (281,24) rectangle (502,119);
\node[font=\footnotesize\bfseries,text=designretained] at (391.5,109)
  {Distance-feasible expressions};
\draw[frame,draw=designretained,fill=designretained!9]
  (360,47) rectangle (495,96);
\node[font=\scriptsize,text=designretained] at (426,87)
  {Preserve counts through $w_\star$};
\foreach \x in {291,368,483}
  \draw[retained] (\x,69) -- (\x,11);
\draw[frame,fill=white] (283,-5) rectangle (501,11);
\node[font=\scriptsize] at (392,3) {Retain verified stage endpoints};
\draw[flow] (242,69) -- (291,69);
\node[note] at (263,84) {Distance\\verified};
\node[font=\scriptsize] at (322,96) {\textit{Spectral refinement}};
\node[font=\scriptsize] at (322,84) {$A_{d_\star}\downarrow$};
\draw[flow,designretained] (291,69) -- (368,69);
\node[font=\scriptsize] at (426,77) {\textit{Decoding-prefix protection}};
\draw[flow,designretained] (368,69) -- (483,69);
\node[font=\scriptsize] at (426,56) {Improve transmitted-path retention};
\node[font=\scriptsize,fill=designretained!3] at (368,36) {Spectral endpoint};
\foreach \x in {291,368,483}
  \fill[designretained] (\x,69) circle (1.7pt);
\end{tikzpicture}}
\caption{Algorithm~\ref{alg:profiles} proposes a profile and
Algorithm~\ref{alg:construction} seeks expressions that meet its distance target.
Solid feedback arrows act on proved exclusions; dashed feedback from an
unresolved suffix changes only the order of profile exchanges.
After distance verification, the two stages aim first to reduce the
target-weight multiplicity and then to reduce transmitted-path loss while
preserving the first stage's spectrum through $w_\star$. Dotted arrows mark
the verified designs retained for comparison.}
\label{fig:design-invariants}
\end{figure*}
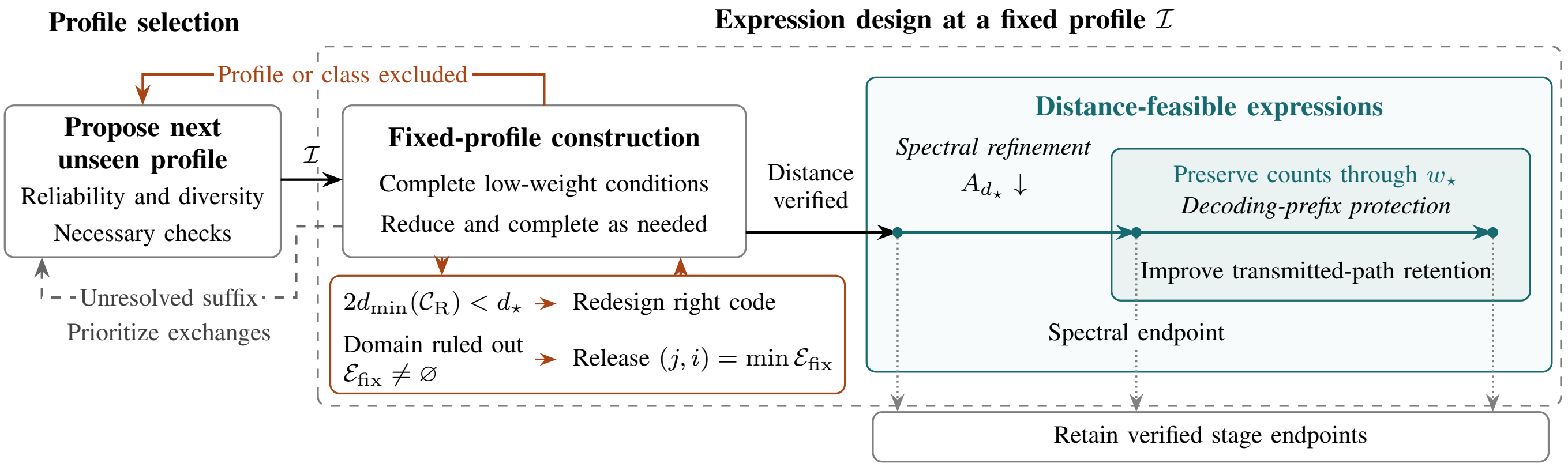

We now apply the exclusion, completion, and preservation results
to design polar subcodes. We first construct dynamic frozen
expressions for a fixed profile, then use this construction to
guide reliability-aware profile selection. At a fixed profile,
\textit{expression refinement} seeks to improve the low-weight spectrum
and then transmitted-path retention while preserving the distance target.
Fig.~\ref{fig:design-invariants} relates these steps and their feedback.

\subsection{Initial Coefficient Domains}
\label{subsec:domain-selection}
An initial admissible domain $\mathcal S_0$ reduces the
coefficient choices and exclusion conditions to be handled first.
We retain high-weight information rows and allow changes only
through selected frozen rows. The resulting \textit{common supercode}
identifies which weight layers below $d_\star$ can occur, so only
those layers enter the \textit{complete directory} and exclusion conditions.
Before specifying this domain, we give a constituent-containment
condition that simplifies the completion test of
Theorem~\ref{thm:conditional-completion}.

\subsubsection{Completion from a retained constituent}
\label{subsec:uniform-extension}
At the split $N=2h$, consider constituent pairs satisfying
\begin{equation}
 \mathcal C_{\mathrm L}\subseteq\mathcal C_{\mathrm R},\qquad
 d_{\min}(\mathcal C_{\mathrm R})\ge\delta.
 \label{eq:retained-left}
\end{equation}
Containment places every \textit{critical} left codeword $\mathbf a$
in the right code. When $d_\star\le2\delta$, the distance constraint
permits only zero and $\mathbf a$ to be supported inside
$\mathcal S_{\mathbf a}$, as shown below. This fixes the number
of solutions of each consistent inside-coordinate equation and
hence the quotient rank $r_{\mathbf a}$
from $w_{\mathbf a}$ and $f=h-k_{\mathrm R}$. Counting the outside
assignments $\mathbf z$ then bounds the forbidden syndromes
uniformly over right codes satisfying \eqref{eq:retained-left}.

Let $A_w^{(i)}=|\{\mathbf m\in\mathcal L_i:w_{\mathbf a}=w\}|$
count weight-$w$ left codewords in \textit{source} group $\mathcal L_i$, and let
$V(p,t)=\sum_{j=0}^{\min(p,t)}\binom pj$ be the Hamming-ball volume for $p,t\ge0$.
For $D=d_{\star}-1$, define
\begin{equation}
 \Lambda_i(D)=\sum_{w=1}^{\min(D,h)}A_w^{(i)}2^{w-1}
 V\!\left(h-w,\left\lfloor\frac{D-w}{2}\right\rfloor\right).
 \label{eq:uniform-column-bound}
\end{equation}
The quantity $\Lambda_i(D)$ is a uniform upper bound on $\lambda_i$, and hence on the number of rejected choices at source $i$, over right codes satisfying \eqref{eq:retained-left}. For fixed $h$ and $D$, it uses only the grouped left counts $A_w^{(i)}$.

\begin{corollary}[Uniform extension from a retained constituent]
\label{cor:uniform-extension}
Fix $\mathcal C_{\mathrm L}$, $k_{\mathrm R}$, $\delta\ge1$, and
$1\le d_{\star}\le2\delta$, and put $f=h-k_{\mathrm R}$.
Consider any length-$h$ right polar subcode of dimension
$k_{\mathrm R}$ satisfying \eqref{eq:retained-left}.
If
\begin{equation}
 \Lambda_i(D)<2^f\quad\text{for every left source }i,
 \label{eq:uniform-extension-test}
\end{equation}
then at least
\begin{equation}
 \prod_{i=0}^{k_{\mathrm L}-1}(2^f-\Lambda_i(D))
 \label{eq:uniform-extension-count}
\end{equation}
distinct outer completions have distance at least $d_{\star}$.
Source-wise conditional means find one in the full outer space
without intersource backtracking.
\end{corollary}
\begin{IEEEproof}
For a nonzero $\mathbf a\in\mathcal C_{\mathrm L}$ of weight
$w_{\mathbf a}\le D$, containment gives $\mathbf a\in\mathcal C_{\mathrm R}$.
If another $\mathbf c\in\mathcal C_{\mathrm R}\setminus\{0,\mathbf a\}$
were supported on $\mathcal S_{\mathbf a}$, then
$\mathbf c+\mathbf a$ would have the complementary support there, giving
\[
 2\delta\le\wt(\mathbf c)+\wt(\mathbf c+\mathbf a)
 =w_{\mathbf a}\le d_{\star}-1<2\delta,
\]
a contradiction. Hence the right subcode supported on
$\mathcal S_{\mathbf a}$ is $\{0,\mathbf a\}$.
Deleting its zero coordinates identifies it with
$\ker\mathbf P_{\mathcal S_{\mathbf a}}$, so rank--nullity gives
\begin{equation}
 k_{\mathbf a}=1,\qquad
 \rk\mathbf P_{\mathcal S_{\mathbf a}}=w_{\mathbf a}-1,\qquad
 r_{\mathbf a}=f-w_{\mathbf a}+1.
 \label{eq:retained-rank}
\end{equation}
There are $V(h-w_{\mathbf a},t_{\mathbf a})$ outside assignments
$\mathbf z$ with $\wt(\mathbf z)\le t_{\mathbf a}$. Their image therefore satisfies
$|\mathcal B_{\mathbf a}(D)|\le V(h-w_{\mathbf a},t_{\mathbf a})$.
Substitution into \eqref{eq:source-bound} yields
\[
 \lambda_i=\sum_{\mathbf m\in\mathcal L_i}
 2^{w_{\mathbf a}-1}|\mathcal B_{\mathbf a}(D)|
 \le\Lambda_i(D)<2^f.
\]
Together with $2d_{\min}(\mathcal C_{\mathrm R})\ge2\delta\ge d_{\star}$,
Theorem~\ref{thm:conditional-completion}(ii) gives the count and construction.
\end{IEEEproof}

Constructing a particular $\mathbf T$ uses the quotient conditions
and conditional means of the chosen constituent pair.
The pair in Example~\ref{ex:n32-completion} satisfies the containment
condition with $\delta=4$ and $f=7$. At $d_\star=6$, its grouped
counts give $\Lambda_0(5)=8$ and $\Lambda_1(5)=16$, so
\eqref{eq:uniform-extension-count} gives at least \num{13440}
completions in the full outer space.

\subsubsection{Initial admissible domain}
For a nonempty profile $\cI$, let
$\mathcal E_{\mathrm{all}}=\{(j,i):j\in\cF,\ i\in\mathcal I_j\}$
be the causal coefficient positions of \eqref{eq:canonical}.
A subset $\mathcal E\subseteq\mathcal E_{\mathrm{all}}$ specifies
the admissible domain
\begin{equation}
 \mathcal S(\mathcal E)
 =\{\mathbf s\in\Omega_{\cI}:s_{ji}=0
       \text{ for }(j,i)\in\mathcal E_{\mathrm{all}}\setminus\mathcal E\}.
 \label{eq:coefficient-domain}
\end{equation}
With the zero baseline, this is the coefficient form of the
admissible increment domain $\mathscr D$ in
Theorem~\ref{thm:shared-elimination}.

Following the subcode--supercode construction of
\cite{miloslavskaya2021recursive}, retain the information rows
whose weights already meet the target. Modify only the remaining
generators, using frozen rows whose weights are at least the
unprecoded distance $d_{\mathrm P}$ in \eqref{eq:profile-bound}. By the same polar-row bound, this prevents the common supercode from introducing weight layers below $d_{\mathrm P}$.
This gives
\begin{equation}
 \begin{aligned}
 \mathcal H_{d_\star}&=\{i\in\cI:2^{\pc(i)}\ge d_\star\},\\
 \mathcal J_{\mathrm{init}}&=\{j\in\cF:j>\min\cI,\
                         2^{\pc(j)}\ge d_{\mathrm P}\},\\
 \mathcal E_0&=\{(j,i):j\in\mathcal J_{\mathrm{init}},\
                       i\in\mathcal I_j\setminus\mathcal H_{d_\star}\}.
 \end{aligned}
 \label{eq:domain-rule}
\end{equation}
Set $\mathcal S_0=\mathcal S(\mathcal E_0)$. The generator
formula \eqref{eq:systematic-input} and the \textit{common supercode}
of \eqref{eq:relaxation} give, for every $\mathbf s\in\mathcal S_0$,
\begin{equation}
 \spn\{\mathbf g_i:i\in\mathcal H_{d_\star}\}
       \subseteq\mathcal C_{\mathbf s}\subseteq\widetilde{\mathcal C}_0,
 \quad
 \widetilde{\mathcal C}_0
       :=\widetilde{\mathcal U}_{\mathcal J_{\mathrm{init}}}\mathbf G_N.
 \label{eq:domain-inclusion}
\end{equation}
The retained span preserves the high-weight information rows,
while the common supercode permits
$\cW=\{w:1\le w<d_\star,\ A_w(\widetilde{\mathcal C}_0)>0\}$
in Theorem~\ref{thm:shared-elimination}, with its existing
\textit{complete directory} and shared-coefficient conditions.

At a recursive node, apply \eqref{eq:domain-rule} to its local
length and information set with $d_\star=\delta$, where $\delta$
is the local target and $d_{\mathrm P}$ is given by \eqref{eq:profile-bound}.
In particular, apply the rule to the length-$h$ right
constituent. For an unprecoded left constituent with
$\mathcal I_{\mathrm L}\subseteq\mathcal H_\delta$ in these
local coordinates, \eqref{eq:domain-inclusion} gives
\[
 \mathcal C_{\mathrm L}
 \subseteq\spn\{\mathbf g_i:i\in\mathcal H_\delta\}
 \subseteq\mathcal C_{\mathrm R}.
\]
After verifying $d_{\min}(\mathcal C_{\mathrm R})\ge\delta$,
Corollary~\ref{cor:uniform-extension} applies to the parent target
$d_\star\le2\delta$ in the full outer space. Restricted outer
domains instead use their actual attainable sets in
Theorem~\ref{thm:conditional-completion}.

\subsection{Obstruction Feedback and Fixed-Profile Construction}
\label{sec:construction}
\label{sec:outer}
\label{subsec:proved-obstructions}
When the initial domain is insufficient, the scope of a proved
obstruction determines whether to redesign a constituent, release
a coefficient, or change the profile. Table~\ref{tab:construction-outcomes}
summarizes these distinctions; the implications below justify
the feedback used in Algorithm~\ref{alg:construction}.

\begin{table*}[!t]
\centering
\caption{Construction Outcomes and Scope of Feedback}
\label{tab:construction-outcomes}
\papertablestyle
\footnotesize
\begin{tabular}{@{}C{52pt}C{164pt}C{100pt}C{128pt}@{}}
\toprule
\textbf{Outcome} & \textbf{Evidence} & \textbf{Scope} & \textbf{Next step}\\
\midrule
\multirow{2}{*}[-6pt]{\textbf{Feasible}}
& Complete distance verification & Verified design & \textbf{Retain} and refine.\\
\addlinespace[3pt]
& Theorem~\ref{thm:conditional-completion}: (i) $\Phi(\boldsymbol\pi)<1$, or (ii) $\lambda_i<2^f$ at every source
& (i) Affine domain extending $\boldsymbol\pi$; (ii) full outer coefficient space & \textbf{Complete} by smaller-mean choices.\\
\midrule
\multirow{3}{*}[-7.5pt]{\textbf{Infeasible}}
& Fixed right code: $2d_{\min}(\mathcal C_{\mathrm R})<d_\star$
& This right constituent & \textbf{Redesign} the right code.\\
\addlinespace[3pt]
& Proved contradiction in \eqref{eq:obstruction-assumptions}, $\mathcal E_{\mathrm{fix}}\ne\varnothing$
& Restricted coefficient domain & \textbf{Release} a coefficient in $\mathcal E_{\mathrm{fix}}$; update the conditions.\\
\addlinespace[3pt]
& Exclusion over $\Omega_{\cI}$ or full-suffix exclusion \eqref{eq:profile-suffix-feedback}
& Profile or proved profile class & \textbf{Change} the information set.\\
\midrule
\multirow{2}{*}[-6pt]{\textbf{Undecided}}
& Sufficient completion bound not met & No feasibility decision & \textbf{Traverse} the exact conditions.\\
\addlinespace[3pt]
& Work limit before a distance decision & Unknown on the unfinished domain
& \textbf{Continue} construction or guide profile exchanges.\\
\bottomrule
\end{tabular}
\par\smallskip
\parbox{\dimexpr444pt+6\tabcolsep\relax}{\footnotesize
For fixed constituents, both completion tests require $2d_{\min}(\mathcal C_{\mathrm R})\ge d_\star$.}
\end{table*}

\subsubsection{Necessary suffix distances}
For a dyadic suffix length $\nu\mid N$, let $\mathbf{u}_{\nu}\in\F^{\nu}$
denote the suffix input and set earlier information bits to
zero. Causality sets all earlier inputs to zero, so
the tensor factorization $\mathbf{G}_{N}=\mathbf{G}_{N/\nu}\otimes \mathbf{G}_{\nu}$ gives
\[
 \begin{aligned}
 (0,\ldots,0,\mathbf{u}_{\nu})\mathbf{G}_{N}
   &=(\mathbf{x}_{\nu},\ldots,\mathbf{x}_{\nu}),\qquad \mathbf{x}_{\nu}=\mathbf{u}_{\nu}\mathbf{G}_{\nu},\\
 \wt((\mathbf{x}_{\nu},\ldots,\mathbf{x}_{\nu}))&=(N/\nu)\wt(\mathbf{x}_{\nu}).
 \end{aligned}
\]
The last row of $\mathbf{G}_{N/\nu}$ is all ones. Hence, for the induced
suffix code $\mathcal{C}^{(\nu)}$,
\begin{equation}
 d_{\min}(\mathcal{C}_{\mathbf{s}})\ge d_{\star}
 \ \Longrightarrow\
 d_{\min}(\mathcal{C}^{(\nu)})\ge\left\lceil\frac{d_{\star} \nu}{N}\right\rceil.
 \label{eq:suffix-distance-necessity}
\end{equation}
In particular, $\nu=N/2$ gives
$d_{\min}(\mathcal C_{\mathrm R})\ge\lceil d_{\star}/2\rceil$.

\subsubsection{Coefficient-independent profile exclusions}
\label{subsec:profile-exclusions}
Let $\mathscr P$ be a declared family of size-$K$ profiles.
An exclusion must hold for all \textit{coefficient configurations} of the
profiles it discards.
If its proof depends only on the information/frozen status of specified
positions, it excludes every profile in $\mathscr P$ with those same
memberships.
For a dyadic suffix of length $\nu$, define $\mathcal T_{\nu}=[N-\nu,N)$,
$\cI_{\nu}=\{i-(N-\nu):i\in\cI\cap \mathcal T_{\nu}\}$, and
$\delta_{\nu}=\lceil d_{\star} \nu/N\rceil$.
By \eqref{eq:suffix-distance-necessity}, every feasible full-code configuration
induces a suffix configuration in $\mathscr F_{\delta_\nu}(\cI_\nu)$.
Consequently,
\begin{equation}
 \begin{split}
 &\mathscr F_{\delta_\nu}(\cI_\nu)=\varnothing\\
 &\quad\Longrightarrow\ \text{exclude all profiles with this suffix}.
 \end{split}
 \label{eq:profile-suffix-feedback}
\end{equation}
The excluded profiles share the entire suffix information set,
independently of earlier information positions and all coefficients.

\subsubsection{Proved obstructions and enlargement}
\label{subsec:domain-feedback}
We relate an infeasibility proof to the fixed coefficients
responsible for it, obtaining a necessary condition for enlargement.
At length $N$, suppose the domain $\mathcal S(\mathcal{E})$ is
infeasible. Let $\mathcal U_{\mathrm{bad}}$ be a finite set of
inputs with $1\le\wt(\mathbf{u}\mathbf{G}_{N})<d_{\star}$. Suppose
\begin{equation}
 \mathbf{H}(\mathbf{s})\mathbf{u}^{\mathrm{T}}\ne0\quad(\mathbf{u}\in\mathcal U_{\mathrm{bad}}),\qquad
 s_{ji}=0\quad((j,i)\in \mathcal E_{\mathrm{fix}})
 \label{eq:obstruction-assumptions}
\end{equation}
are jointly infeasible over the full canonical space
$\Omega_{\cI}$, where $\mathcal E_{\mathrm{fix}}\subseteq \mathcal{E}_{\mathrm{all}}\setminus \mathcal{E}$.
Every feasible configuration must reject all inputs in
$\mathcal U_{\mathrm{bad}}$, by Theorem~\ref{thm:shared-elimination}.
Any feasible configuration must therefore change at least one
fixed-zero assumption:
\begin{equation}
 \bigvee_{(j,i)\in \mathcal E_{\mathrm{fix}}}(s_{ji}=1).
 \label{eq:release-feedback}
\end{equation}
An empty $\mathcal E_{\mathrm{fix}}$ proves infeasibility for the entire profile.

A complete failed domain supplies
$\mathcal E_{\mathrm{fix}}=\mathcal{E}_{\mathrm{all}}\setminus\mathcal E$;
a smaller set may be used if the constraints in
\eqref{eq:obstruction-assumptions} remain jointly infeasible.

\subsubsection{Fixed-profile construction}
\label{sec:workflow}
\label{subsec:fixed-profile-correctness}
Algorithm~\ref{alg:construction} solves each domain directly by
$\proc{SolveDomain}$ or through its constituent factorization,
then uses proved obstructions to enlarge an infeasible domain.
Each recursive trial applies its prescribed constituent restrictions at
the suffix target. It reports a verified suffix configuration, an
unresolved suffix, or a proof of infeasibility over the full suffix domain.
A failed restricted domain leaves the suffix unresolved if the trial
stops there. If construction is deferred at any stage, including
when a work limit interrupts a search or verification, the call
immediately returns \textit{unresolved} with the available suffix outcomes.
Only a completed, unsuccessful domain search reaches coefficient
release.

\begin{algorithm}[!t]
\caption{Fixed-profile expression construction with domain feedback}
\label{alg:construction}
\small
\begin{algorithmic}[1]
\Require $N=2^{n}$, profile $\cI$, target $d_{\star}\ge1$
\Require Direct or constituent-factorized solving; optional ordered recursive trials and deferral rule
\Ensure Suffix outcomes and a verified $\mathbf{s}$ meeting $d_\star$, \textit{unresolved}, or $\bot$ for $\Omega_{\cI}$
\State \textbf{On deferral, return} \textit{unresolved} with the available suffix outcomes
\State \textbf{if} $\cI=\varnothing$ \textbf{then return} $\mathbf s=0$
\State \textbf{if} $d_\star>N$ \textbf{then return} $\bot$
\State \textbf{if} $d_{\mathrm P}\ge d_\star$ \textbf{then return} $\mathbf s=0$ \Comment{(\ref{eq:profile-bound})}
\State $D\gets d_\star-1$, $\mathcal E\gets\mathcal E_0$ \Comment{(\ref{eq:domain-rule})}\label{line:construction-initial}
\State Run optional recursive trials and report suffix outcomes \Comment{(\ref{eq:suffix-distance-necessity})}\label{line:construction-trials}
\State \textbf{if} a verified $\mathbf s$ is found \textbf{then return} $\mathbf s$
\For{$t=0,\ldots,|\mathcal E_{\mathrm{all}}\setminus\mathcal E_0|$}
 \State $\mathbf s\gets\bot$
 \If{using constituent factorization, with $N>1$}\label{line:construction-factor}
  \State $h\gets N/2$; form the local half-profiles $\mathcal I_{\mathrm L},\mathcal I_{\mathrm R}$
  \ForAll{$\mathbf s_{\mathrm L}\in\Omega_{\mathcal I_{\mathrm L}}$, $\mathbf s_{\mathrm R}\in
       \mathscr F_{\lceil d_\star/2\rceil}(\mathcal I_{\mathrm R})$}\label{line:construction-pairs}
   \State Restrict $\mathbf T$ by $(\mathbf s_{\mathrm L},\mathbf s_{\mathrm R},\mathbf T)\in\mathcal S(\mathcal E)$
   \State \textbf{if} this outer domain is empty \textbf{then continue}
   \State $\mathbf T\gets$ Algorithm~\ref{alg:outer} for this pair and outer domain\label{line:construction-outer}
   \If{$\mathbf T\ne\bot$}
    \State $\mathbf s\gets(\mathbf s_{\mathrm L},\mathbf s_{\mathrm R},\mathbf T)$; \textbf{break}
   \EndIf
  \EndFor
 \Else
  \State $\mathbf s\gets\proc{SolveDomain}(N,\cI,d_\star,\mathcal S(\mathcal E))$ \Comment{Thm.~\ref{thm:shared-elimination}}\label{line:construction-direct}
 \EndIf
 \If{$\mathbf{s}\ne\bot$}
  \State Verify $A_1=\cdots=A_{d_\star-1}=0$\label{line:construction-verify}
  \State \Return $\mathbf{s}$
 \EndIf
 \State Obtain a proved set $\mathcal E_{\mathrm{fix}}$ from \eqref{eq:obstruction-assumptions}\label{line:construction-feedback}
 \State Use $\mathcal E_{\mathrm{all}}\setminus\mathcal E$ if no smaller proved set is available
 \State \textbf{if} $\mathcal E_{\mathrm{fix}}=\varnothing$ \textbf{then return} $\bot$ for $\Omega_{\cI}$
 \State $(j,i)\gets\min \mathcal E_{\mathrm{fix}}$ in lexicographic order
 \State $\mathcal E\gets\mathcal E\cup\{(j,i)\}$ \Comment{(\ref{eq:release-feedback})}\label{line:construction-release}
 \State Update the complete exclusion conditions on $\mathcal S(\mathcal E)$
\EndFor
\end{algorithmic}
\end{algorithm}

The complete domain solver of Section~\ref{subsec:exact-solvers}
preserves feasibility by Theorems~\ref{thm:shared-elimination} and
\ref{thm:polar-spectrum}. Factoring has the same coverage:
every canonical configuration decomposes uniquely as
$\mathbf s=(\mathbf s_{\mathrm L},\mathbf s_{\mathrm R},\mathbf T)$,
and every feasible one satisfies \eqref{eq:suffix-distance-necessity}.
Algorithm~\ref{alg:construction} uses Algorithm~\ref{alg:outer} to test
outer feasibility for each admissible pair, stopping at the first success.
Source-compatible reduction preserves the existence of a completion for each pair.
The search imposes no separate distance target on the left code.
Equal local spectra need not give the same parent checks.

Enlargement changes only the fixed-zero constraints defining
$\mathcal S(\mathcal E)$.
A failed domain either proves $\mathscr F_{d_\star}(\cI)=\varnothing$
when $\mathcal E_{\mathrm{fix}}=\varnothing$, or releases one
still-fixed coefficient, so the domains increase strictly. After at most
$|\mathcal E_{\mathrm{all}}\setminus\mathcal E_0|$ releases,
the main loop reaches $\mathcal S(\mathcal E)=\Omega_{\cI}$ unless
it has already returned. Including this final domain query, the
bounded loop decides fixed-profile feasibility when all queries
run to completion.

Line~\ref{line:construction-verify} uses complete enumeration or
\eqref{eq:full-outer-spectrum} for distance verification.

\begin{example}[Completing a specified profile at length 256]
\label{subsec:fixed-design-example}
We illustrate one execution of the fixed-profile construction.
Following the RM--polar principle of combining row weight and
bit-channel reliability \cite{li2014rmpolar}, we order indices by
decreasing polar-row weight and rank ties by decreasing
Gaussian-approximation (GA) reliability \cite[Sec.~III]{trifonov2012efficient}.
Let $\cI_{\mathrm{RM-GA}}(N,K)$ contain the first $K$ indices in this order.
We use the binary-input additive white Gaussian noise (AWGN) channel
at design $E_b/N_0=2$~dB, normalized at rate $K/N$, with increasing
index breaking remaining ties. The profile used here is
\[
 \begin{split}
 \cI={}&\bigl(\cI_{\mathrm{RM-GA}}(256,128)\\
       &\quad\setminus\{31,195,225,226,228,232,240\}\bigr)\\
       &\cup\{90,92,101,102,105,142,149\}.
 \end{split}
\]
This is the profile of the recursively designed $[256,128,24]$
code in \cite{miloslavskaya2021recursive}. Here $d_{\mathrm P}=16$ and
$(k_{\mathrm L},k_{\mathrm R})=(39,89)$, with $f=128-89=39$. Set $d_\star=24$
and initialize all coefficients to zero.

\emph{1) Right-constituent requirement.}
The zero-left branch repeats the right code, so
\eqref{eq:suffix-distance-necessity} requires right distance 12.
Apply the initial-domain rule \eqref{eq:domain-rule} to this
length-128 profile at target 12. The retained span
$\spn\{\mathbf g_i:i\in\mathcal H_{12}\}$ has dimension 64 and
distance 16. Its containing supercode $\widetilde{\mathcal C}_0$
has dimension 96 and only weight 8 below 12, with
$A_8=\num{74288}$. Solving the complete conditions of
Theorem~\ref{thm:shared-elimination} to eliminate this layer
constructs the dimension-89 right code with distance at least 12.

\emph{2) Complete outer conditions.}
At the parent node, the recursive trial in
line~\ref{line:construction-trials} of Algorithm~\ref{alg:construction}
fixes the unprecoded left code and the constructed right code. It allows
all $39\times39$ outer coefficients to vary, rather than restricting them
to the full-length initial domain.
Enumeration of the left code gives 920 nonzero
codewords below 24, all of weight 16. Thus \eqref{eq:outer-data} gives
$t_{\mathbf a}=\lfloor(23-16)/2\rfloor=3$: outside weights
0--3 produce full weights 16, 18, 20, and 22, all of which
must be excluded. The joint observation rank is \num{1082} on
\num{1521} coefficients. By Proposition~\ref{prop:global-completion},
adding combinations of the 439 independent kernel directions to a
representative recovers all configurations with its observation.

\emph{3) Coefficient choices.}
The test $\lambda_i<2^f$ of Theorem~\ref{thm:conditional-completion}(ii) is inconclusive for the last three source groups.
For the following completion count, we apply source-wise conditional
means to the original coefficients without basis reduction. Along this
execution, setting the first 4, 5, and 6 entries to zero in destination
order for \textit{sources} 125, 126, and 127 gives current-group means
$\Phi_i(\boldsymbol\pi)$ of 0.916058, 0.989367, and 0.958744, respectively.
For the last step, fix the constituent pair and the first 38 accepted
columns from this execution. In the 39-entry column
for \textit{source} 127, let $\boldsymbol\pi$ set the first six entries to
zero in destination order. This leaves 33 free bits, with
$\Phi_{38}(\boldsymbol\pi)=\num{4021265}/\num{4194304}<1$.
Theorem~\ref{thm:conditional-completion}(i) now guarantees
completion from this assignment without backtracking. The earlier
groups are unchanged by the last column, so at least
\begin{equation}
 2^{33}\left(1-\frac{\num{4021265}}{\num{4194304}}\right)
 =\num{354383872}
 \label{eq:example-completion-count}
\end{equation}
distinct assignments of these 33 original bits meet the target
for the fixed constituent pair and earlier choices.

\emph{4) Distance verification.}
Complete enumeration of the resulting design excludes weights
1--23. In the branch $\mathbf m=0$, the right check $u_{147}=u_{142}$
is the only frozen expression involving $u_{142}$. Setting only
information bit $u_{142}=1$ therefore gives input support $\{142,147\}$
and codeword $\mathbf g_{142}+\mathbf g_{147}$.
The two rows have weight 16 and their supports intersect in four
positions, giving weight $16+16-2\times4=24$. Hence each counted
completion has exact distance 24.
The generator for \textit{source} 102 has coefficient $s_{131,102}=1$
on the weight-8 row $\mathbf g_{131}$. The polar rows form a basis,
and $\operatorname{RM}(4,8)$ is spanned by the rows of weight at least 16.
By uniqueness of the polar-row expansion, this codeword lies outside
the RM supercode containing the reference code. It witnesses that the
completed code is not contained in that supercode.
\end{example}

\subsection{Design Criteria and Feasible Candidates}
\label{subsec:design-criteria}
Specify $N,K,d_{\star}$, a reliability order $\prec$ on $[N]$,
nonnegative costs $\ell_{i}$ with
smaller values preferred, and an output count $Q\ge1$ of
distinct profiles. The profile family $\mathscr P$ of
Section~\ref{subsec:proved-obstructions} defaults to all size-$K$
subsets of $[N]$.
A declared cost ceiling or polar-row-weight floor may
restrict it. Define
\begin{equation}
 c_{\mathrm{rel}}(\cI)=\sum_{i\in\cI}\ell_{i},\qquad
 \mathscr P_{\star}=
 \{\cI\in\mathscr P:\mathscr F_{d_{\star}}(\cI)\ne\varnothing\}.
 \label{eq:profile-reliability-objective}
\end{equation}
Example~\ref{subsec:fixed-design-example} specifies the RM--GA
order used below.

Use the truncated spectrum $\boldsymbol\Sigma_{w_\star}(\mathcal C_{\mathbf s})$
of Section~\ref{sec:model}, with a common integer cutoff
$w_\star\ge d_\star$. On verified candidates, its entries below
$d_\star$ vanish, so comparisons depend only on
$A_{d_\star},\ldots,A_{w_\star}$.

\paragraph{Prefix loss}
Here a \textit{decoding prefix} consists of successive input bits along
an SCL path, distinct from the \textit{partial coefficient assignment} in
Section~\ref{subsec:support-feasibility}.
SCL uses each frozen constraint at its position and cannot
recover a pruned path. Unlike the conditional-entropy criterion
in \cite[Sec.~III-A]{coskun2022information}, we measure finite-list path loss.

Fix the design channel, operating point, list size $L$,
and $n_{\mathrm{obs}}$ observations $\mathscr O$ with known messages.
Use shared observations under all-zero transmission on the
symmetric channel, or common noise draws for other messages.
For a received vector $\mathbf{y}$, let $\tau(\mathbf{y};\cI,\mathbf{s})$ be the first index at which the transmitted
path leaves the list, with $\tau=+\infty$ if it survives. Define
\begin{equation}
 \widehat{p}_{\mathrm{pr}}(\cI,\mathbf{s})
   =\frac1{n_{\mathrm{obs}}}\sum_{\mathbf{y}\in\mathscr O}\ind{\tau(\mathbf{y};\cI,\mathbf{s})<\infty}.
 \label{eq:empirical-prefix-loss}
\end{equation}
The first-loss histogram retains the positions of these events.
For a decoder whose output belongs to its final list, the two
error events in \cite[Sec.~III-A]{choi2025sparse} give
\begin{equation}
 \mathrm{FER}={}\PP\{\tau<\infty\}
 +\PP\{\tau=\infty,\ \widehat{\mathbf{x}}\ne \mathbf{x}\},
 \label{eq:fer-event-decomposition}
\end{equation}
where $\mathbf{x},\widehat{\mathbf{x}}$ are the transmitted and decoded codewords.
The first event forces an error; the second selects another codeword
despite retaining the transmitted path. Neither probability is
determined by reliability cost or the low-weight spectrum alone.

\paragraph{Feasible candidates and comparison}
Let $\mathcal R$ collect verified pairs $(\cI,\mathbf s)$ satisfying the distance
target. The recorded design criteria are
\begin{equation}
 \bigl(c_{\mathrm{rel}}(\cI),\boldsymbol\Sigma_{w_\star}(\mathcal C_{\mathbf s}),
       \widehat p_{\mathrm{pr}}(\cI,\mathbf s)\bigr).
 \label{eq:joint-design-criteria}
\end{equation}
The reliability cost guides profile proposals and selection.
For a fixed profile, refinement first seeks to reduce $A_{d_\star}$ and then
targets prefix loss while preserving the resulting spectrum through
$w_\star$. Spectrum comparisons use the same cutoff, and prefix-loss
comparisons use the same design observations. Independent SCL simulations
evaluate selected codes separately from the design observations.

\subsection{Expression Refinement}
\label{subsec:scl-refinement}
Starting from a verified $(\cI,\mathbf s)$, the operation
$\proc{Refine}$ first reduces the multiplicity at the target weight
while preserving the distance target, then seeks better transmitted-path
retention while preserving the resulting spectrum through $w_\star$.

For the protection cutoff $w_\star$, define the coefficient pairs
\begin{equation}
 \mathcal E_{\mathrm{pre}}=\{(j,i):j\in\cF,\ i\in\cI,\ i<j<i_{w_\star}\}.
 \label{eq:spectral-prefix-domain}
\end{equation}
Changing only these pairs leaves $\mathcal E_{w_\star}$ fixed,
so Proposition~\ref{prop:reduction} gives
\begin{equation}
 \{\mathbf x\in\mathcal C_{\mathbf s'}:\wt(\mathbf x)\le w_\star\}
 =\{\mathbf x\in\mathcal C_{\mathbf s}:\wt(\mathbf x)\le w_\star\}.
 \label{eq:prefix-spectrum-invariance}
\end{equation}
This protects the codeword set through $w_\star$ across coefficient changes on the fixed profile.
The restriction $j<i_{w_\star}$ places the modified frozen constraints
before $i_{w_\star}$ in the SCL decoding order, where they can distinguish
competing \textit{decoding prefixes}.

First, \textit{spectral refinement} accepts a proposal only if
$\mathbf s'\in\mathscr F_{d_\star}(\cI)$ and
\begin{equation}
 A_{d_\star}(\mathcal C_{\mathbf s'})<A_{d_\star}(\mathcal C_{\mathbf s}).
 \label{eq:spectral-refinement-accept}
\end{equation}
The default proposals flip individual causal coefficients outside
$\mathcal E_{\mathrm{pre}}$. When only outer coefficients change,
Corollary~\ref{cor:affine-distance-domain} at $D=d_\star-1$ gives
a preservation test for the fixed constituent pair. If a constituent
changes, we test the complete exclusion conditions of
Theorem~\ref{thm:shared-elimination}.

Second, \textit{decoding-prefix protection} varies coefficients in a
spectrum-preserving domain. In $\mathcal E_{\mathrm{pre}}$,
\eqref{eq:prefix-spectrum-invariance} protects the low-weight codewords
regardless of how proposals are scored.
One possible rule accepts an update when the estimated fraction of frames
in which the transmitted path leaves the list decreases:
\begin{equation}
 \widehat p_{\mathrm{pr}}(\cI,\mathbf s')<\widehat p_{\mathrm{pr}}(\cI,\mathbf s).
 \label{eq:prefix-refinement-accept}
\end{equation}
With either displayed strict-decrease rule, repeated single-coefficient
improvements terminate because the configuration space is finite.

\subsection{Reliability-Aware Profile Construction}
\label{subsec:profile-design}
Algorithm~\ref{alg:profiles} selects information positions by reliability
and structural diversity, using Algorithm~\ref{alg:construction} to construct
$\mathbf s\in\mathscr F_{d_\star}(\cI)$ for each new profile.
The following rules specify a concrete default.

\paragraph{Initial profile}
Apply the available proved coefficient-independent exclusions of
Section~\ref{subsec:proved-obstructions}, including consequences of
\eqref{eq:suffix-distance-necessity}. Propose the first surviving
profile in $\mathscr P$ in lexicographic rank order under $\prec$,
if one exists. This a priori screening precedes the fixed-profile
test of distance attainability.

\paragraph{Profile exchanges}
The default neighbourhood consists of single exchanges
$\cI'=(\cI\setminus\{a\})\cup\{b\}\in\mathscr P$, with
$a\in\cI$ and $b\in\cF$. An exchange improves reliability
when $\ell_b<\ell_a$.
\paragraph{Using construction feedback}
Following the outcome scopes in Table~\ref{tab:construction-outcomes},
use the suffix outcomes reported by Algorithm~\ref{alg:construction}.
Let $\nu$ be the length of the smallest unresolved dyadic suffix and
$\mu$ the length of the largest verified shorter suffix. Use $\mu=0$ and
$\mathcal T_0=\varnothing$ when no shorter suffix is verified.
The suffix-directed exchanges leave the information set within the verified
suffix $\mathcal T_{\mu}$ unchanged and alter at least one information position in
$\mathcal T_{\nu}\setminus\mathcal T_{\mu}$. An exchange within this
difference changes the frozen positions available there; moving an information
position from it to outside $\mathcal T_{\nu}$ reduces the suffix dimension
by one. Without an unresolved suffix, this group is empty.

Order the proposals in three successive groups: the suffix-directed
exchanges just defined, the remaining reliability-improving
exchanges, and all remaining admissible single
exchanges. Within each group, prefer smaller $c_{\mathrm{rel}}(\cI')$.
The last group permits
reliability cost to increase when needed to cross a structural obstacle.

\paragraph{Diverse profile output}
Two profiles belong to the same \textit{suffix class} when their
information counts in every proper dyadic suffix agree. These classes
diversify the output without imposing a feasibility condition.
Among the profiles represented in $\mathcal R$, select up to $Q$ in
successive rounds over the nonempty classes, taking a lowest-cost
remaining profile from each class.

\begin{algorithm}[!t]
\caption{Reliability-prioritized profile and expression design}
\label{alg:profiles}
\small
\begin{algorithmic}[1]
\Require $N,K,d_\star$, reliability order $\prec$, costs $(\ell_i)$, family $\mathscr P$, count $Q$
\Require Ordered proposal and refinement rules; design observations with $L,w_\star$
\Ensure Feasible candidate set $\mathcal R$ and up to $Q$ distinct profiles
\State $\mathcal R\gets\varnothing$; propose the first rank-ordered profile not excluded, if one exists
\While{an unprocessed proposal remains}
 \State Take the next unseen profile $\cI$ in the prescribed order
 \If{$\cI\notin\mathscr P$ or a proved coefficient-independent exclusion applies}
  \State \textbf{continue}
 \EndIf
 \State Run Algorithm~\ref{alg:construction} on $(N,\cI,d_\star)$ with its deferral rule
 \State Use the reported verified and unresolved suffixes to prioritize exchanges
 \If{a verified $\mathbf s$ is returned}
  \State Retain $(\cI,\mathbf s)$ in $\mathcal R$
  \State Apply \textit{spectral refinement}, then \textit{decoding-prefix protection}
  \State Retain the verified stage endpoints
 \EndIf
 \State Incorporate coefficient-independent exclusions from Section~\ref{subsec:profile-exclusions}
 \State Propose unseen exchanges in the prescribed order
\EndWhile
\State \Return $\mathcal R$ and the class-wise profile selection above
\end{algorithmic}
\end{algorithm}

\paragraph{Feasibility of the design procedure}
\label{subsec:profile-guarantees}
The candidate set satisfies the invariant
\[
 \mathcal R\subseteq
 \{(\cI,\mathbf s):\cI\in\mathscr P,\ \mathbf s\in\mathscr F_{d_\star}(\cI)\}.
\]
Algorithm~\ref{alg:construction} supplies feasible configurations on profiles
in $\mathscr P$, and the acceptance conditions of
Section~\ref{subsec:scl-refinement} preserve their feasibility.
Exclusions preserve feasible profiles by Section~\ref{subsec:profile-exclusions}.
Each successful fixed-profile call contributes a pair to $\mathcal R$
before refinement. If every profile in $\mathscr P$ is visited and each
fixed-profile problem is completed, the profiles represented in
$\mathcal R$ are exactly $\mathscr P_\star$; otherwise it returns the
verified designs found.

\begin{example}[From constituent feedback to a feasible profile]
\label{subsec:profile-design-example}
For $[256,64]$ and $d_\star=48$, we trace the exchanges leading from the
initial profile to a feasible design, omitting trials on other branches.
Restrict information positions to polar rows
of weight at least 32 and use the RM--GA order of
Example~\ref{subsec:fixed-design-example}. For a tentative information
set, let $k_\nu=|\cI\cap\mathcal T_\nu|$.
The suffix-distance condition imposes
\[
 \begin{array}{c|ccc}
 \nu&8&16&32\\\hline
 \delta_\nu=\lceil48\nu/256\rceil&2&3&6\\
 \text{upper bound on }k_\nu&7&11&21
 \end{array}.
\]
For distance at least 2 the length-8 code cannot be the full
space, giving $k_8\le7$. At length 16, the sets of $1+16$
vectors within distance one of distinct codewords are disjoint,
giving $2^{k_{16}}(1+16)\le2^{16}$, hence $k_{16}\le11$.
For the third bound, local position zero is frozen, so the
length-32 code is even and the all-one vector is a parity check.
The 32 single-error syndromes then differ from
all triple-error syndromes when the distance is at least 6.
Triples with the same syndrome must be disjoint, since otherwise
their difference would be a codeword of weight 2 or 4. Each
syndrome therefore represents at most $\lfloor32/3\rfloor=10$ triples.
All these patterns have syndrome bit one for the all-one check,
leaving $2^{31-k_{32}}$ possible syndromes. Hence
\[
 2^{31-k_{32}}\ge32+\frac{\binom{32}{3}}{10}=528>512,
 \qquad k_{32}\le21.
\]
Start with the empty information set. Scan the RM--GA order,
accepting a position if none of these three suffix bounds is
violated, until 64 positions have been accepted.
This skips $248,244,242,241,227$, in that order.
Relative to the unconstrained first 64 positions, the next
accepted positions are $179,\allowbreak121,\allowbreak173,\allowbreak199,\allowbreak118$. Hence
\[
 \begin{split}
 \cI_0={}&\bigl(\cI_{\mathrm{RM-GA}}(256,64)
           \setminus\{227,241,242,244,248\}\bigr)\\
         &\cup\{118,121,173,179,199\}.
 \end{split}
\]
Here $d_{\mathrm P}=32$ and $(k_{\mathrm L},k_{\mathrm R})=(12,52)$.
This procedure determines the starting profile without a known
expression design. Its distance-48 feasibility remains to be
established by the following constituent constructions.

\emph{1) Local obstruction.}
Equation~\eqref{eq:suffix-distance-necessity} assigns targets
6, 12, and 24 to the right suffixes of lengths 32, 64, and 128.
Start at length 32. The initial domain in \eqref{eq:domain-rule}
has 20 free coefficients. Its 19 shared conditions cannot be
satisfied simultaneously in that domain. This rules out only the
20-coefficient domain, not $\cI_0$. Here we defer further domain
enlargement for $\cI_0$ and report its length-32 suffix as unresolved.
This feedback prioritizes exchanges in that suffix as in
Section~\ref{subsec:profile-design}.

\emph{2) Changing a frozen position.}
Move global information position $233$ to $227$, or local
position 9 to 3. Position 9 now admits the previously unavailable
check $u_9=u_3+u_5+u_6$. Solving the shared conditions in the
resulting 23-coefficient domain eliminates all weights below 6
for its 21 information positions. Set $\cI_1=(\cI_0\setminus\{233\})\cup\{227\}$.
The retained span selected by $\mathcal H_6$ is unchanged:
both exchanged local rows have weight 4.

\emph{3) Completing the larger suffixes.}
At $\cI_1$, the length-32 suffix is verified, but the length-64
trial is unresolved at target 12. The feedback rule uses
$(\nu,\mu)=(64,32)$. The exchange $211\to158$ moves one
information position from the length-64 suffix to its left sibling,
keeping the verified length-32 information set. Thus
$\cI_2=(\cI_1\setminus\{211\})\cup\{158\}$ reduces $k_{64}$
from 37 to 36 and changes the length-128 split from $(15,37)$
to $(16,36)$. In this execution, solving the complete common-supercode
conditions of Theorem~\ref{thm:shared-elimination} in the two larger
initial domains, with their internal coefficients free, gives suffix
codes of distance at least 12 and 24, respectively, as verified by
complete enumeration.
These solutions include $s_{211,205}=1$ in the length-64 suffix
and $s_{167,158}=1$ in the left half of the length-128 right suffix
of the full code.

\emph{4) Outer completion.}
At length 256, use the unprecoded 12-dimensional left constituent
and fix the verified length-128, dimension-52 right constituent.
At this split,
the cross-half part of $\mathcal E_0$ in \eqref{eq:domain-rule} leaves sources $\{118,121,122,124\}$ and 12 destination frozen positions, hence 48 outer coefficients. Every polar row $\mathbf g_i$ occurring in $\mathbf B(\mathbf s)\mathbf G_{256}$ has weight at least 32.
The resulting code lies in $\operatorname{RM}(3,8)$, whose weights
strictly between 32 and 48 are absent \cite{kasami1970weights}.
Only weight 32 therefore needs elimination. The left constituent has 44
weight-32 codewords. For each, keep the low-weight inputs satisfying the
unchanged right checks and collect their residuals on the 12 editable
checks. The 44 left messages give 11 nonzero \textit{source patterns}
$\mathbf q$ on the four editable sources. Taking the union of residuals
within each pattern gives 32 forbidden increments per pattern
(Theorem~\ref{thm:shared-elimination}). All 48 entries of
$\mathbf V\in\F^{12\times4}$ are free. For each nonzero
$\mathbf q$, uniformly chosen coefficients make $\mathbf V\mathbf q^{\mathrm T}$
uniform on $\F^{12}$, by the equal-preimage argument of
Theorem~\ref{thm:conditional-completion}(i). Hence
\[
 \Phi(())=11\frac{32}{2^{12}}
 =\frac{11}{128}<1.
\]
Theorem~\ref{thm:conditional-completion}(i) guarantees that
successive smaller-mean choices can complete this outer domain
without backtracking. In the constructed code, setting only
information bit $u_{121}=1$ gives input support $\{121,155\}$.
The rows $\mathbf g_{121},\mathbf g_{155}$ each have weight 32 and
intersect in eight positions, so their sum has weight
$32+32-2\times8=48$, establishing $d_{\min}=48$.
The exchanges enable the shorter constructions, after which the
mean bound guarantees a completion of the remaining outer coefficients.
\end{example}

\subsection{Computational Complexity}
\label{subsec:complexity}
We separate condition generation, reduction, completion, and
code evaluation. The algebraic bounds count elementary binary operations.

\paragraph{Condition generation}
The \textit{directory} cost is at least its output size.
Let $T_{\mathrm{gen}}$ include enumeration of the critical messages
$\mathbf m\in\mathcal M_D$ and their left codewords $\mathbf a(\mathbf m)$,
quotient formation, and generation of the \textit{forbidden sets}.
The quotient formula of Theorem~\ref{thm:polar-spectrum} bounds each set by
\[
 |\mathcal B_{\mathbf{a}}(D)|\le
 \min\{2^{r_{\mathbf{a}}},V(|\mathcal{O}_{\mathbf{a}}|,t_{\mathbf{a}})\},
\]
where $V$ is the ball volume of Section~\ref{subsec:uniform-extension}.
This preparation can be exponential even when completion takes few decisions.

\paragraph{Completion in reduced coordinates}
Let $T_{\mathrm{red}}$ be the cost of forming and reducing $\mathbf{L}_{D}\mathbf{R}$.
For columns constrained independently to $\mathcal D_i$, write $p_{i}$ for the retained
parameter count at \textit{source} $i$, $P_{i}$ for the number of
partial assignments evaluated there, and $C_{i}$ for the maximum cost of
evaluating the group mean $\Phi_i(\boldsymbol\pi)$ at one such assignment.
This mean sums only over messages in $\mathcal L_i$. Then
\begin{equation}
 T_{\mathrm{out}}\le T_{\mathrm{gen}}+T_{\mathrm{red}}+\sum_{i} P_{i} C_{i}.
 \label{eq:outer-completion-cost}
\end{equation}
Binary elimination tests membership of each explicitly listed
$\mathbf b\in\mathcal B_{\mathbf a}(D)$ in
$\mathcal Y_{\mathbf m}(\boldsymbol\pi)$, giving the intersection
sizes for $\Phi_i(\boldsymbol\pi)$. Thus $C_i$ and $T_{\mathrm{red}}$
are polynomial in the explicit matrices and forbidden-set representation.

Under Theorem~\ref{thm:conditional-completion}(ii), every \textit{source}
is completed without backtracking. The same holds for independently
restricted columns when their actual group means are below one after
every accepted earlier-column assignment. One initial mean evaluation
and at most three evaluations per fixed bit give $P_i\le3p_i+1$.
Proposition~\ref{prop:global-completion}
then gives
\[
 \sum_{i} p_{i}=\rho_{\mathbf{R}}=\rk(\mathbf{L}_{D}\mathbf{R}),\qquad
 \sum_{i} P_{i}\le3\rho_{\mathbf{R}}+k_{\mathrm{L}}.
\]
This is a polynomial bound in the explicit condition size,
not in code length alone. Outside
this regime, the binary decision tree on the reduced coordinates can have
$2^{\rho_{\mathbf{R}}+1}-1$ nodes, so completion can be exponential.
Affine domains with cross-column restrictions use the global means and corresponding
global parameter count, rather than these source-wise bounds.

\paragraph{Complete design and evaluation}
Total cost sums condition generation, completion, and verification
over the visited domains and constituent pairs. The full design
spaces contain $2^{p_{\cI}}$ configurations per profile and at most
$\binom NK$ profiles. Spectral and SCL evaluations add their costs
for each evaluated candidate.

\section{Simulation Results and Analysis}
\label{sec:example}
We first examine the distances obtained by the construction of
Section~\ref{sec:reliability-design}. We then compare codes of equal
minimum distance to see how the choice of information positions
and frozen expressions affects finite-list decoding.
\newcommand{\FERSpectrumPairs}{%
$[64,32]$ & 8 & $\to$ & 12 & \num{2048} & \num{5024} & 0.343692 & 0.281875 & 447.42 & 726.86\\
$[128,64]$ & 8 & $\to$ & 16 & \num{10648} & \num{31768} & 0.261333 & 0.261333 & 176.63 & 894.06\\
$[256,64]$ & 32 & $\to$ & 48 & \num{95904} & \num{128480} & 0.113066 & 0.194171 & 82.59 & 82.59\\
$[256,128]$ & 16 & $\to$ & 24 & \num{340882} & \num{366656} & 0.221656 & 0.194145 & 64.60 & 510.01\\
$[256,171]$ & 8 & $\to$ & 12 & \num{17920} & \num{19200} & 0.487618 & 0.475578 & 135.29 & 588.20\\
$[\num{1024},512]$ Proposed & 16 & $\to$ & 24 & \num{16128} & \num{488} & 0.048328 & 0.063305 & 239.48 & \num{1169.44}\\
$[\num{1024},512]$ Proposed-LM & 16 & $\to$ & 24 & \num{512} & \num{488} & 0.053795 & 0.063305 & 252.80 & \num{1836.08}\\
$[\num{1024},512]$ Proposed-D32 & 16 & $\to$ & 32 & \num{96886} & \num{104854} & 0.066972 & 0.071716 & 378.23 & \num{2554.67}\\%
}

\begin{table*}[!t]
\centering
\caption{Distance, Minimum-Weight Multiplicity, Reliability, and Construction Time}
\label{tab:selected-designs}
\papertablestyle
\setlength{\tabcolsep}{3.25pt}
\begin{tabular}{@{}C{125pt}C{15pt}@{}C{11pt}@{}C{17pt}C{40pt}C{43pt}C{42pt}C{44pt}C{55pt}C{53pt}@{}}
\toprule
\multirow{2}{*}[-2.8pt]{$[N,K]$ and variant} & \multicolumn{3}{c}{\multirow{2}{*}[-2.8pt]{$d_{\mathrm P}\to d_{\min}$}} & \multicolumn{2}{c}{Exact $A_{d_{\min}}$}
& \multicolumn{2}{c}{$P_{\mathrm{GA}}$ at 2.25~dB} & \multicolumn{2}{c}{Proposed construction time (s)}\\
\cmidrule(lr){5-6}\cmidrule(lr){7-8}\cmidrule(lr){9-10}
& & & & Proposed & Reference & Proposed & Reference & w/o refinement & w/ refinement\\
\midrule
\FERSpectrumPairs
\bottomrule
\end{tabular}
\par\smallskip
\begin{minipage}{\linewidth}\footnotesize
$d_{\mathrm P}$ refers to the proposed profile; both codes in each
pair have the listed $d_{\min}$. Reference is SC-ML
\cite{miloslavskaya2020short} at $[64,32]$, RecursCode
\cite{miloslavskaya2021recursive} at $N=128,256$, and Number-polar
\cite{miloslavskaya2024compact} at $N=\num{1024}$.
The Number-polar parameters are $(13,1,4)$ for $d_{\min}=24$
and $(15,6,1)$ for $d_{\min}=32$.
\textbf{w/ refinement} gives the full construction wall time.
\textbf{w/o refinement} gives the subtotal from the same replay after
deducting logged spectral refinement, prefix protection, and their joint
feedback stages. It includes the reliability-guided profile search and
coefficient construction that produce a distance-qualified code, as
well as separately timed assembly and verification. Proposed-LM and
Proposed-D32 each add their continuation time to the corresponding
time of the shared Proposed run.
Runs were timed sequentially on an Intel Core i7-12700K with 32~GiB RAM.
Independent spectrum and FER evaluations follow construction.
\end{minipage}
\end{table*}

\subsection{Simulation Setup and Code Characteristics}
\label{subsec:experiment-setup}
All codes use binary phase-shift keying on the binary-input AWGN channel, with $E_b/N_0$
normalized by the payload rate $K/N$. Each proposed code and its
reference use the same SCL decoder and list size. We use $L=32$
for the comparisons at lengths $64$ to $256$.

For each $[N,K]$, we evaluate the design selected in
Section~\ref{sec:reliability-design} on independent noise samples.
\begin{itemize}
\setlength{\itemsep}{1pt}
\setlength{\parsep}{0pt}
\item \textbf{Proposed} denotes the main design. At $[\num{1024},512]$,
\textbf{Proposed-LM} uses a different profile with fewer weight-24
codewords at the same minimum distance, \textbf{Proposed-D32} has minimum distance 32,
and \textbf{Proposed-NP} is the main design before
\textit{decoding-prefix protection}.
\item \textbf{SC-ML}, \textbf{RecursCode}, and
\textbf{Number-polar} are the precoded-polar references from
\cite{miloslavskaya2020short}, \cite{miloslavskaya2021recursive}, and
\cite{miloslavskaya2024compact}, respectively. We use the GA-based
\textbf{Number-polar} codes listed in Table~\ref{tab:selected-designs}.
\item \textbf{PAC-RMGA, PAC-SP} use impulse vector
$(1,0,1,1,0,1,1)$ in increasing-delay order \cite[Sec.~VII]{arikan2019sequential}.
The former uses RM--GA profiles designed at 2~dB and the latter uses
the \textbf{Proposed} profile at $[\num{1024},512]$.
\item \textbf{CRC4-GA, CRC6-5G, CRC11-5G} add 4, 6, or 11
CRC bits for CRC-aided SCL decoding \cite{tal2015list}. CRC4-GA uses
$X^4+X^3+X^2+X+1$ and a GA profile designed at 1~dB.
The other two use the 5G CRC polynomials and reliability order in
\cite[Secs.~5.1 and~5.3.1.2]{etsi2026nr}.
\item \textbf{eBCH polar subcode} uses the method of
\cite{trifonov2015polar}, with GA selection at 2~dB and the
corresponding distances in Table~\ref{tab:selected-designs} as
design lower bounds for $N\le256$.
\item \textbf{Normal approx., Meta-converse} denote the normal
approximation and meta-converse bound for the same binary-input
AWGN channel, $N$, and payload $K$
\cite{polyanskiy2010finite,erseghe2016finite}.
\end{itemize}

\begin{figure}[!tp]
\centering
\includegraphics[width=4.05in]{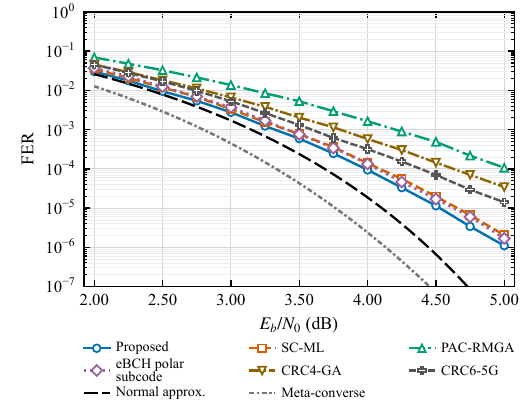}
\caption{FER of the $[64,32]$ designs at $L=32$, with binary-input AWGN benchmarks. Curve labels are defined in Section~\ref{subsec:experiment-setup}.}
\label{fig:fer-n64}
\end{figure}
\begin{figure}[!tp]
\centering
\includegraphics[width=4.05in]{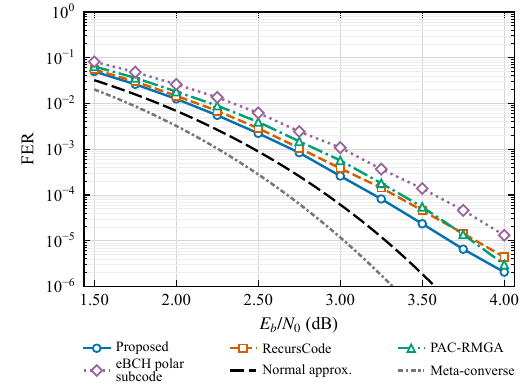}
\caption{FER of the $[128,64]$ designs at $L=32$. Proposed and RecursCode share the same profile.}
\label{fig:fer-n128}
\end{figure}
\begin{figure}[!tp]
\centering
\includegraphics[width=4.05in]{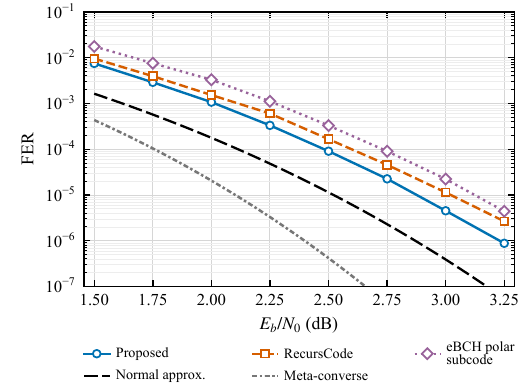}
\caption{FER of the $[256,64]$ designs at $L=32$.}
\label{fig:fer-n256k64}
\end{figure}
\begin{figure}[!tp]
\centering
\includegraphics[width=4.05in]{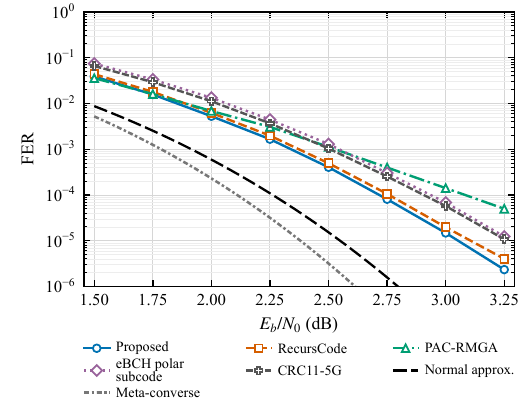}
\caption{FER of the $[256,128]$ designs and comparison codes at $L=32$.}
\label{fig:fer-n256k128}
\end{figure}
\begin{figure}[!tp]
\centering
\includegraphics[width=4.05in]{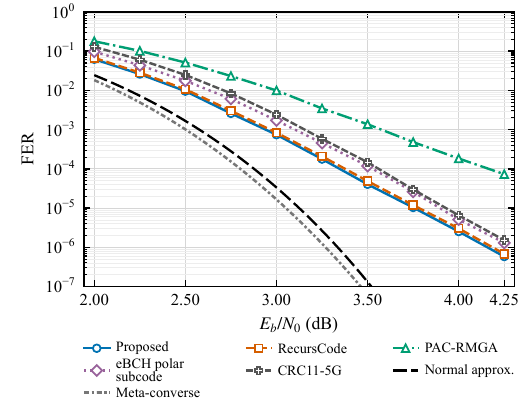}
\caption{FER of the $[256,171]$ designs at $L=32$.}
\label{fig:fer-n256k171}
\end{figure}
\begin{figure}[!tp]
\centering
\includegraphics[width=4.05in]{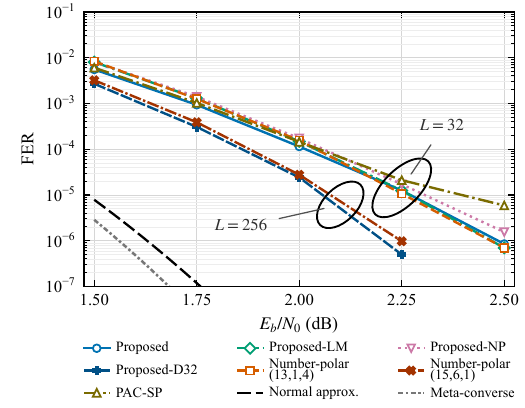}
\caption{FER of the $[\num{1024},512]$ designs. Ellipses identify the $L=32$ and $L=256$ curve families. Proposed-D32 and Number-polar $(15,6,1)$ use $L=256$ and have distance 32. The other code curves use $L=32$. Both binary-input AWGN benchmarks apply to either list size.}
\label{fig:fer-n1024}
\end{figure}

Table~\ref{tab:selected-designs} compares information sets at a common
$E_b/N_0$ of 2.25~dB. We compute the bit-error estimates
$p_i^{\mathrm{GA}}$ by polar GA recursion \cite[Sec.~III]{trifonov2012efficient}
with the check-node approximation of \cite[p.~661]{chung2001ga}.
Writing $\ell_i=-\log(1-p_i^{\mathrm{GA}})$ gives the profile-ranking proxy
\begin{equation}
 P_{\mathrm{GA}}(\cI)=1-\prod_{i\in\cI}(1-p_i^{\mathrm{GA}})
 =1-\exp\{-c_{\mathrm{rel}}(\cI)\}.
 \label{eq:ga-reliability-cost}
\end{equation}
A smaller $P_{\mathrm{GA}}(\cI)$ indicates a more reliable set of
information positions according to the criterion of
Section~\ref{subsec:design-criteria}.

In each of the six main runs, the reliability-guided search finds
a code meeting the minimum-distance target before expression refinement.
The measured time without refinement ranges from 1.1 to 7.5 minutes
and does not increase monotonically with block length
(Table~\ref{tab:selected-designs}). For example, the $[256,64]$ design
takes less time than either the $[64,32]$ or the $[\num{1024},512]$ main design.
Subsequent refinement
seeks to reduce the minimum-weight multiplicity and improve path
retention. On five of the selected
information sets the frozen expressions raise the minimum distance to
$1.5d_{\mathrm P}$, and at $[128,64]$ they raise it from 8 to 16.
The distance-32 variant at length \num{1024} also doubles its
unprecoded distance. Each reference in the table has the same minimum
distance as its proposed counterpart. We can therefore compare how these
codes perform once the distance target has been met.

\subsection{Information-Set Reliability and Minimum-Weight Multiplicity}
\label{subsec:selected-designs}
For the $[256,64]$ design of Example~\ref{subsec:profile-design-example},
both indicators in Table~\ref{tab:selected-designs} favor \textbf{Proposed}
over \textbf{RecursCode}. Its information set has a smaller $P_{\mathrm{GA}}(\cI)$,
it has fewer codewords of weight 48, and its FER is lower
throughout their comparison in Fig.~\ref{fig:fer-n256k64}.
The two indicators favor different designs at $[64,32]$ and $[256,128]$.
In both cases \textbf{Proposed} has fewer minimum-weight words despite a larger
$P_{\mathrm{GA}}(\cI)$, and its FER lies below that of \textbf{SC-ML} and
\textbf{RecursCode}, respectively (Figs.~\ref{fig:fer-n64} and~\ref{fig:fer-n256k128}).
The \textbf{Proposed} $[256,128]$ code uses a different information set from
Example~\ref{subsec:fixed-design-example}. It attains the same distance
of 24 as \textbf{RecursCode} even though some of its codewords lie outside
the $\operatorname{RM}(4,8)$ supercode containing \textbf{RecursCode}. For the
higher-rate $[256,171]$ pair, the multiplicities differ
by only 6.7\%, and their FER curves remain close
(Fig.~\ref{fig:fer-n256k171}).

The $[128,64]$ comparison isolates the effect of the frozen expressions.
\textbf{Proposed} and \textbf{RecursCode} use the same information positions and both have
minimum distance 16. Changing the expressions reduces $A_{16}$ from
\num{31768} to \num{10648}, or by 66.5\%. The separation between the two
FER curves is more pronounced at higher SNR in Fig.~\ref{fig:fer-n128}. At
3.5~dB the FER of \textbf{Proposed} is about half that of \textbf{RecursCode}.

All $[\num{1024},512]$ codes in Fig.~\ref{fig:fer-n1024} carry 512 payload
bits without a CRC. The pair with minimum distance 32 uses $L=256$,
and the other codes use $L=32$. When the information positions can also
change, the ordering by multiplicity can differ from the ordering by FER.
\textbf{Proposed-LM} has only 512 words of weight 24,
compared with \num{16128} for \textbf{Proposed}, but uses less reliable
information positions according to $P_{\mathrm{GA}}(\cI)$.
At 2.0~dB, \textbf{Proposed} has the lower FER
($1.15\times10^{-4}$ versus $1.49\times10^{-4}$).
\textbf{Number-polar} $(13,1,4)$ also has a small $A_{24}$ and follows
\textbf{Proposed-LM} closely at this SNR. At the two higher SNRs the curves
are close, with \textbf{Proposed-LM} and \textbf{Number-polar} slightly below
\textbf{Proposed}. This change in ordering illustrates why
Section~\ref{sec:reliability-design} weighs information-set reliability
alongside multiplicity. In the separate distance-32 comparison,
\textbf{Proposed-D32} has both a smaller $A_{32}$ and a smaller
$P_{\mathrm{GA}}(\cI)$ than \textbf{Number-polar} $(15,6,1)$ and achieves
lower FER.

\subsection{Prefix Protection at a Fixed Low-Weight Spectrum}
\label{subsec:prefix-design-example}
Fig.~\ref{fig:fer-n1024} also compares three ways to encode the same
information positions at $[\num{1024},512]$. \textbf{PAC-SP} uses a convolutional
precoder and has higher FER than \textbf{Proposed} over the displayed range.
The comparison with \textbf{Proposed-NP} isolates the effect of
\textit{decoding-prefix protection}. Starting from \textbf{Proposed-NP}, we change
only the early frozen expressions allowed by \eqref{eq:spectral-prefix-domain}
to obtain \textbf{Proposed}. With $w_\star=24$ and $i_{w_\star}=688$,
\eqref{eq:prefix-spectrum-invariance} guarantees that both codes contain
the same codewords through weight 24. Exact enumeration gives their
common truncated WEF as
\begin{equation}
 W_{\mathcal C,\le24}(X)=1+\num{16128}X^{24}.
 \label{eq:n1024-protected-spectrum}
\end{equation}

For \textbf{Proposed}, we compare candidate expressions at each editable
frozen position on design frames whose transmitted path is still in the
incoming list. Each expression prescribes a bit for every surviving
path and thereby determines its metric update. We weight each incorrect
prefix by the exponential of its negative updated metric, normalize
each frame by the total weight in its incoming list, and prefer candidates
with a smaller sum. For \textbf{Proposed}, this local score determines which
expression is used at each editable frozen
position. We measure whole-frame path loss and FER by decoding with
the completed expressions.

At $L=32$, \textbf{Proposed} has lower FER than \textbf{Proposed-NP}
in Fig.~\ref{fig:fer-n1024} with the low-weight codeword set held fixed.
Table~\ref{tab:prefix-diagnostics} separates losses of the transmitted
path from wrong final selections when that path survives, as in
\eqref{eq:fer-event-decomposition}. At both SNRs,
most of the FER reduction comes from fewer losses of the transmitted
path, with a smaller decrease in final-selection errors. The prefix
changes allowed by Section~\ref{subsec:scl-refinement} thus improve
path retention while preserving the low-weight codewords.
\begin{table}[!t]
\centering
\caption{SCL Error Components for $[\num{1024},512]$, $L=32$}
\label{tab:prefix-diagnostics}
\papertablestyle
\setlength{\tabcolsep}{3pt}
\begin{tabular}{@{}C{32pt}C{48pt}C{48pt}C{48pt}C{48pt}@{}}
\toprule
\multirow{2}{*}[-2.8pt]{\shortstack{$E_{b}/N_{0}$\\(dB)}}
& \multicolumn{2}{c}{Proposed-NP} & \multicolumn{2}{c}{Proposed}\\
\cmidrule(lr){2-3}\cmidrule(lr){4-5}
& Pruning & Selection & Pruning & Selection\\
\midrule
2.25 & \num{1.33e-5} & \num{4.66e-6} & \num{9.08e-6} & \num{3.83e-6}\\
2.50 & \num{1.19e-6} & \num{4.16e-7} & \num{4.75e-7} & \num{3.56e-7}\\
\bottomrule
\end{tabular}
\par\smallskip
\begin{minipage}{248pt}\footnotesize\raggedright
At each SNR, we replayed the noise frames common to both codes
in an earlier snapshot of the FER runs in Fig.~\ref{fig:fer-n1024}.
Each entry is the fraction of those frames with the corresponding
error event in \eqref{eq:fer-event-decomposition}.
\end{minipage}
\end{table}

\section{Conclusion}
\label{sec:conclusion}
The \textit{common supercode} yields necessary and sufficient conditions
for a single coefficient assignment to exclude all codewords of the
specified weights. With the constituent codes fixed, the polar
decomposition gives a minimum-rank linear map of the coefficients
linking the two halves that determines the truncated spectrum.
The map reduces the search to representative assignments while
allowing exact recovery of every feasible assignment. We give
sufficient conditions for assigning the coefficients without
backtracking and use construction feedback to guide both the release
of fixed coefficients and the exchange of information positions.

The resulting designs attain their minimum-distance targets with
competitive FER. After the target is met, the code can still be
adapted to a finite-list decoder by choosing early frozen expressions
that improve path retention while preserving the low-weight spectrum.
\section*{Acknowledgment}
The authors used OpenAI's GPT-6 to help check mathematical rigor in Sections~II--V, polish the language, and prepare the accompanying code repository for open release so that readers can study and reproduce the algorithms and experiments. The authors independently developed the technical content and take responsibility for the entire paper.
\bibliographystyle{IEEEtranDOI}
\bibliography{references}
\end{document}